\documentclass[%
reprint,onecolumn,
aps,
showkeys,
floatfix,
]{revtex4-2}

\usepackage{booktabs}
\usepackage{nicefrac, xfrac}
\usepackage{mathrsfs}
\usepackage{xcolor}
\usepackage{subfigure}
\usepackage{bbm,bm}
\usepackage{booktabs}
\usepackage{cancel}
\usepackage[colorlinks=true, linkcolor=blue, urlcolor=blue, citecolor=blue]{hyperref}
\usepackage{natbib}

\usepackage{amssymb,amsmath,amsthm}
\usepackage{mathtools}
\usepackage{graphicx}
\usepackage{multirow}
\usepackage{soul}
\usepackage{appendix}
\usepackage[normalem]{ulem}

\usepackage{bigints}
\usepackage[rightcaption]{sidecap}
\usepackage{xcolor}
\usepackage[normalem]{ulem}
\usepackage{algorithm}
\usepackage{algpseudocode}
\usepackage{smartdiagram}
\algnewcommand\algorithmicforeach{\textbf{for each}}
\algdef{S}[FOR]{ForEach}[1]{\algorithmicforeach\ #1\ \algorithmicdo}

\numberwithin{equation}{section}
\renewcommand*{\thesection}{\arabic{section}}
\usepackage{enumerate}
\usepackage{multirow}
\renewcommand{\theenumi}{\roman{enumi}}%

\usepackage[colorlinks=true,linkcolor=blue,
urlcolor=blue,citecolor=blue]{hyperref}

\usepackage{mathrsfs}
\begin{document}

\title{Nonmodal stability analysis of the plane Poiseuille flow in a 
multilayer porous-fluid channel }
\author{Supriya Karmakar} 
\author{Priyanka Shukla}\email{priyanka@iitm.ac.in}
\affiliation{Department of Mathematics,\\Indian Institute of Technology Madras, Chennai 600036, India
}

\begin{abstract}
The stability of plane Poiseuille flow of a viscous Newtonian fluid in a multilayer channel with anisotropic porous walls is analyzed using the classical modal analysis, the energy method, and the non-modal analysis. The influence of porous wall parameters such as depth ratio (ratio of porous layer thickness to fluid layer thickness) and anisotropic permeability (in terms of mean-permeability and anisotropy parameter) on flow instability are investigated. The modal stability analysis and energy method show that the anisotropy parameter can stabilize the flow, whereas the depth ratio and mean permeability effects can cause destabilization. Furthermore, the energy budget analysis reveals that the energy production term transfers energy to the disturbance from the base flow through the Reynolds stress, amplifying the kinetic energies in all layers and, hence, enhancing the growth rates of the unstable modes. A significant disparity is observed between the critical Reynolds number obtained through modal analysis and the one determined by the energy method, which confirms the growth of transient perturbation kinetic energy. Specifically, transient growth and response functions are examined to understand the flow response to initial conditions and external excitation (receptivity analysis). It turns out that there is substantial transient growth at a sub-critical Reynolds number. These transient growths are greatly enhanced by increasing the mean permeability or the depth ratio and reducing the anisotropy parameter. The optimal perturbations leading to the maximum transient amplification are determined for various parameters, including counter-rotating vortices (rolls) and streaks. The present analysis shows that the energy amplification develops from the combined effects of (i) the lift-up mechanism and (ii) the two-dimensional Reynolds stress mechanism, resulting in the formation of tilted rolls and streaks with significantly amplified perturbation energy.
\end{abstract}

% \begin{keywords}
%     {Multi-layer flow, porous flow, anisotropic permeability, linear stability analysis, transient growth, transition-to-turbulence.}
% \end{keywords}

\maketitle 

% -------------------------------------------------------
\section{Introduction}
% ---------------------------------------------------------
Fluid flow through systems comprised of multiple layers of porous
media holds significant importance across various disciplines due to its wide range of applications, for instance, geological systems~\citep{berkowitz2002characterizing,voermans_ghisalberti_ivey_2017}, petroleum engineering~\citep{Allen1984collocation},  industry and technology~\citep{blest1999curing}, cross-flow filtration systems~\citep{nassehi1998modelling,kilic2018numerical} and biomedical engineering~\citep{chang1989velocity,majdalani2002two} etc. Numerous efforts have been made in such flow systems to explore pure hydrodynamic instabilities encompassing a broad range of parameters linked with porous layer(s), from laminar to turbulent flows. This attention stems from recognizing the importance of wall permeability and transpiration in such systems ~\citep{Berman1953laminar,Schlichting2000}. Stability analysis provides essential insights into the behavior of fluid systems, thereby facilitating the prediction and control of various phenomena encountered in engineering and natural processes~\citep{Roberts20}.

The classical approach of linear (modal) stability, relying on the normal mode analysis, has been a cornerstone in hydrodynamic stability studies for many decades~\citep{drazin2004hydrodynamic}. In this approach, each mode exhibits an exponential time dependence, consequently, the base flow is regarded as unstable if an eigenvalue of the linear operator is located in the unstable complex half-plane~\citep{schmid2001stability}. The model analysis is useful for comprehending long-term stability behavior of the flow, as it provides insights into the dominant modes that determine the stability of the system.

While modal stability analysis effectively captures the instabilities of certain fluid systems, such as Rayleigh--B\'enard convection and Taylor--Couette flow~\citep{drazin2004hydrodynamic}, several flow problems remain inconclusive. Particularly in the case of Poiseuille and Couette flows, the modal stability analysis fails to match most experimental observations. For instance, the modal stability analysis reveals that the onset of linear instability occurs at $Re=5772$ for the classical plane Poiseuille flow, whereas plane Couette flow and pipe Poiseuille flow remain always linearly stable (i.e., $Re=\infty$). In contrast, experimental studies display that the transition occurs for plane Poiseuille flow at $Re\approx 1000$~\citep{orszag_kells_1980}, for plane Couette flow at $Re \approx 360$~\citep{tillmark_alfredsson_1992} or 370~\citep{Malerud1995measurement}, and for pipe Poiseuille flow at $Re\approx 2000$~\citep{wygnanski_champagne_1973,wygnanski_sokolov_friedman_1975}. 
Owing to limitations in modal analysis, particularly its discrepancies with experimental studies, the non-modal stability theory has emerged to address fluid flow stability~\citep[see e.g.,][]{butler1992_three,gustavsson_1991,reddy_henningson_1993,reddy1993pseudospectra,threfeten1993Hydrodynamic,schmid2001stability}.

Note that, the modal stability analysis identifies instability by determining the eigenvalues of the linear stability operator. These eigenvalues only describe the long-term behavior of perturbations and do not capture their short-term characteristics. In contrast, the non-modal stability analysis concentrates on the transient amplification of disturbances rather than the eigenvalues of the system~\citep{schmid2001stability,schmid2007annual}. Therefore, the non-modal analysis sheds light on the short-term amplification, which may not be detected by modal analysis alone, by examining the sensitivity of the flow to the initial conditions and external excitation such as free-stream turbulence, wall roughness, etc~\citep{schmid2001stability}. Moreover, even if all eigenvalues of the linear system are distinct and remain well within the complex stable half-plane, significant input amplification can occur if the underlying linear operator is non-normal~\citep{schmid2007annual}.

Within the framework of non-modal stability analysis,~\citet{reddy_henningson_1993} investigated the energy growth of plane Poiseuille and Couette flows and discussed the conditions under which there is no energy growth. A similar study by~\citet{threfeten1993Hydrodynamic} addressed the general concepts of non-modal stability, including transient energy growth and the response to external excitation in the form of harmonic force. The transient behavior of pipe Poiseuille flow was presented by~\citet{schmid_henningson_1994}.

\subsection{A note on multi-layer porous-fluid channel flow}
% -------------------------------------------------------------------

The modeling of fluid flow through the porous-fluid multi-layer systems typically involves two approaches: the one-domain approach and the two-domain approach. In the former approach, the porous-fluid system is treated as a pseudo-continuum,  assuming continuously position-dependent properties of the porous materials, 
and the entire flow region is modeled using a single equation.
On the other hand, the latter approach utilizes two sets of governing equations to characterize the fluid and porous regions, which are coupled by appropriate interface conditions.  
Both these approaches have been extensively used to analyze the stability of the flows through porous-fluid systems.  
For instance, using two domain approach, the stability of plane Poiseuille flow (PPF) in a channel containing a fluid layer overlying a porous layer is studied by~\cite{chang1989velocity,chang2006instability,hill_straughan_2008Poiseuille,hill2009instability,liu2008instability}, and the stability of three-layer channel flow confined between two porous layer adjacent to the walls is carried out by 
~\cite{tilton2008linear,karmakar2022stability,Karmakar2023instability,karmakar2024linear}.
Several researchers have also adopted the one-domain approach to study heat transfer, instability, transport phenomena, etc.~\cite[e.g.,][]{arquis1984conditions,Beckermann1988NaturalCI,gobin1998double,Samanta_Goyeau_Ruyer-Quil_2013,Ghosh2019modal,Mukhopadhyay_Cellier_R_Chhay_Ruyer-Quil_2022,valdes2021novel}. Compared to the one-domain approach, the two-domain approach is more effective in capturing the essential information in the bulk- and inter-regions of the system~\citep{VALDESPARADA2013Velocity}.

Apart from modeling approach,
selecting a suitable transport equation for modeling flows through porous media is critical. This consideration involves accounting for experimentally inspired and theoretically verified permeabilities and porosities of the porous medium ~\citep{Bejan}.
~\citet{Beaver1970experiment} were the first to experimentally verify the destabilizing impact of wall permeability in a parallel channel, where one of the walls acted as a porous medium. Later,~\citet{Sparrow1973breakdown} performed an experiment to find the critical Reynolds number for a similar setup. They further validated their findings numerically by analyzing the two-dimensional Darcy flow 
with the Beaver--Joseph interface condition~\citep{beavers1967boundary}. 
Several modal stability  
studies~\citep[e.g.,][]{chang2006instability,hill_straughan_2008Poiseuille,liu2008instability,tilton2006destabilizing,tilton2008linear,hill2009instability,Hill2009advances} employ the two-domain approach by assuming either Darcy's law along with Beaver--Joseph conditions~\citep{beavers1967boundary} or the volume-averaged Navier--Stokes equations~\citep{Whitaker1996} along with the jump shear stress conditions~\citep{OCHOATAPIA1995a,OCHOATAPIA1995b} for modeling. These studies indicate that the wall permeability and the thickness of the porous wall(s) have a destabilizing effect on PPF, as these factors lead to a drop in the critical Reynolds numbers of the least stable modes.

Although the modal stability has been extensively studied, the non-modal stability analysis in the multi-layer porous channel remains largely limited~\citep{quadrio2013effects,Ghosh2019modal}.  
For instance,~\citet{quadrio2013effects} 
performed the non-modal 
analysis for PPF in a confined isotropic porous channel.
They did not consider the inertial term associated to the porous layer from the volume averaged Navier--Stokes equation (VANS), and studied only the transient energy amplification of the initial conditions. 
They suggested that the transient growth may is pronounced in the presence of porous walls.
Note that
the ignored inertial term~\citep{quadrio2013effects}
in the VANS equations is crucial for the modal stability analysis
to capture
the destabilization of short-wave modes in highly permeable porous media~\citep{karmakar2022stability}.
Recently,~\citet{Ghosh2019modal} investigated both the modal and non-modal stability of PPF in a channel overlying a porous layer using the one-domain modeling approach 
They demonstrated that the optimal mechanism of transient amplification follows the exact mechanisms as classical PPF~\citep{schmid2007annual} for a wide range of porosities. 

\subsection{Present work}
The anisotropic variation of permeability is a crucial factor in determining the stability characteristics of the flow. 
To analyze such variations,
~\citet{deepu_dawande_basu_2015,karmakar2022stability,karmakar2024linear,Karmakar2023instability}
examined anisotropic permeability effects on the linear stability of PPF in an anisotropic porous channel.
Furthermore, recently,
\citet{Samanta2022Nonmodal} employed Darcy's law to characterize the flow through anisotropic and inhomogeneous porous media and
% modelled the flow through a porous medium using Darcy's law and 
investigated the modal and nonmodal stability for a similar
porous-fluid-porous
configuration. Although Darcy's law is extensively used in the porous flow community, it can only capture the fluid's creeping motion, i.e. primarily applicable for low permeable porous substrate~\citep{Bejan}.  
In this context, we conduct modal and nonmodal stability analyses of {\it porous-fluid-porous PPF in a confined anisotropic porous channel}, extending the findings of~\citet{karmakar2022stability}.
The main objective
here is to provide physical insights into the effects of anisotropic permeability and the thickness of porous walls on the long- and short-term stability characteristics, which has not been reported to the best of authors' knowledge.
Within the framework of modal stability analysis, the characteristics of instability throughout a broad spectrum of wavenumbers, including the mode of instability and the pattern of secondary flow, are examined. By solving the Orr--Sommerfeld type eigenvalue problem, we identified two distinct modes of instability, each of which manifests itself as a minimum in the neutral stability curve. Thus, in specific parameter regimes, the instability exhibits bi-modal behavior. 
Critical energy Reynolds number curves (below which instability diminishes for any instance) are obtained to gain comprehension regarding the kinetic energy of the perturbations. Our primary emphasis is on the non-modal instability.
%of the current configuration. 
Our analysis offers a concrete understanding of how the aforementioned system parameters impact non-modal behaviors, such as response to external stimuli and the transient energy amplification of initial perturbations. 
The present work aims to provide valuable observations for optimizing the equipment design for dealing with multi-layer porous-fluid systems commonly encountered in diverse industrial and natural processes.  It is worth noting that there is a need for more research that focuses on non-modal stability analysis for the flow through a multi-layer porous-fluid system.
 
The manuscript is organized as follows. The mathematical formulations of the physical problem, and the base state solution are presented in \S\ref{sec:Math_formulation}. The linear stability problem 
is formulated in~\S\ref{sec:linear_stability_analysis}. 
The modal and energy stability analyses are presented in~\S\ref{subsec:modal_stability} and~\S\ref{subsec:energy_stability}. Results and discussions from the non-modal stability analysis are given in~\S\ref{sec:Nonmodal_PFP}. The instability mechanisms and conclusions are detailed  in~\S\ref{sec:growth_mechanism} and~\S\ref{sec:conclusion}.

% --------------------------------------
\section{Mathematical formulation}
\label{sec:Math_formulation}
% ----------------------------------------------------
\begin{figure}
    \centering
    \includegraphics[scale=0.5]{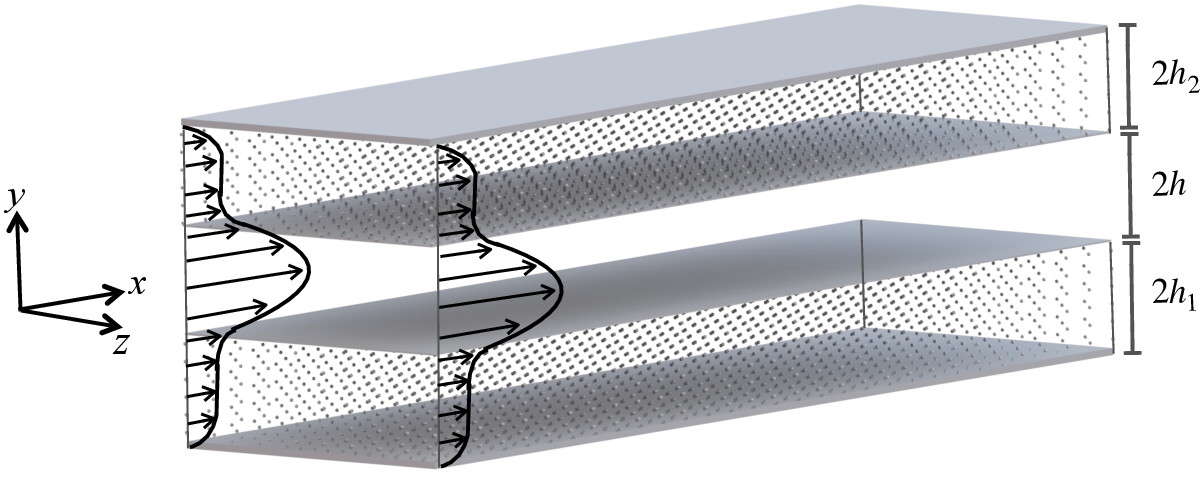}
    \caption{Flow configuration for plane Poiseuille flow of a viscous Newtonian fluid in a confined porous channel.}
    \label{fig:Schematic}
\end{figure}
We consider a plane Poiseuille flow (PPF) of an incompressible Newtonian fluid in a confined porous channel, as shown schematically in figure~\ref{fig:Schematic}. The physical set-up is the same as that of studied by~\cite{karmakar2022stability}. 
The width of fluid layer is $2h$, and the width of lower and upper porous layers are $2 h_1$ and $2 h_2$, respectively.   
A homogeneous longitudinal pressure gradient $dp/dx$ drives the flow.

\subsection{Governing equations}

The Navier--Stokes and continuity equations govern the flow in the fluid layer
% \begin{subequations}
\begin{align}
    \begin{aligned}
        \rho \left( \frac{\partial \boldsymbol{u}}{\partial t}+ \boldsymbol{u} \cdot \nabla \boldsymbol{u} \right)&=-\nabla p+\mu \nabla^2 \boldsymbol{u},\\
        \nabla \cdot {\boldsymbol{u}} &=0,
    \end{aligned}
    \label{eqn:gov_fluid}
\end{align}
% \end{subequations}
where $\boldsymbol{u}=(u,v,w)^{\top},\;p,\; \rho$ and $\mu$ denote the velocity, pressure, density and viscosity of the fluid, respectively. Here the superscript $\top$ denotes the transpose. 
The volume-averaged Navier--Stokes and continuity equations govern the flow through porous layers, which read
\begin{align}
    \begin{aligned}
        \frac{\rho}{\epsilon_j}  \left(\frac{\partial  
       \boldsymbol{u}_j }{\partial t}+\frac{  \boldsymbol{u}_j\boldsymbol{.} \nabla   \boldsymbol{u}_j}{\epsilon_j} \right)   &= - \nabla    p_j^f   + \frac{\mu}{\epsilon_j} \nabla^2    \boldsymbol{u}_j -   \frac{\mu}{\boldsymbol{K}}   \boldsymbol{u}_j,\\
       \nabla \cdot  \boldsymbol{u}_j  &=0,
    \end{aligned}
  \label{eqn:gov_porous}
\end{align}
where $\boldsymbol{u}_j=( u_j,v_j,w_j )^{\top},\; p_j^f$ and $\epsilon_j$ are the superficial volume-averaged velocity, intrinsic volume-averaged pressure, and the porosity, respectively~\citep{Whitaker1986,Whitaker1996}. The subscripts $j=1$ and 2 refer to the quantities in the lower $y\in \left[ -h-2h_1,-h \right]$ and upper $y\in \left[ h,h+2h_2 \right]$ porous layers, respectively. The second order diagonal permeability tensor $\boldsymbol{K}=\text{diag}\left( k_{jx},k_{jy},k_{jz} \right)$ is defined with $k_{jx},\; k_{jy}$ and $k_{jz}$ being the permeabilities along the streamwise ($x$), normal ($y$), and spanwise ($z$)-directions, respectively.

At the lower $y=-h-2h_1$ and upper $y=h+2h_2$ impermeable rigid walls, no-slip and no-penetration
conditions are imposed, respectively, i.e.
\begin{align}
    \boldsymbol{u}_j=0.
    \label{eqn:bd_noslip}
\end{align}
The boundary conditions at
the lower and upper porous-fluid 
interfaces
at $y=\mp h$ are
\begin{subequations}
    \begin{align}
        \boldsymbol{u}_j&=\boldsymbol{u},
        \label{eqn:interface_vel}
        \\
        -p_j^f+\frac{2\mu}{\epsilon_j}\frac{\partial v_j}{\partial y}&=-p+{2 \mu} \frac{\partial v}{\partial y},
        \label{eqn:interface_normal1}
        \\
        -p_j^f+\frac{2\mu}{\epsilon_j}\frac{\partial w_j}{\partial y}&=-p+{2 \mu} \frac{\partial w}{\partial y},
        \label{eqn:interface_normal2}
        \\
        \frac{1}{\epsilon_j} \frac{\partial u_j}{\partial y}- \frac{\partial u}{\partial y}&=\pm \frac{\tau_j}{\sqrt{k_{jx}}} u,
        \label{eqn:interface_shear1}
        \\
        \frac{1}{\epsilon_j} \frac{\partial w_j}{\partial y}- \frac{\partial w}{\partial y}&=\pm \frac{\tau_j}{\sqrt{k_{jz}}} w.
        \label{eqn:interface_shear2}
    \end{align}
    \label{eqn:interface}% 
\end{subequations}
The conditions in~\eqref{eqn:interface} represent the continuity of velocity~\eqref{eqn:interface_vel} and normal stress~\eqref{eqn:interface_normal1}--\eqref{eqn:interface_normal2} and jump shear stress~\eqref{eqn:interface_shear1}--\eqref{eqn:interface_shear2}, respectively.
The positive and negative signs in~\eqref{eqn:interface_shear1} and~\eqref{eqn:interface_shear2} are associated with the lower ($j=1$) interface at $y=-h$ and the upper ($j=2$) interface  at $y=h$, respectively.
Note that shear stress exhibits a jump proportional to the interface coefficient $\tau_j$, which accounts for momentum transfer at the interface and is dependent on the local structure of the fluid-porous interface~\citep{OCHOATAPIA1995a,OCHOATAPIA1995b}.

The condition in which the stress exerted by the free-flowing fluid is completely transmitted to the fluid that saturates the porous matrix is denoted as $\tau_j=0$, so-called Brinkman condition~\citep{Neale1974practical}. On the other hand, when $\tau_j \neq 0$, it signifies the amount of stress the solid matrix can sustain at the interface. The pioneering work of~\citet{OCHOATAPIA1995a,OCHOATAPIA1995b} establishes a practical interval of -1 to 1.5 for the interface coefficient $\tau_j$. This range is consistent with the findings of~\citet{beavers1967boundary} on various porous materials such as foametal A, foametal B, foametal C and aloxite. 
The study by~\citet{carotenuto2015predicting} provided experimental evidences for the absence of stress jump in flow over a porous layer and through a confined porous channel.
In the context of stability analysis of a similar configuration,~\citet{tilton2008linear} demonstrated that variations in the interface coefficients, $\tau_j$, have minimal effect on the linear stability boundaries.
In this study, we assume a state at the interface where there are no jumps ($\tau_j=0$) for the sake of simplicity.

\subsection{Dimensionless governing equations and boundary conditions}

For non-dimensionalization, 
half of the height of the fluid layer, $h$, and mean  fluid layer velocity, $U_m$, %as the 
are used as
reference length and velocity scales, respectively. 
The following dimensionless variables are introduced in (2.1)--(2.4)
\begin{align*}
    \left( x^{\ast},y^{\ast},z^{\ast} \right)=\frac{\left( x,y,z \right)}{h}, \; t^{\ast}=\frac{tU_m}{h}, \; \boldsymbol{u}^{\ast}=\frac{\boldsymbol{u}}{U_m},\; p^{f^\ast}=\frac{p^{f}}{\rho U_m^2},\; \boldsymbol{u}_j^{\ast}=\frac{\boldsymbol{u}_j}{U_m},\; p_j^{f^\ast}=\frac{p_j^{f}}{\rho U_m^2}.
\end{align*}
After dropping the asterisk, the non-dimensional form of~\eqref{eqn:gov_fluid} and~\eqref{eqn:gov_porous} read as
\begin{align}
    \begin{aligned}
  \frac{\partial \boldsymbol{u}}{\partial t}+ \boldsymbol{u} \cdot \nabla \boldsymbol{u}& =-\nabla p+\frac{1}{Re} \nabla^2 \boldsymbol{u}, %\quad &&
  \\
    \nabla \cdot \boldsymbol{u}&=0,
        \\
   \frac{\partial \boldsymbol{u_j}  }{\partial t}+\frac{\boldsymbol{u_j} \cdot \nabla  \boldsymbol{u_j} }{\epsilon_j}   &= - \epsilon_j \nabla   p_j^f   + \frac{1}{Re} \nabla^2  \boldsymbol{u_j} -  \frac{\epsilon_j}{\sigma_j^2 Re} \left( 1,\xi_{j1},\xi_{j2} \right) \odot \boldsymbol{u}_j,
 %  \quad & &
   \\
   \nabla \cdot \boldsymbol{u}_j&=0,
    \end{aligned}
    \label{eqn:gov_fluid_porous_nd}
\end{align}
where $Re\!=\!\rho U_m h/\mu$ and $\sigma_j\!=\!\sqrt{k_{jx}}/h$ are the Reynolds number and dimensionless permeability (referred to the mean permeability), respectively; $\xi_{j1}=k_{jx}/k_{jy}$ and $\xi_{j2}=k_{jx}/k_{jz}$ are the wall-normal and the spanwise anisotropy parameters, respectively. The symbol $\odot$ denotes %the Hadamard product or 
the element-wise product of two vectors. 
%[i.e. $\left(1,\xi_{j1},\xi_{j2}\right) \odot \boldsymbol{u}_j=\left(u_j,\xi_{j1}v_j,\xi_{j2}w_j\right)$]. 
Indices $j=1$ and 2 correspond to the lower $-1-2d_1<y<-1$ and the upper $1<y<1+2d_2$ porous layers, respectively, with $d_j=h_j/h$ being the thickness ratio of the $j^{\text{th}}$ porous layer to the fluid layer. An increase in mean permeability increases streamwise permeability, thereby increasing the total fluid volume in porous layers. Altering $\sigma_j$ %(increasing or decreasing) 
corresponds to varying the channel half-height in the experimental investigations~\citep{beavers1967boundary,Sparrow1973breakdown}, as $\sigma_j$ increases when $h$ decreases and vice versa. As the anisotropy parameters $\xi_{j1}$ and $\xi_{j2}$ decrease (or $k_{jy}$ and $k_{jz}$ increase), normal and spanwise permeability increases. This increase in $k_{jy}$ and $k_{jz}$, for a given streamwise permeability $k_{jx}$, increases the total volume of fluid streaming through the porous layer. In other words, a decrease in anisotropy ($\xi_{j1}$ and/or $\xi_{j2}$), assuming an identical mean permeability, increases the fluid volume in the porous layer. However, in the current study, we limit ourselves to an isotropic porous media along the $y$ and $z$ directions, i.e., $k_{jy}=k_{jz}$~\citep{chen1991onset,karmakar2022stability}.

The dimensionless form of boundary conditions~\eqref{eqn:bd_noslip}--\eqref{eqn:interface} are
\begin{align}
    \boldsymbol{u}_j=0,\quad\mbox{at}\quad  y=\mp \left(1+2d_j\right),   \label{eqn:bd_noslip_nd}
\end{align} 
and 
\begin{align}
    \begin{aligned}
    \boldsymbol{u}_j&=\boldsymbol{u},\\
    -p_j^f+\frac{2}{\epsilon_j Re} \frac{\partial v_j}{\partial y}&=-p+\frac{2}{Re} \frac{\partial v_j}{\partial y},\quad -p_j^f+\frac{2}{\epsilon_j Re} \frac{\partial w_j}{\partial y}=-p+\frac{2}{Re} \frac{\partial w_j}{\partial y},\\
    \frac{\sigma_j}{\epsilon_j} \frac{\partial u_j}{\partial y}- \sigma_j \frac{\partial u}{\partial y}&=\pm \tau_j u, \qquad \qquad \frac{\sigma_j}{\epsilon_j} \frac{\partial w_j}{\partial y}- \sigma_j \frac{\partial w}{\partial y}=\pm \tau_j \sqrt{\xi_{j2}}w,
    \end{aligned}
    \label{eqn:interface_nd}
\end{align}
along the lower and upper porous-fluid interfaces at $y=\mp 1$.

% ---------------------------------------------
\subsection{Base state}
\label{subsec:base_flow}
% -------------------------------------------------------------------

The base flow is plane Poiseuille flow (PPF), which is steady $\left[ \partial \left(\cdot\right)/\partial t=0\right]$, fully developed $\left[ \partial \left(\cdot\right)/\partial x=0\right]$ and uni-directional 
(velocity component is along the  $x$-direction) driven by a uniform longitudinal pressure gradient. 
With these assumptions, the flow variables reduce 
to
\begin{align}
    \begin{aligned}
    \boldsymbol{u}_j(x,y,z,t)&=\boldsymbol{U}_j(y)=\left[ U_j(y),0,0 \right]^{\top}, & p_j^f(x,y,z,t)&=P_j(x),
        \\
        \boldsymbol{u}(x,y,z,t)&=\boldsymbol{U}(y)=\left[ U(y),0,0 \right]^{\top},  & p_j(x,y,z,t)&=P(x),%
    \end{aligned}
\end{align}
and the governing equations~\eqref{eqn:gov_fluid_porous_nd} simplify to
\begin{align}
    \frac{d^2U}{dy^2}=Re \frac{dP}{dx} \quad \text{and}\quad \frac{d^2U_j}{dy^2}-\frac{\epsilon_j}{\sigma_j}U_j=\epsilon_j Re \frac{dP}{dx}.
    \label{eqn:base_flow}
\end{align}
It should be noted that the pressure gradient is constant and identical across all three layer, i.e., ${d P_1}/{dx}={d P_2}/{dx}={d P}/{dx}$. The boundary conditions~\eqref{eqn:bd_noslip_nd} at $y=\mp (1+2d_j)$ reduces to
\begin{align}
    U_j=0,
    \label{eqn:base_noslip}
\end{align}
and the interface conditions~\eqref{eqn:interface_nd} at $y=\mp 1$ becomes
\begin{align}
    U_j=U,\quad P_j=P \quad \text{and} \quad \frac{\sigma_j}{\epsilon_j} \frac{dU_j}{dy}-\sigma_j \frac{dU}{dy}=\pm \tau_j U.
    \label{eqn:base_interface}
\end{align}
The base state equations~\eqref{eqn:base_flow}--\eqref{eqn:base_interface} are analytically solvable for finite porous wall thickness~\citep{tilton2008linear,karmakar2022stability}.
% ~\citep[see Appendix~A of][]{karmakar2022stability}.
It is observed that the velocities $U_j$ at the porous layers contain terms such as $e^{\pm 1/\sigma_j}$, which leads to rounding errors in computations for small $\sigma_j$ values. 
To mitigate such difficulties,
we follow the approach of~\citet{tilton2008linear}, who used semi-infinite porous walls (porous layers are presumed infinitely spanned, $d_j \rightarrow \infty$). 
Under the assumption of semi-infinite porous walls, the base state is
\begin{align}
\begin{aligned}
    U&=C\left[ y^2+\frac{B_1-B_2}{A_1-A_2}y+ \frac{B_1A_2-B_2A_1}{A_1-A_2} \right],
    \\
    U_j&=C\left[ \left( 2\sigma_j^2+1 \pm \frac{B_2-B_1}{A_1-A_2}+  \frac{B_1A_2-B_2A_1}{A_1-A_2} \right) e^{(1\mp y)\frac{\sqrt{\epsilon_j}}{\sigma_j}} -2\sigma_j^2 \right],
\end{aligned}
\label{eqn:base_state_sol}
\end{align}
where
\begin{align}
\begin{aligned}
    C&=\frac{3(A_1-A_2)}{A_1-A_2+3(B_1A_2-B_2A_1)},\\
    A_j&=\pm 1\pm \sigma_j \left( \frac{1}{\sqrt{\epsilon_j}} - \tau_j\right)^{-1},\\
    B_j&=(2\sigma_j^2+1)+2\sigma_j(1+\sigma_j \tau_j) \left(\frac{1}{\sqrt{\epsilon_j}}-{\tau_j} \right).
\end{aligned}
\label{eqn:base_state_coef}
\end{align}
Note that the base state~\eqref{eqn:base_state_sol} is independent of the anisotropy parameter $(\xi_{j1},\xi_{j2})$, and depends on $\epsilon_j,\;\sigma_j,\; \tau_j$ and $d_j$. In~\eqref{eqn:base_state_sol} and~\eqref{eqn:base_state_coef}, 
%the plus or minus sign from 
$\pm$ or $\mp$ is employed first for $j=1$ and then for $j=2$.

% -----------------------------------------------------
\section{Linear stability analysis}
\label{sec:linear_stability_analysis}
% ------------------------------------------------------
\subsection{Disturbance equations}

We decompose each flow variable into a steady base-flow term (denoted by an upper-case letter) and
an unsteady small disturbance term (denoted by a lower-case letter with tilde):
\begin{align}
    \begin{aligned}
        \boldsymbol{u}_j=\boldsymbol{U}_j+\widetilde{\boldsymbol{u}}_j,\; \boldsymbol{u}=\boldsymbol{U}+\widetilde{\boldsymbol{u}},\; p_j^f=P_j+\widetilde{p}_j^f,\; \text{and} \; p=P+\widetilde{p}. 
        \label{eqn:stability_decomposition}
    \end{aligned}
\end{align}
The linearized perturbation equations are derived by substituting the above decomposition~\eqref{eqn:stability_decomposition} into the dimensionless equations~\eqref{eqn:gov_fluid_porous_nd}--\eqref{eqn:interface_nd},
the base-flow terms are subtracted out, and the nonlinear terms in the disturbances are dropped. 
The linear disturbance equations read
\begin{align}
    \begin{aligned}
        \frac{\partial \widetilde{\boldsymbol{u}}}{\partial t} + \left( \boldsymbol{U} \cdot \nabla \widetilde{\boldsymbol{u}} + \widetilde{\boldsymbol{u}} \cdot \nabla \boldsymbol{U} \right)&= -\nabla \widetilde{p} + \frac{1}{Re} \nabla^2 \widetilde{\boldsymbol{u}}, \\
        \nabla \cdot \widetilde{\boldsymbol{u}} &=0,
    \end{aligned}
    \label{eqn:pert_channel}
\end{align}
in the fluid layer spanning $y \in \left( -1,1\right)$ and 
\begin{align}
    \begin{aligned}
    \hspace{-1em}    \frac{\partial \widetilde{\boldsymbol{u}}_j}{\partial t}+\frac{1}{\epsilon_j} \left(  \boldsymbol{U}_j \cdot \nabla \widetilde{\boldsymbol{u}}_j + \widetilde{\boldsymbol{u}}_j \cdot \nabla \boldsymbol{U}_j \right)&=-\epsilon_j \nabla \widetilde{p}_j^f +\frac{1}{Re} \nabla^2 \widetilde{\boldsymbol{u}}_j - \frac{\epsilon_j}{\sigma_j^2 Re} \left(  1,\xi_{j1},\xi_{j2} \right) \odot \widetilde{\boldsymbol{u}}_j,\\
        \nabla \cdot \widetilde{\boldsymbol{u}}_j&=0,%
    \end{aligned}%
    \label{eqn:pert_porous}%
\end{align}
in the porous layers spanning $y \in \left(-1-2d_1,-1 \right) \cup \left( 1,1+2d_2 \right)$, respectively.
The boundary conditions at 
the rigid walls $y=\mp \left( 1+2d_j \right)$,
\begin{align}
    \widetilde{\boldsymbol{u}}_j=0,
    \label{eqn:pert_bd_noslip}
\end{align}
and at the lower ($y=-1$) and upper ($y=1$) interfaces,
\begingroup
\allowdisplaybreaks
\begin{align}
    \begin{aligned}
        \widetilde{\boldsymbol{u}}_j&=\widetilde{\boldsymbol{u}},\\
    -\widetilde{p}_j^f+\frac{2}{\epsilon_j Re} \frac{\partial \widetilde{v}_j}{\partial y}&=-\widetilde{p}+\frac{2}{Re} \frac{\partial \widetilde{v}}{\partial y},\quad & -\widetilde{p}_j^f+\frac{2}{\epsilon_j Re} \frac{\partial \widetilde{w}_j}{\partial y}&=-\widetilde{p}+\frac{2}{Re} \frac{\partial \widetilde{w}}{\partial y},
    \\
    \frac{\sigma_j}{\epsilon_j} \frac{\partial \widetilde{u}_j}{\partial y}- \sigma_j \frac{\partial \widetilde{u}}{\partial y}&=\pm \tau_j \widetilde{u}, \quad &\frac{\sigma_j}{\epsilon_j} \frac{\partial \widetilde{w}_j}{\partial y}- \sigma_j \frac{\partial \widetilde{w}}{\partial y} &=\pm \tau_j \sqrt{\xi_{j2}}\widetilde{w}.
    \end{aligned}
    \label{eqn:pert_bd_interface}
\end{align}
\endgroup

\subsection{Fourier modes}

We simplify~\eqref{eqn:pert_channel}--\eqref{eqn:pert_bd_interface} by 
expressing disturbance variables 
as the superposition of Fourier modes in the $x$- and $z$-directions, 
\begin{align}
\widetilde{\boldsymbol{X}}(x,y,z,t)={\boldsymbol{X}}(y,t) e^{\mathrm{i} ( \alpha x + \beta z)},
\label{eqn:Fourier_mode}
\end{align}
where $\widetilde{\boldsymbol{X}}(x,y,z,t)=\big[ \widetilde{\boldsymbol{u}},\widetilde{\boldsymbol{u}}_j,\widetilde{p},\widetilde{p}_j^f \big]^\top$, ${\boldsymbol{X}}(x,y,z,t)=\big[ \overline{\boldsymbol{u}},\overline{\boldsymbol{u}}_j,\overline{p},\overline{p}_j^f \big]^\top$, and $\alpha$ and $\beta$ are positive numbers representing the streamwise and spanwise wavenumbers, respectively. 
Furthermore, pressure is eliminated by taking the divergence of momentum equation and using the continuity equation, which gives equation for normal velocity.
Next, the normal vorticity 
is introduced as
\begin{align}
  ( \widetilde{\eta},  \widetilde{\eta}_j )=\left( \frac{\partial \widetilde{u}}{\partial z} - \frac{\partial \widetilde{w}}{\partial x}, \frac{\partial \widetilde{u}_j}{\partial z} - \frac{\partial \widetilde{w}_j}{\partial x} \right)
    &=\mathrm{i} \left( \beta \overline{u} -\alpha \overline{w} , \beta \overline{u}_j -\alpha \overline{w}_j\right) e^{\mathrm{i}\left( \alpha x + \beta z \right)} \nonumber \\
    &=\mathrm{i}\left( \overline{\eta} , \overline{\eta}_j \right) e^{\mathrm{i}\left( \alpha x + \beta z \right)},
\end{align}
which provides an equation connecting velocity and vorticity. 
The coupled velocity-vorticity equations in the fluid layer reads
\begin{subequations}
    \begin{align}
    -\left( k^2-D^2\right) \frac{\partial \overline{v}}{\partial t}&=\frac{1}{Re} \left( k^2-D^2 \right)^2\overline{v} + \mathrm{i} \alpha U \left( k^2-D^2\right)\overline{v} + \mathrm{i}\alpha (D^2U)\overline{v},\\
    -\frac{\partial \overline{\eta}}{\partial t}&= \beta (DU) \overline{v} + \mathrm{i}\alpha U \overline{\eta} + \frac{1}{Re} \left( k^2-D^2 \right)\overline{\eta},
\end{align}
\end{subequations}
and in the porous layers equations read
 \begingroup
\allowdisplaybreaks
\begin{subequations}
    \begin{align}
        -\sigma_j^2 \left( k^2-D^2 \right) \frac{\partial \overline{v}_j}{\partial t}&=\frac{\sigma_j^2}{Re}  \left( k^2-D^2 \right)^2\overline{v}_j + \frac{\mathrm{i}\alpha \sigma_j^2}{\epsilon_j} \left[ (D^2U_j) + U_j  \left( k^2-D^2 \right)\right]\overline{v}_j \nonumber \\ 
        & \quad +\frac{\epsilon_j}{Re} \bigg[ \xi_{j1}k^2- \frac{(\alpha^2+\beta^2 \xi_{j2})}{k^2}D^2\bigg] \overline{v}_j + \frac{\epsilon_j}{Re} \frac{\mathrm{i}\alpha \beta}{k^2} \left( 1-\xi_{j2}\right)D \overline{\eta}_j,\\
        -\sigma_j^2 \frac{\partial \overline{\eta}_j}{\partial t}&= \left[\frac{\beta \sigma_j^2}{\epsilon_j}(DU_j) + \frac{\epsilon_j}{Re} \frac{\mathrm{i} \alpha \beta}{k^2} \left( 1-\xi_{j2} \right)\right]\overline{v}_j + \bigg[ \frac{\sigma_j^2}{Re} \left( k^2 - D^2\right) \nonumber \\ 
        &\quad  + \frac{\mathrm{i}\alpha \sigma_j^2}{\epsilon_j}U_j + \frac{\epsilon_j}{Re} \frac{\beta^2+\alpha^2 \xi_{j2}}{k^2}\bigg] \overline{\eta}_j,
    \end{align}
    \label{eqn:LSA_eqns}
\end{subequations}
\endgroup
where $D \equiv \partial / \partial y,\;D^{p} \equiv \partial^{p}/\partial y^{p}$, and $k^2=\alpha^2+\beta^2$. 
The boundary conditions are
\begin{align}
\overline{v}_j=D\overline{v}_j=\overline{\eta}_j&=0,\quad \mbox{at}\quad y=\mp (1+2d_j),
\label{eqn:LSA_BCs}
\end{align}
and along the porous--fluid interfaces $y=\mp 1$,
\begingroup
\allowdisplaybreaks
\begin{align}
\begin{aligned}
  \overline{v}_j &=\overline{v}&, %\quad 
    \\
    D\overline{v}_j&=D\overline{v}, %\quad 
    \\
    \overline{\eta}_j&=\overline{\eta},\\
    \frac{\sigma_j}{\epsilon_j}D\overline{\eta}_j - \sigma_j D\overline{\eta}&= \pm \frac{\tau_j}{k^2} \left[ \mathrm{i}\alpha \beta \left( 1 - \sqrt{\xi_{j2}}\right) D\overline{v} + \left( \beta^2+\alpha^2 \sqrt{\xi_{j2}} \right) \overline{\eta}\right],\\
    \frac{\sigma_j}{\epsilon_j} D^2\overline{v}_j - \sigma_j D^2 \overline{v}&=\pm \frac{\tau_j}{k^2} \left[ \left( \alpha^2 + \beta^2 \sqrt{\xi_{j2}}\right)D \overline{v} - \mathrm{i}\alpha \beta \left( 1-\sqrt{\xi_{j2}} \right)\overline{\eta}\right],\\
    \sigma_j^2\left( 1-\frac{1}{\epsilon_j} \right)\frac{\partial^2 \overline{v}}{\partial t \partial y}&=-\frac{\sigma_j^2}{\epsilon_j Re} D^3 \overline{v}_j + \mathrm{i}\alpha \sigma_j^2 \left( DU-\frac{DU_j}{\epsilon_j^2} \right)\overline{v}-\mathrm{i}\alpha \sigma_j^2 \left( U-\frac{U_j}{\epsilon_j^2} \right)D \overline{v} \\
    & \quad +\frac{1}{Re}\frac{\alpha^2 + \beta^2 \xi_{j2}}{k^2} D \overline{v} + \frac{\sigma_j^2 \beta^2}{Re} \left(  1-\frac{1}{\epsilon_j}\right) D \overline{v} + \frac{\sigma_j^2}{Re} D^3\overline{v} \\
    & \qquad + \frac{2k^2 \sigma_j^2}{Re}\left( \frac{1}{\epsilon_j}-1 \right)D \overline{v} + \frac{\mathrm{i}\alpha \beta \sigma_j^2}{Re} \left( 1-\frac{1}{\epsilon_j} \right) \overline{\eta} - \frac{1}{Re}\frac{\mathrm{i}\alpha \beta (1-\xi_{j2})}{k^2} \overline{\eta}. 
\end{aligned}
\label{eqn:LSA_InterfaceCond}
\end{align}
\endgroup

The above system of equations~\eqref{eqn:LSA_eqns}--\eqref{eqn:LSA_InterfaceCond} represent the Orr-Sommerfeld and Squire system, which can be recast into a matrix initial value problem, 
\begin{align}
    \frac{\partial {\boldsymbol{q}}}{\partial t}=-\mathrm{i}\mathscr{L} {\boldsymbol{q}},\qquad {\boldsymbol{q}}_0={\boldsymbol{q}}(y,t=0)
    \label{eqn:initial_value_problem}
\end{align}
where $\mathscr{L}$ is a $6 \times 6$ block-diagonal matrix operator, 
${\boldsymbol{q}}_0$ is the temporal initial condition and $\boldsymbol{q} 
\equiv {\boldsymbol{q}(y,t)}=
\begin{bmatrix}
    \overline{v}_2& \overline{\eta}_2 &\overline{v} &\overline{\eta}&\overline{v}_1&\overline{\eta}_1 
\end{bmatrix}^{\top}$ is the column vector of unknown. 
The solution of~\eqref{eqn:initial_value_problem} is expressed as 
\begin{equation}
    \boldsymbol{q}(y,t)=\boldsymbol{q}_0 \exp{\left( -\mathrm{i} \mathscr{L}t\right)}.
   \label{eqn:Solinitial_value_problem}  
\end{equation}
The velocity-vorticity formulation~\eqref{eqn:initial_value_problem} can be transformed for use with primitive fluctuations as
\begin{align}
    \begin{pmatrix}
        \overline{u} \\ \overline{v} \\ \overline{w}
    \end{pmatrix} = \frac{1}{k^2} 
    \begin{pmatrix}
        \mathrm{i} \alpha D & \beta \\
        k^2   & 0\\
        \mathrm{i} \beta D  &  \alpha
    \end{pmatrix}
    \begin{pmatrix}
        \overline{v} \\ \overline{\eta} 
    \end{pmatrix}
    \quad \textnormal{and} \quad 
    \begin{pmatrix}
        \overline{u}_j \\ \overline{v}_j \\ \overline{w}_j
    \end{pmatrix} = \frac{1}{k^2} 
    \begin{pmatrix}
        \mathrm{i} \alpha D & \beta \\
        k^2   & 0\\
        \mathrm{i} \beta D  & \alpha
    \end{pmatrix}
    \begin{pmatrix}
        \overline{v}_j \\ \overline{\eta}_j
    \end{pmatrix},
\end{align}
for the fluid layer and porous layers, respectively.

\subsection{Modal stability analysis: Eigenvalue problem}

In modal stability analysis, the solution of linear system~\eqref{eqn:initial_value_problem}
is assumed in the separable form
\[\boldsymbol{q}(y,t)=\widehat{\boldsymbol{q}}(y) \exp{( - \mathrm{i} \omega t )},\] 
which reduces the Orr--Sommerfeld and Squire system~\eqref{eqn:initial_value_problem} into
an eigenvalue problem
\begin{align}
    %\mathscr{L} \widehat{\boldsymbol{q}}=\omega \widehat{\boldsymbol{q}}, \quad \mbox{or}\quad
  %  \mathscr{A} 
    \mathscr{L}
    \widehat{\boldsymbol{q}}=\omega %\mathscr{B} 
    \widehat{\boldsymbol{q}}.\label{eqn:modal_eqn}
\end{align}
with complex wave frequency $\omega=\omega_r+\mathrm{i} \omega_i$ and 
$\widehat{\boldsymbol{q}}=\begin{bmatrix}
\widehat{v}_2 & \widehat{\eta}_2 & \widehat{v} & \widehat{\eta} & \widehat{v}_1 &\widehat{\eta}_1 
\end{bmatrix}^\top$
being the eigenvalue and eigenfunction, respectively. 
The sign of the imaginary part $\omega_i$
of $\omega$
determines the growth (or decay) rate.  
If $\omega_i>0$, the amplitude grows exponentially with time, and if $\omega_i<0$, it decays. 
The above eigenvalue problem is solved using the Chebyshev-spectral collocation method, see Ref.~\cite{karmakar2022stability}
for more detail.

The two porous layers are assumed to be made of identical materials and have the same height. Consequently, the subscript 
$j$ can be omitted from the flow parameters associated with the porous layers.
Furthermore, previous studies~\citep{tilton2008linear,karmakar2022stability} have established that the impact of porosity and the interface coefficient 
$\tau$ on stability is minimal.
Therefore, in this investigation, we set the values of $\epsilon$ and $\tau$ as 0.6 and 0, respectively. Additionally, for simplicity, we assume that the permeabilities $\xi_1, \xi_2$ along the $y$- and $z$-directions are identical, i.e. $\xi_1=\xi_2=\xi$.

It is worth to mention that in the present modal analysis, 
the Forchheimer drag term $-\frac{\mu}{\boldsymbol{K}}\boldsymbol{F} \cdot \boldsymbol{u}$ in the VANS equation is not considered~\citep{Whitaker1996}.
~\citet{tilton2008linear} proposed a rationale
of neglecting the Forchheimer drag term (as the Forchheimer coefficient $ F \ll 1 $) in comparison to the Darcy drag term.
Their analysis confirmed that neglecting this term is justified for porous media composed of packed granular material by using a modified Ergun equation~\citep{Whitaker1996} for a similar \textit{porous-fluid-porous} configuration. 
In the present study, we repeated this approach by varying the anisotropy parameter and depth ratios, and found that within the parameter space under consideration, the Forchheimer drag term remains negligible.

% -------------------------------------------------------------
\section{Modal (long-term) stability results}
\label{subsec:modal_stability}
% -------------------------------------------------------------

~\citet{karmakar2022stability} have conducted a detailed investigation into the modal stability of PPF using the same setup. Thus, here we present a brief overview of their results.  
They showed that Squire modes remain damped and that Squire's theorem~\citep{Squire1933} 
is valid in porous-fluid-porous channels.
Henceforth, the Reynolds number for modal instability arising from two-dimensional infinitesimal disturbances is less than that of three-dimensional disturbances.
Their investigation unveiled the presence of fluid, porous, and center modes. While fluid modes have destabilizing effects, the other modes exhibit stability across a wide range of %considered 
parameters. 
Their findings indicate that the system tends to stabilize through various means, including reducing the mean permeability $\sigma$, increasing the anisotropy parameter $\xi$, or decreasing the depth ratio $d$. 
It is worth noting that their study is limited to short-wave instability exclusively; however, the present work encompasses both the short-wave and long-wave instabilities. 
To summarize modal behavior, we present below bi-modal stability curves, spectrum, and eigenfunctions for typical sets of parameters.  
The present numerical results for linear (modal) stability have been validated by comparison with published data for classical PPF~\citep{orszag_1971}, where the critical parameters are $(Re_c, \alpha_c) = (3848.17, 1.02)$. For the case of anisotropic permeability $(\sigma, \xi) = (0.0002, 10)$ and depth ratio of $d = 1$, the critical Reynolds number $Re_c$ is found to be 3843.92, occurring at a wavenumber of $\alpha_c = 1.02$. These results demonstrate excellent agreement between the present results with nearly impermeable walls and those of classical plane Poiseuille flow.

\begin{figure}
    \centering
    \includegraphics[scale=0.43]{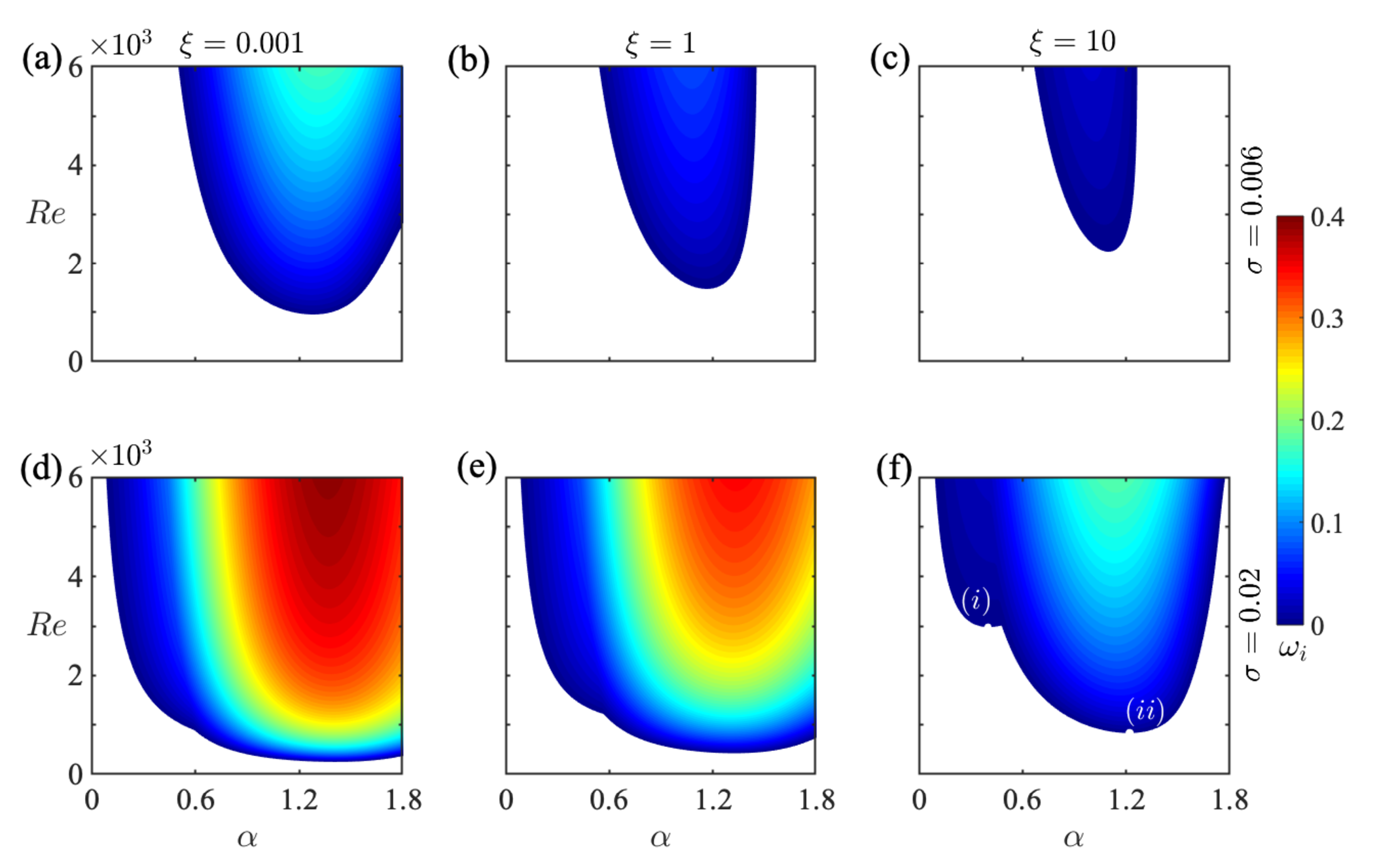}
    \caption{\small The contour of constant growth rate ($\omega_i$) in jet-scale for $\sigma=0.006$ (1st row, panels a--c) and $\sigma=0.02$ (2nd row, panels d--f), and for $\xi=0.001$ (1st column, panels a, d), $\xi=1$ (2nd column, panels b, e), and $\xi=10$ (3rd column, panels c, f)
    in the $(\alpha, Re)$-plane when $d=1$. 
    The flow is stable ($\omega_i<0$) outside the shaded region. 
    }
    \label{fig:temporal_growth}
\end{figure}
 
\begin{figure}
    \centering
    \includegraphics[scale=0.25]{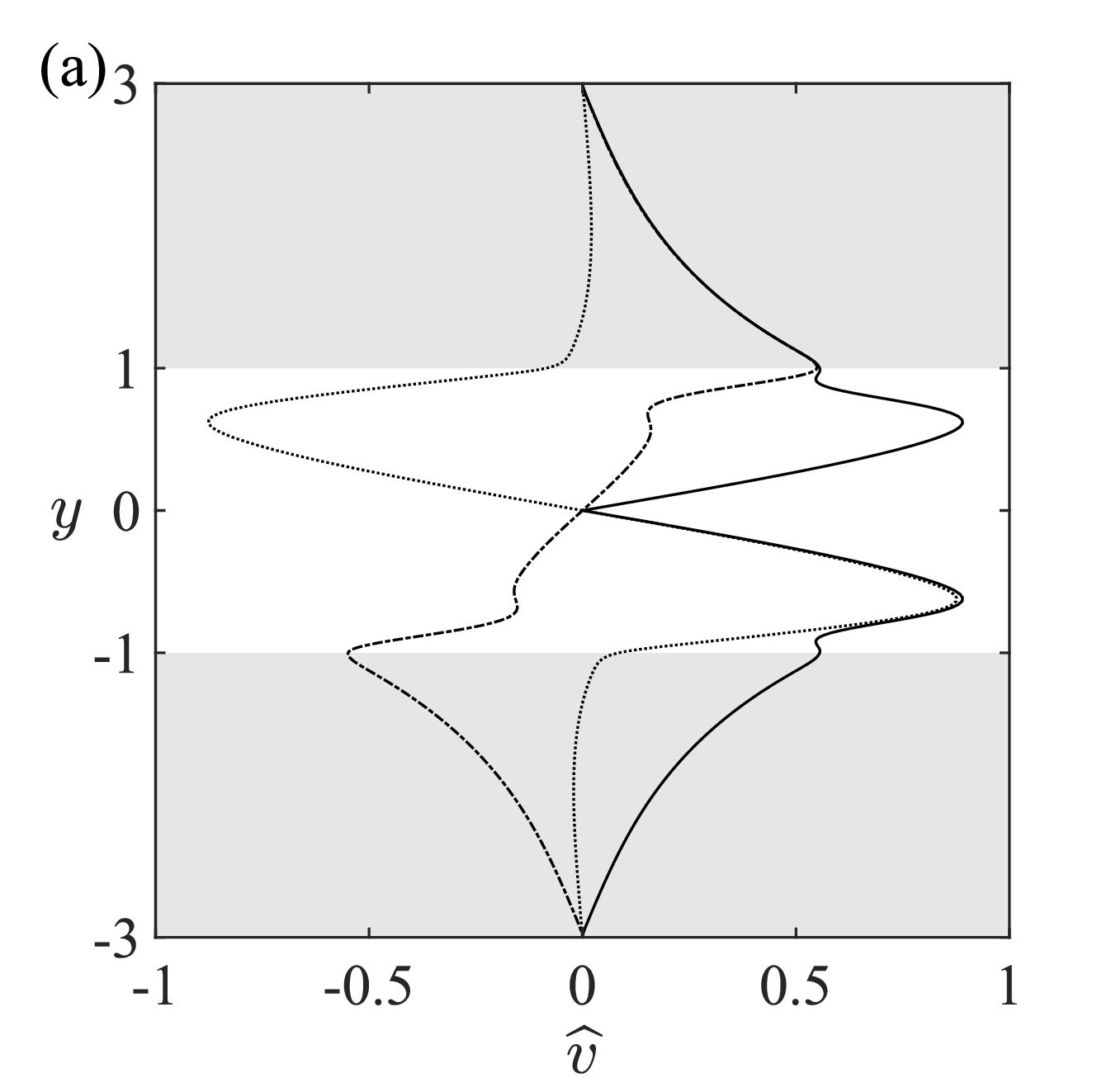}
    \includegraphics[scale=0.25]{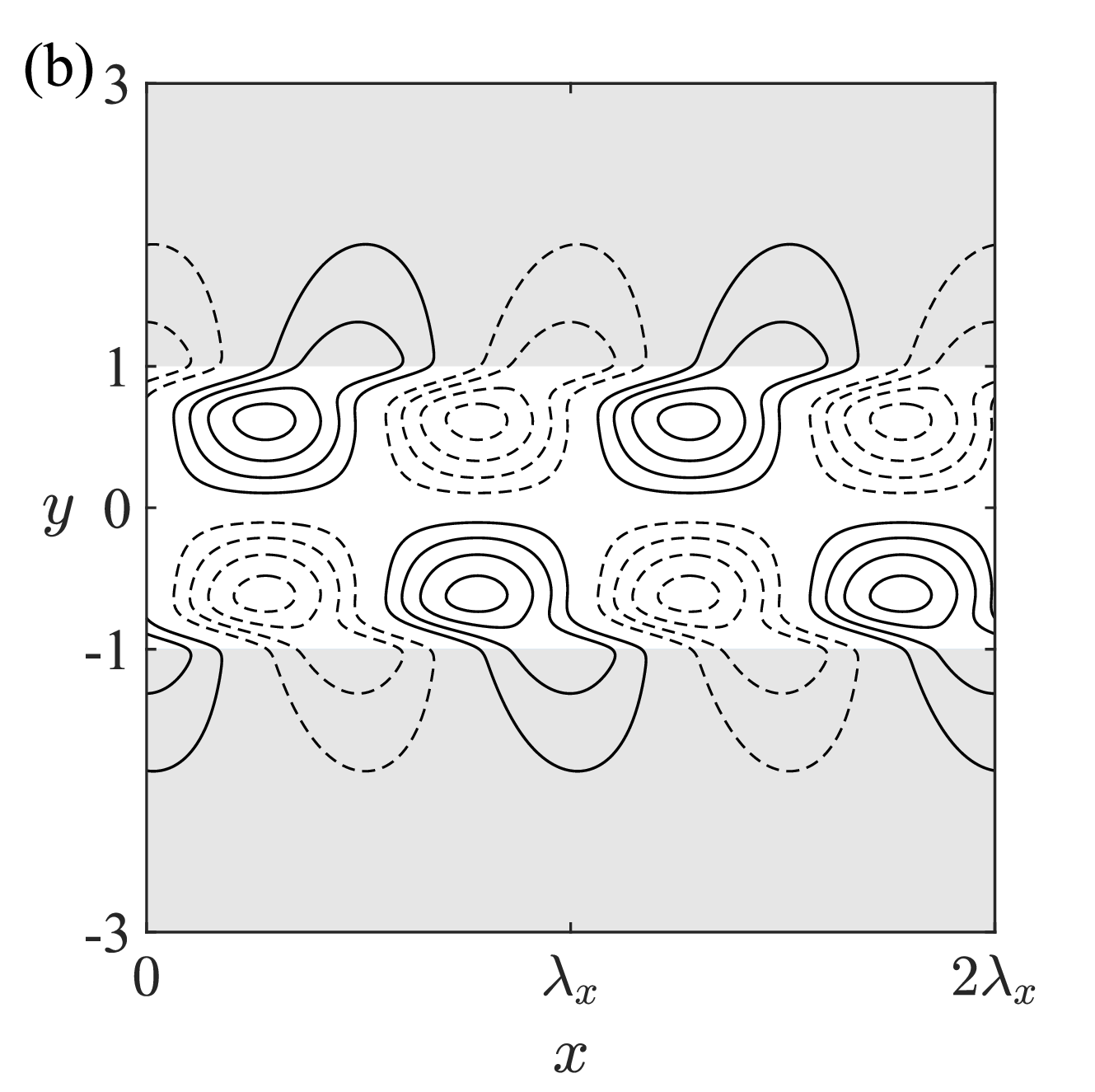}\\
    \includegraphics[scale=0.25]{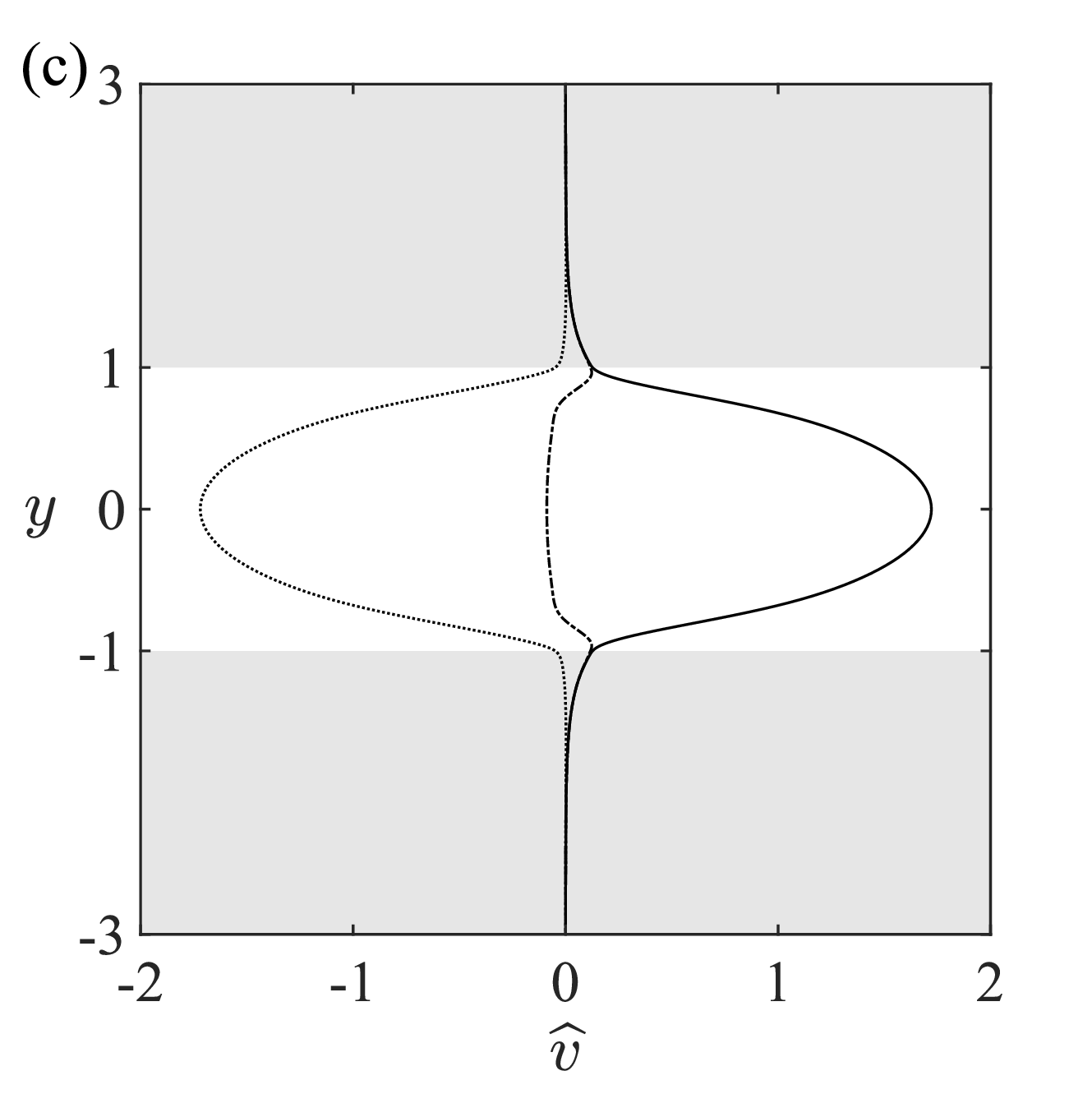}
    \includegraphics[scale=0.25]{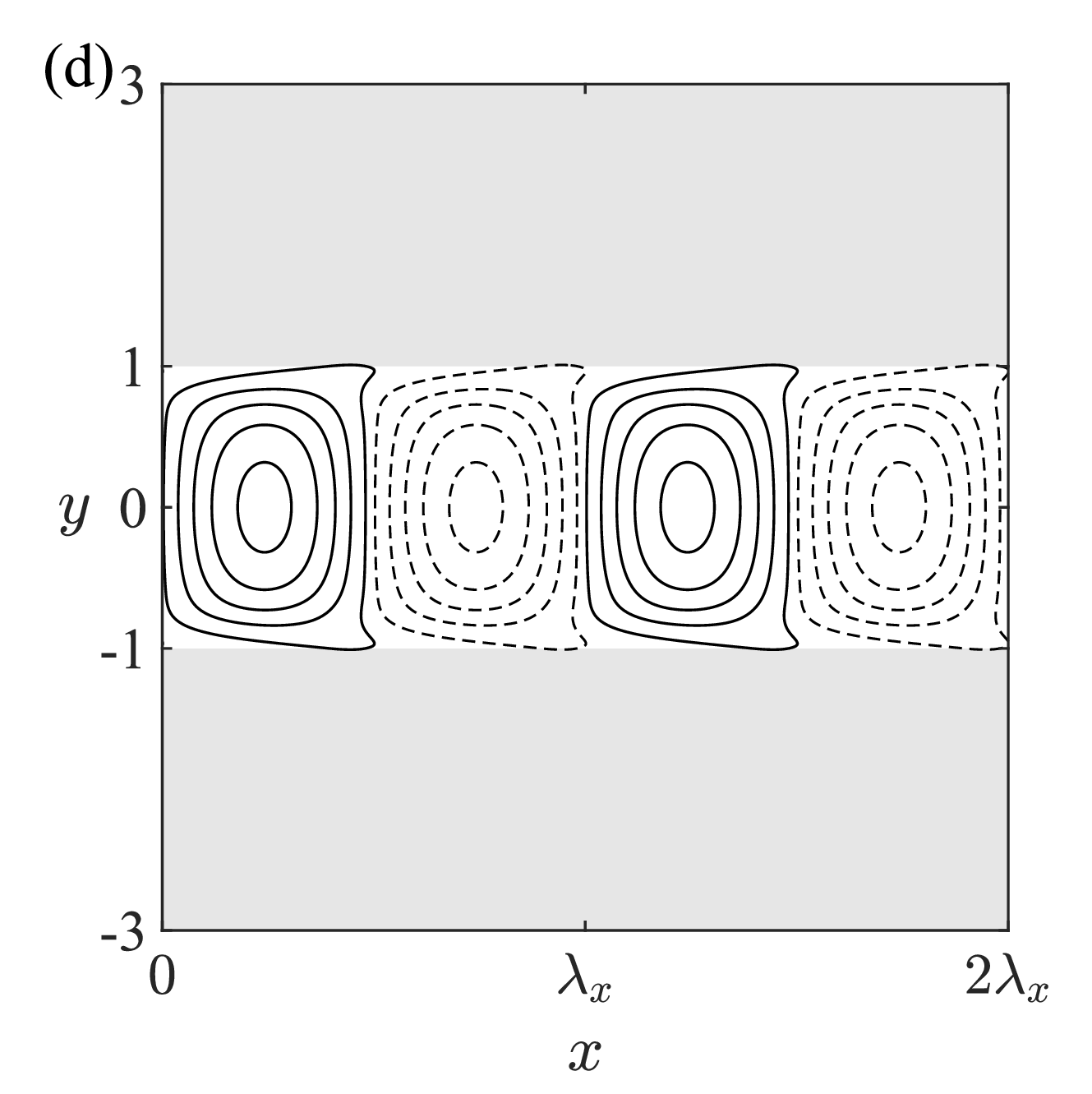}
    \caption{\small The eigenfunctions $\widehat{\boldsymbol{v}}$ and the corresponding patterns of the stream function $\widetilde{\psi}$ corresponding to the least stable mode at critical parameters $(i)$ and $(ii)$ marked in figure~\ref{fig:temporal_growth}
    with $(\alpha,Re)=$ (a) (0.4, 2977.1), (b) (1.22, 833.45), respectively.
    }
    \label{fig:bimodal_stream}
\end{figure}

The contours of constant growth rate in the $(\alpha,Re)$-plane are depicted in figure~\ref{fig:temporal_growth} for three distinct values of the anisotropy parameter $\xi$ of 0.001, 1, and 10 in columns, as well as 
%two distinct values of the mean permeability $\sigma$, specifically 0.006 and 0.02 (column-wise). 
two mean permeability $\sigma$ values of 0.006 and 0.02 in rows. The exponential growth region experiences a substantial expansion as $\sigma$ increases and $\xi$ decreases, respectively. In addition, the growth rate of the least stable mode is enhanced due to the above alteration on the anisotropic permeability. It is seen from the first row that for small mean permeability $\sigma=0.006$, the neutral stability curves ($\omega_i=0$) are uni-modal for all $\xi$'s.  %Observe that 
As $\sigma$ approaches $\sigma=0.02$, in the second row, the unstable regions with the uni-modal shape of neutral curves nearly enclose the $(\alpha,Re)$--plane for small values of anisotropy parameters $\xi \le 1$ (panels d and e). In contrast, 
%At this $\sigma$, 
at $\xi=10$, 
%in the case where the relative magnitude of cross-stream permeability is minimal ($\xi=10$), 
it is observed that the neutral stability curve exhibits two minima 
at intermediate- and short-wavenumber regimes, indicating the occurrence of bi-modal instability (see panel f).  
These two minima, corresponding to each lobe, are marked by `$(i)$' and `$(ii)$' in panel (f), occur at $(\alpha, Re)=(0.4, 2977.1)$ and (1.22, 833.45), respectively. 
To understand instability behavior, eigenfunctions and streamfunctions are shown next in figure~\ref{fig:bimodal_stream}.

Let us examine the normal velocity eigenfunction $\widehat{\boldsymbol{v}}$ corresponding to
the least stable mode
with ($Re, \alpha$) 
being the
points $(i)$ and $(ii)$ of figure~\ref{fig:temporal_growth}(f). 
It is seen in figures~\ref{fig:bimodal_stream}(a) and~\ref{fig:bimodal_stream}(c) that the amplitude of normal disturbance velocity varies significantly in the fluid layer at both critical points $(i)$ and $(ii)$, which classify these modes as fluid modes. The 
velocity eigenfunctions are markedly different in both critical points.  
Figures~\ref{fig:bimodal_stream}(b) and~\ref{fig:bimodal_stream}(d) display the contours of the stream function $\widetilde{\psi}$ 
corresponding to the modes shown on panels (a) and (c) of figure~\ref{fig:bimodal_stream}, respectively, with
the solid line representing the positive  stream function
and dashed negative stream function. 
The core of solid and dashed curves  corresponds to the maximum and minimum values of the stream function, respectively. 
In panels (b) and (d), the $x$-axis has been scaled using twice the streamwise
wavelength $\lambda_x=2\pi/\alpha$.
Therefore, to see the true inclination of vortices, these figures should be stretched by a factor of $4 \pi/\alpha$ in the streamwise direction.
The wavelengths $\lambda_x$ for the first and second fluid modes are 
$5.15$ and 15.71, respectively. 
The stream function distribution exhibits a pattern of periodic bands consisting of small and large vortices that are uniformly formed and arranged to align with the flow direction. 
It should be noted that in fluid mode 1, the contours of the stream function are confined solely to the fluid layer. However, in the case of fluid mode 2, the stream function contours extend into the porous walls. %In both fluid modes, 
Note that the formation of a vortex core occurs only within the fluid layer, and this generation of vortices has a significant influence on flow instability.

It has also been confirmed that for a given $(\sigma,\xi)$, there is a specific cutoff depth ratio ($d_c$) beyond which the critical Reynolds number remains constant. Hence, for a fixed constant anisotropic permeability, increasing the relative thickness of the porous layers beyond the critical depth ratio $d_c$ does not influence the linear stability results.

% --------------------------------------------
% ---------------------------------------
\section{Energy stability method, and energy Reynolds number}
\label{subsec:energy_stability}
% ----------------------------------------

\subsection{Concept of energy stability method}

In contrast to modal stability analysis, 
the energy stability method is based on the conditions for no energy growth of
 % {\color{red}disturbances of arbitrary amplitudes} 
 infinitesimal disturbances, specifically ensuring no growth of disturbance kinetic energy.
% . In this sub-section, we are concerned with no growth of disturbance kinetic energy. 
% Energy methods based on the conditions for no energy growth for disturbances of arbitrary amplitudes. In this sub-section, we are concerned with the circumstance in which there is no disturbance kinetic energy growth.
In other words, disturbance kinetic energy $\mathcal{E}_V$ monotonically decreases with time, i.e.~$d\mathcal{E}_V/dt<0$,  implying that the disturbance is monotonically stable. 
%or a monotonically stable $d\mathcal{E}_V/dt<0$ solution, where $\mathcal{E}_V$ is the disturbance kinetic energy 
%{\color{cyan}(see appendix~\ref{appn:norm})}. 
Using the general properties of the inner product and the fact that the quantity $\| \boldsymbol{q}(y,t)\|^2$ is proportional to the disturbance kinetic energy $\mathcal{E}_V$, 
% (see appendix~\ref{appn:TransientGrowth_and_optimal})
an equation for the rate of change of energy can be derived, which reads
%we find:
\begin{align}
    \frac{d \mathcal{E}_V}{dt}=\frac{1}{2k^2} \frac{d \| \boldsymbol{q}\|^2}{dt}= \frac{1}{k^2} \operatorname{Im}{\langle \mathscr{L}\boldsymbol{q},\boldsymbol{q}  \rangle},
    \label{eqn:rate_energy}
\end{align}
where $\Vert \cdot \Vert$ is energy norm defined in appendix~\ref{appn:TransientGrowth_and_optimal}.

If the right-hand side of~\eqref{eqn:rate_energy} is negative, there is no energy growth. 
The inner product $\langle \mathscr{L}\boldsymbol{q},\boldsymbol{q}  \rangle$ on the r.h.s. of~\eqref{eqn:rate_energy} 
%can be equivalently 
%expressed %in terms of 
%as a imaginary part of
is equivalent to  
numerical range $W(\mathscr{L})$ of the linear operator $\mathscr{L}$,   
defined as
\begin{align}
    W(\mathscr{L})=\left\lbrace
    \langle \mathscr{L} \boldsymbol{q} , \boldsymbol{q} \rangle \,\,:\,\, \|\boldsymbol{q}\|=1\right\rbrace. 
       % \boldsymbol{x}^{\ast} \mathbb{L} \boldsymbol{x}\:\textnormal{with}\:\: \boldsymbol{x} \in \mathbb{C}^n \, \textnormal{and}\, \boldsymbol{x}^{\ast}\boldsymbol{x}=1 \right\rbrace,
\end{align}
%on the complex vector space. 
If the numerical range of $\mathscr{L}$ contains an element with a positive imaginary part, 
it will lead to a positive rate of disturbance kinetic energy and hence energy growth. 
Note that the numerical range
contains the convex hull of the eigenvalues of 
the linear operator, and if the operator is normal, then the
numerical range is %equal to
exactly same as
the convex hull of its eigenvalues~\citep{horn_johnson_1991}. 
Associated with the numerical range, one can define the numerical abscissa (also known as numerical radius) as the supremum of the imaginary part of the numerical range, i.e.
\begin{align}
    \varphi \left(\mathscr{L}\right)= \sup \{ \operatorname{Im}   \langle \mathscr{L} \boldsymbol{q} , \boldsymbol{q} \rangle  \,\,:\,\, \|\boldsymbol{q}\|=1\}.
    \label{eqn:numerical_abscissa}
\end{align}
%which is defined as
From~\eqref{eqn:rate_energy} and~\eqref{eqn:numerical_abscissa}, 
the numerical abscissa is proportional to the rate at which perturbation kinetic energy grows or decays~\citep{reddy_henningson_1993,Threfeten2005spectra}. 
To what follows, there is energy-growth (no energy-growth) if and only if the numerical range 
of the linear operator $\mathscr{L}$ lies in the upper (lower) half-plane or, if and only if the numerical abscissa is positive (negative).

Another way ensure no energy growth is to examine the eigenvalues of anti-symmetric part $\left( \mathscr{L}-\mathscr{L}^{\ast} \right)/2$ of $\mathscr{L}$, where $\mathscr{L}^{\ast}$ is the adjoint operator associated with $\mathscr{L}$. If all eigenvalues are situated in the lower half-plane, energy growth will not occur~\citep{reddy_henningson_1993}. 
By employing these definitions, one can determine the {\it critical energy Reynolds number}, denoted as $Re_e$. 
The critical energy Reynolds number signifies the threshold at which the kinetic energy of disturbances experiences a monotonic decay. 
Numerous articles discuss an additional approach to calculate the critical energy Reynolds number by employing the variational method to solve the associated optimization problem~\citep{reddy_henningson_1993,schmid2001stability,nouar_bottaro_brancher_2007}. 
This classical approach, frequently cited in the literature, determines the critical energy Reynolds number as the minimum value at which perturbation kinetic energy grows by solving the optimization problem~\cite{reddy_henningson_1993}.

In this work, the energy Reynolds number $Re_e$ is computed using the numerical abscissa. 
Specifically, for each fixed pair of wavenumber $(\alpha, \beta)$, with other parameters being fixed, numerical abscissa $\varphi$ is calculated for a wide range of  Reynolds number $Re$.  
Then, a lowest Reynolds number $Re=Re_\varphi$ is identified such that $\varphi>0$ for each pair $(\alpha, \beta)$. 
We list all values of $Re_\varphi$ for each pair of $(\alpha, \beta)$. The minimum of this list is the critical energy Reynolds number $Re_e$, i.e.
$Re_e=\min_{\alpha,\beta} Re_\varphi$, where $Re_\varphi$ is a function of $\alpha$ and $\beta$. 

% ------------------------------------------------
\subsection{Discrepancies in the modal and energy stability analysis}
\label{subsec:discrepancies}
% -----------------------------------------------------------------------------

\begin{figure}
    \centering
    \includegraphics[scale=0.3]{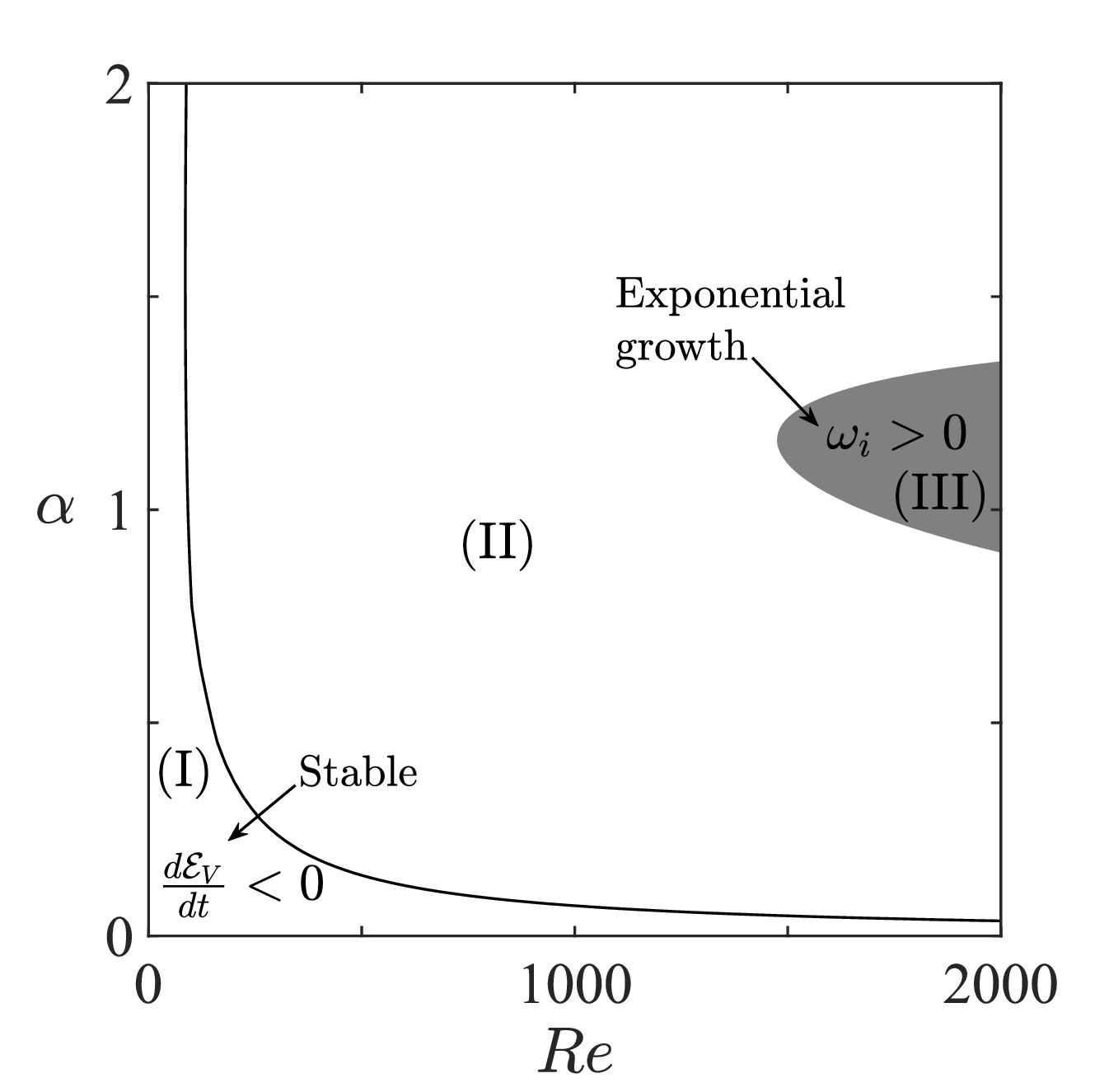}
    \caption{\small Stability boundary for $\beta=0,\; d=1,\; \sigma=0.006$ and $\xi=1$ in the $(Re,\alpha)$--plane. The exponential growth zone and the boundary of the stable zone are represented, respectively, by the shaded region and the solid line.}
    \label{fig:temporal}
\end{figure}

The critical Reynolds number for the classical plane Poiseuille flow, corresponding to the centerline velocity,
is a well-established value of 5772.22 when determined through linear stability analysis; however, it reduces to 49.6 when evaluated using the energy method. 
This clearly highlights the disparities between modal stability analysis and energy stability analysis for the case of classical plane Poiseuille flow. 
In the present set-up involving PPF within a porous-fluid-porous channel, we encounter a similar disparity as in the case of %modal stability versus energy stability analyses of 
classical PPF. %as explain below. 

Figure~\ref{fig:temporal} illustrates the neutral stability curve and
the zero contour of the numerical abscissa $\varphi$ 
in $(Re, \alpha)$-plane, 
for $\beta=0$, $\sigma=0.006$, $\xi=1$, and $d=1$. 
%The flow is stable (unstable) outside (inside shaded region) the zero growth rate curve. 
While the neutral stability curve provides the boundary between exponentially (linearly) stable and unstable regions, vanishing numerical abscissa $\varphi=0$ separates the monotonically stable and unstable regions.  
These boundaries divides the $(Re,\alpha)$-plane into three instability zones: I, II, and III.
In zone I, the numerical abscissa $\varphi$ and the growth rate of the least stable mode are both negative;
thus, the flow is monotonically and exponentially stable.
In zone II, numerical abscissa is positive $\varphi>0$, but the growth rate $\omega_i$ is negative, leading to 
a monotonically unstable flow 
that is nonetheless exponentially stable.
Therefore, transient growth of the disturbances occurs in zone II.    
In zone III, both the growth rate of the least stable mode and numerical abscissa are positive, i.e., $\omega_i>0, \,\varphi>0$, resulting in the transient and exponential growth of the perturbations. In this region, the flow admits long- and short-term growth of the perturbations. 
The above observation confirms that transient growth leads to short-time energy growth of the perturbations.

\begin{table}
\centering
 \begin{tabular}{ c@{\hskip 0.1in} |ccc @{\hskip 0.3in} cc @{\hskip 0.2in} ccc@{\hskip 0.3in}}
 \hline
 \multirow{2}{2em}{}&\multicolumn{3}{c}{$(d,\xi)=(1,1)$}& \multicolumn{2}{c}{$(d,\sigma)=(1,0.02)$} & \multicolumn{3}{c}{$(\sigma,\xi)=(0.02,1)$} \vspace{0.1 cm} \\
 \cline{2-9}
    & $\sigma=0.006$ & $\sigma=0.01$ & $\sigma=0.02$ & $\xi=0.001$ & $\xi=10$ & $d=1/8$  & $d=2$ & $d=4$ \vspace{0.1 cm} \\[0.5em]
    \cline{1-9}
    % \hline
    \vspace{-0.5 cm}\\
    $Re_c$ & 1474.93 & 898.20 & 424.06 & 249.46  & 833.41 & 979.85 & 421.12 & 421.13\\
    % $Re_e(\alpha_e,0)$ & 86.12  & 85.75  & 84.07  & 78.21  &  85.47 &  85.40 & 84.06 &  84.06 \\[0.5em]
    $Re_e$ & 33.46 & 33.40 & 33.15 & 15.34 & 34.36 & 33.20 & 33.15 & 33.15\\
    \hline    
\end{tabular}
\caption{\small 
Comparison 
of the critical Reynolds number derived from modal analysis $Re_c$ with that obtained through energy stability analysis $Re_e$ 
for wide range of parameters. }
\label{table:modal_Reynolds}
\end{table}

Next we present a comparison of the critical Reynolds numbers derived from linear stability analysis ($Re_c$) and energy methods ($Re_e$), illustrating how these values vary across different system parameters $\sigma$, $\xi$ and $d$ in table~\ref{table:modal_Reynolds}.
Recall that the instability occurs at a lower Reynolds number than predicted by the linear stability analysis because the boundary of the stable zone (zone I) lies well behind the exponential growth zone (zone III), see figure~\ref{fig:temporal}.
It is seen from  table~\ref{table:modal_Reynolds} that the critical energy Reynolds number $Re_e$ is always lesser than critical Reynolds number $Re_c$ of linear theory, and 
(i) for a fixed $d$ and $\xi$ value, $Re_c$ and $Re_e$ both decrease with increasing $\sigma$ (columns 2--4), (ii) for a fixed $\sigma$ and $\xi$ value $Re_c$ and $Re_e$ both decreases with increasing $d$ (columns 5--6) and (iii) for a fixed $d$ and $\sigma$ value $Re_c$ and $Re_e$ both increase with increasing $\xi$ (columns 7--9). 
Thus, increasing 
$\sigma$ and $d$ destabilizes the system, whereas decreasing 
$\xi$ destabilizes the system. 

It is also noted that
the difference between $Re_e$ and $Re_c$ is caused by the non-normality of the Orr--Sommerfeld and Squire operators, $\mathscr{L}$. 
In the case of classical plane Poiseuille flow and plane Couette flow, the linear stability operator is non-normal, and thus 
$Re_e$ and $Re_c$ are significantly different for these flows. The linear stability operator for  Rayleigh-B{\'e}nard convection is normal,  leading to comparable $Re_e$ and $Re_c$~\citep{schmid2001stability,drazin2004hydrodynamic}.   
The conventional modal stability analysis characterizes only the asymptotic behavior of the 
perturbation and, therefore, fails to capture the short-term 
characteristics of the flow. 
Next, we look at the non-normal stability to characterize the flow's transient behavior.

% --------------------------------------------------------------------
\section{Non-modal stability analysis}
\label{sec:Nonmodal_PFP}
% --------------------------------------------------------------------

The 
non-modal stability analysis %we 
focuses on two primary aspects: (i) the system's response to external excitation
and (ii) assessing the transient energy amplification of initial conditions.
The response and growth functions characterize these two facets. 
While the response function $R$ represents the maximum possible amplification due to the external excitations, the growth function $G(t)$ delineates the kinetic energy growth of disturbances at any given time $t$, capturing the transient behavior of the initial conditions.

% ----------------------------
\subsection{Response to external excitation: Receptivity analysis}
\label{subsec:response_external_force}
% ----------------------------------

Receptivity analysis concerns with understanding how a fluid system 
reacts to external disturbances. 
These external disturbances can manifest in various ways, such as free-stream turbulence, acoustic perturbations, wall roughness, body forces, etc. The concept of receptivity is often explained through a resonance argument.
The system is most responsive when the frequencies of the external forces align closely with the system's natural frequencies or eigenvalues. 
This argument holds for normal systems (when $\mathscr{L}$ is normal), where the system's response aligns with theoretical predictions.
However, in the case of non-normal systems (when $\mathscr{L}$ is non-normal), the flow system may exhibit a significant amplitude response to external forcing even when the frequency of the forcing is far away from that of the system's eigenvalues. This phenomenon is termed pseudo-resonance~\citep{threfeten1993Hydrodynamic}.

To understand the pseudo-resonance, the temporal initial value problem~\eqref{eqn:initial_value_problem} is recast by introducing an external harmonic force $\boldsymbol{F}$:  
\begin{align}
    \frac{\partial {\boldsymbol{q}}}{\partial t}=-\mathrm{i}\mathscr{L} {\boldsymbol{q}}+ \boldsymbol{F},
    \label{eqn:initial_value_problem_force}
\end{align}
where $\boldsymbol{F}=
\widehat{\boldsymbol{F}}_q \, e^{-\mathrm{i} f t}$ with $\widehat{\boldsymbol{F}}_q$ and $f$ being the amplitude and frequency of the imposed forcing, respectively.
The amplitude of the particular solution of~\eqref{eqn:initial_value_problem_force} can be expressed as
\begin{align}
    \widehat{\boldsymbol{q}}_p=\left(f \mathcal{I} -  \mathscr{L} \right)^{-1}\widehat{\boldsymbol{F}}_q,
    \label{eqn:initial_value_problem_force_PI}
\end{align}
where $\mathcal{I}$ is the identity operator and $\left(f \mathcal{I} - \mathscr{L} \right)^{-1}$ is defined as resolvent operator~\citep{threfeten1993Hydrodynamic,threfeten1997pseudospectra,schmid2001stability,schmid2007annual}.

Following~\citet{threfeten1993Hydrodynamic},  
the response function $R(f)$ is the ratio of the energy of the amplitude of the particular solution $ \Vert \widehat{\boldsymbol{q}}_p\Vert $ to the energy of the amplitude of the external forcing $ \Vert \widehat{\boldsymbol{F}}_q\Vert $, optimized across all possible amplitude forcing profiles:
\begin{align}
    R(f) \equiv R(f,\alpha,\beta)=\max_{\widehat{\boldsymbol{F}}_q \ne 0} \frac{\Vert \widehat{\boldsymbol{q}}_p\Vert }{\Vert \widehat{\boldsymbol{F}}_q \Vert } =\Vert  \left( f \mathcal{I} - \mathscr{L} \right)^{-1}\Vert.
    \label{eqn:responce_function}
\end{align}
Thus, the response function $R$ is equal to the resolvent norm.
The eigenvalues of the operator $\mathscr{L}$ are complex, 
which implies that 
the resolvent norm can tend towards 
infinity when $f$ is matched with these eigenvalues. This phenomenon is called resonance. 
However response function can be substantially larger even if the forcing frequencies $f$ do not match the eigenvalues or natural frequencies of the operator $\mathscr{L}$, previously described as a pseudo-resonance phenomenon. 
The pseudospectrum can analyze the pseudo-resonance, as it captures regions where the resolvent norm is large.

% ----------------------------------
\subsubsection{Pseudospectra}
\label{subsec:pseudospectra_Nrange}
% -----------------------------------

To investigate the response function with complex frequencies, the $\varepsilon$-pseudospectra come into play~\citep{threfeten1993Hydrodynamic,embree2005spectra}.
For any $\varepsilon \ge 0$, the \textit{$\varepsilon$-pseudospectra} of the linear operator $\mathscr{L}$ is defined as
\begin{align}
    \mathrm{sp}_{\varepsilon}(\mathscr{L})
    %\sigma_{\varepsilon}(\mathscr{L})
    =\lbrace \omega \in \mathbb{C}: \| \left( \omega \mathcal{I} - \mathscr{L} \right)^{-1}  \| \ge \varepsilon^{-1}  \rbrace,
\end{align}
which represent all eigenvalues of the operator that are $\varepsilon$-close to it. 
Alternatively, it can be stated that a complex number $\omega \in \mathrm{sp}_{\varepsilon}(\mathscr{L})$ if and only if $\omega$ is an eigenvalue of the perturbed operator $\mathscr{L}+\mathscr{E}$, where $\mathscr{E}$ is satisfying $\|\mathscr{E}\|\le \varepsilon$~\citep{threfeten1993Hydrodynamic}. The concept of $\varepsilon$-pseudospectra is used to measure the sensitivity of eigenvalues of the operator $\mathscr{L}$ to small perturbations $\mathscr{E}$ in the operator. 
It quantifies the extent to which the eigenvalues are affected by these perturbations.
Note that the
pseudospectra are nested sets
with $\varepsilon=0$ being the eigenvalues of the operator $\mathscr{L}$. 
The pseudospectra 
are much larger than the spectra for a non-normal operator $\mathscr{L}$ even if $\varepsilon \ll 1$~\citep{reddy_henningson_1993,reddy1993pseudospectra}. It is preferable to compute the {\it pseudospectra} rather than the {\it spectra} when the underlying operator is non-normal~\citep{threfeten1997pseudospectra,threfeten1993Hydrodynamic}. 
In general, the pseudospectra is illustrated numerically using the contour plot in complex $\omega$-plane with the level contours corresponding to the $\varepsilon$ value. 

\begin{figure}
    \centering
    \includegraphics[scale=0.3]{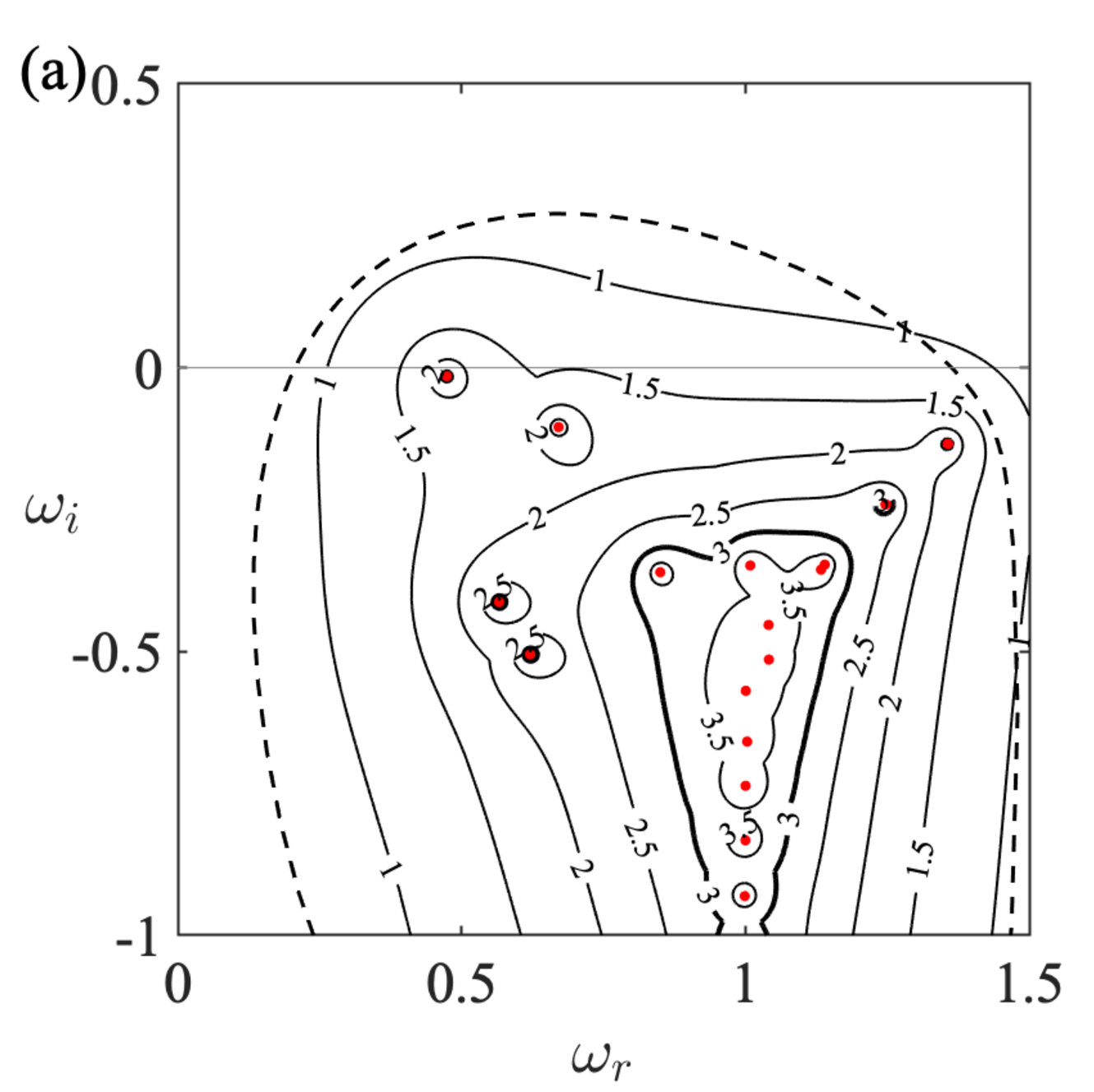}
    \includegraphics[scale=0.3]{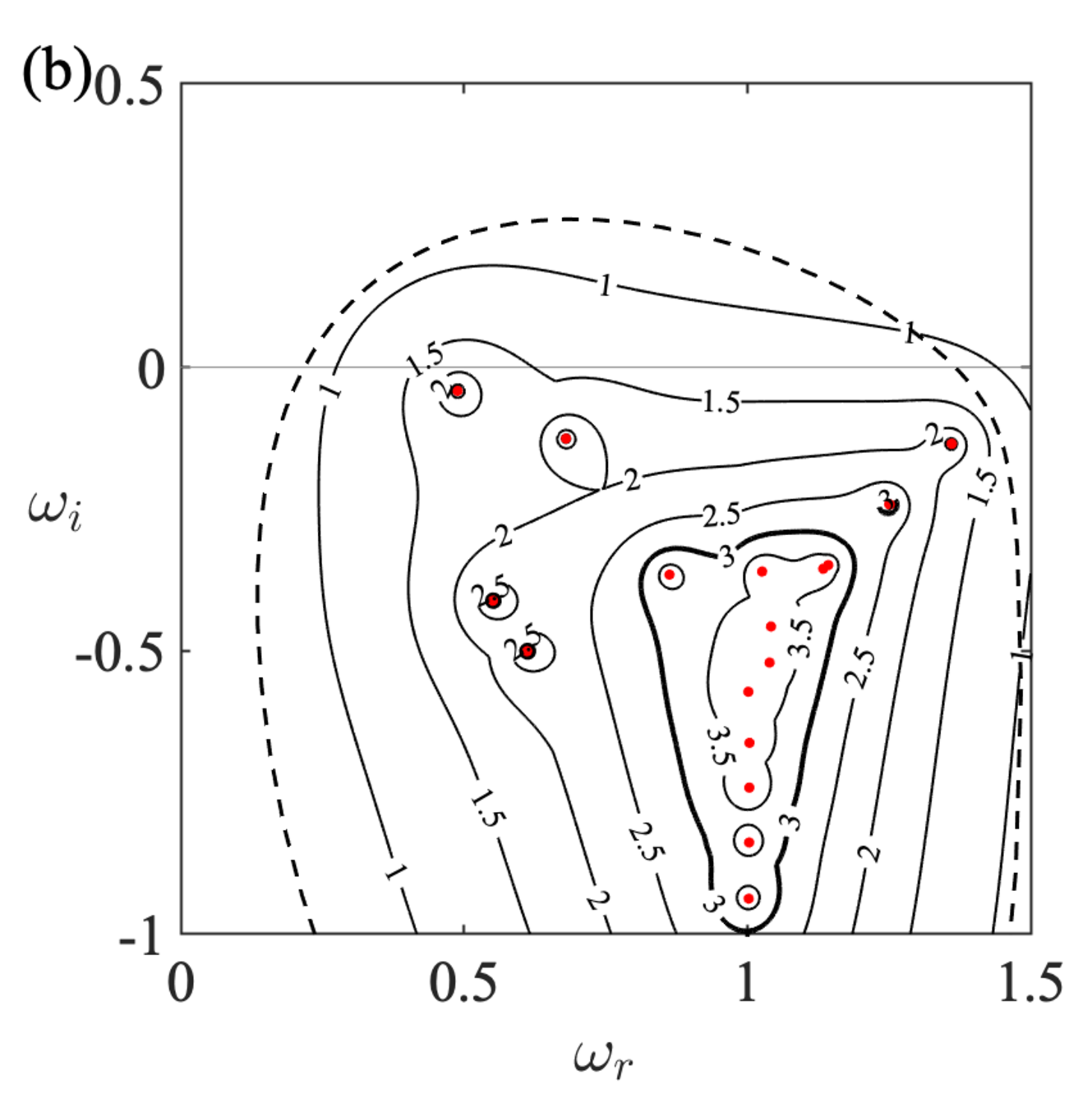}
    \caption{\small [$Re=1000,\,d=1$,\, $\alpha=1$, $\beta=0$ (spanwise independent perturbation)] Eigenvalues (red dots), $\varepsilon$-pseudospectra (solid labeled lines) and the boundary of numerical range (dashed line) for $(\sigma,\,\xi)=$ (a) (0.006,\,0.001) and (b) (0.002,\,10). Numerical abscissa $\varphi=$0.2715 and 0.2613 for panels a and b, respectively. Here contour label denotes $p$ value of $\epsilon=10^{-p}$.}
    \label{fig:pseudo_anisotropy2}
\end{figure}

\begin{figure}[!htbp]
    \centering
    \includegraphics[scale=0.3]{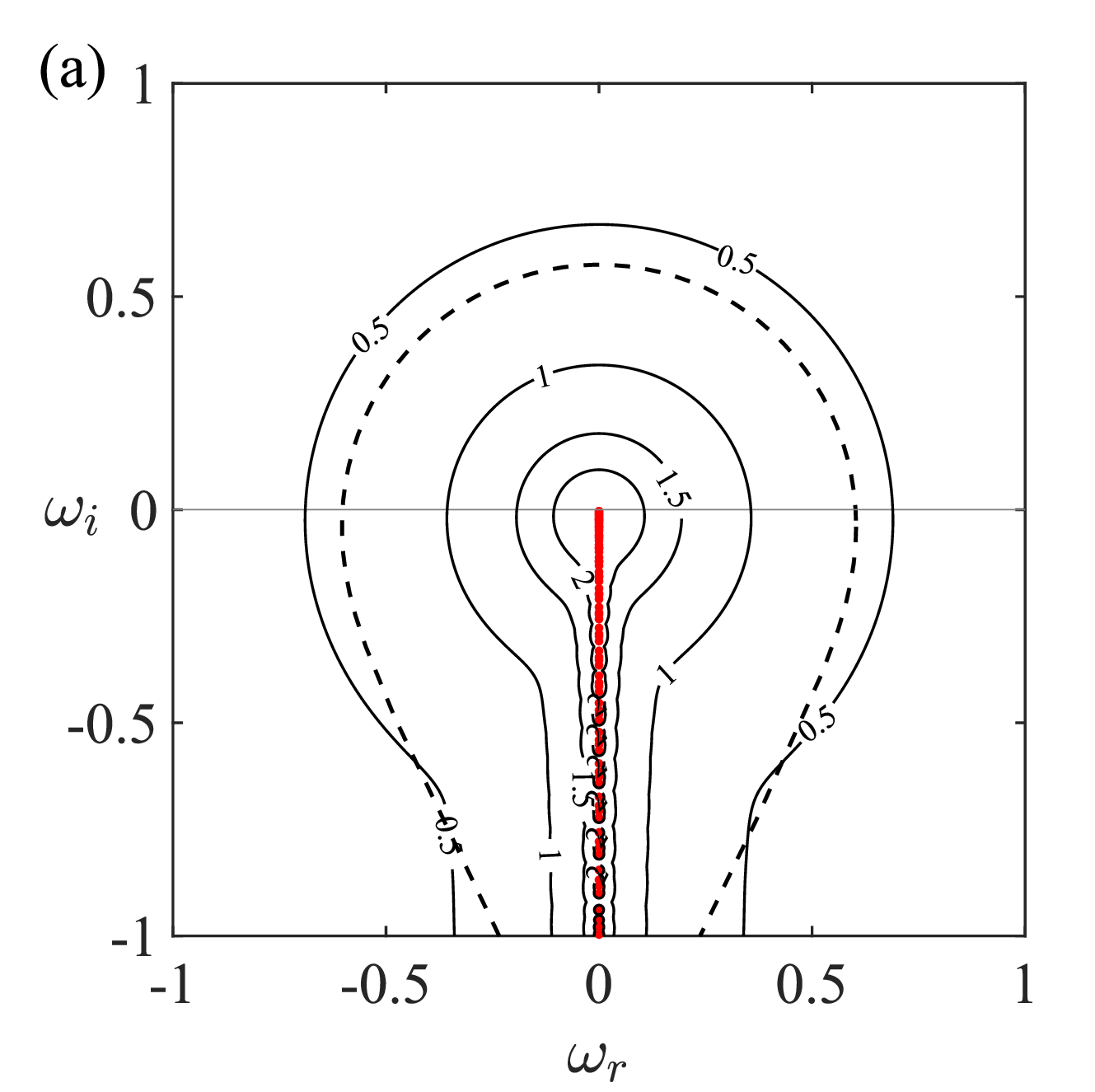}
    \includegraphics[scale=0.3]{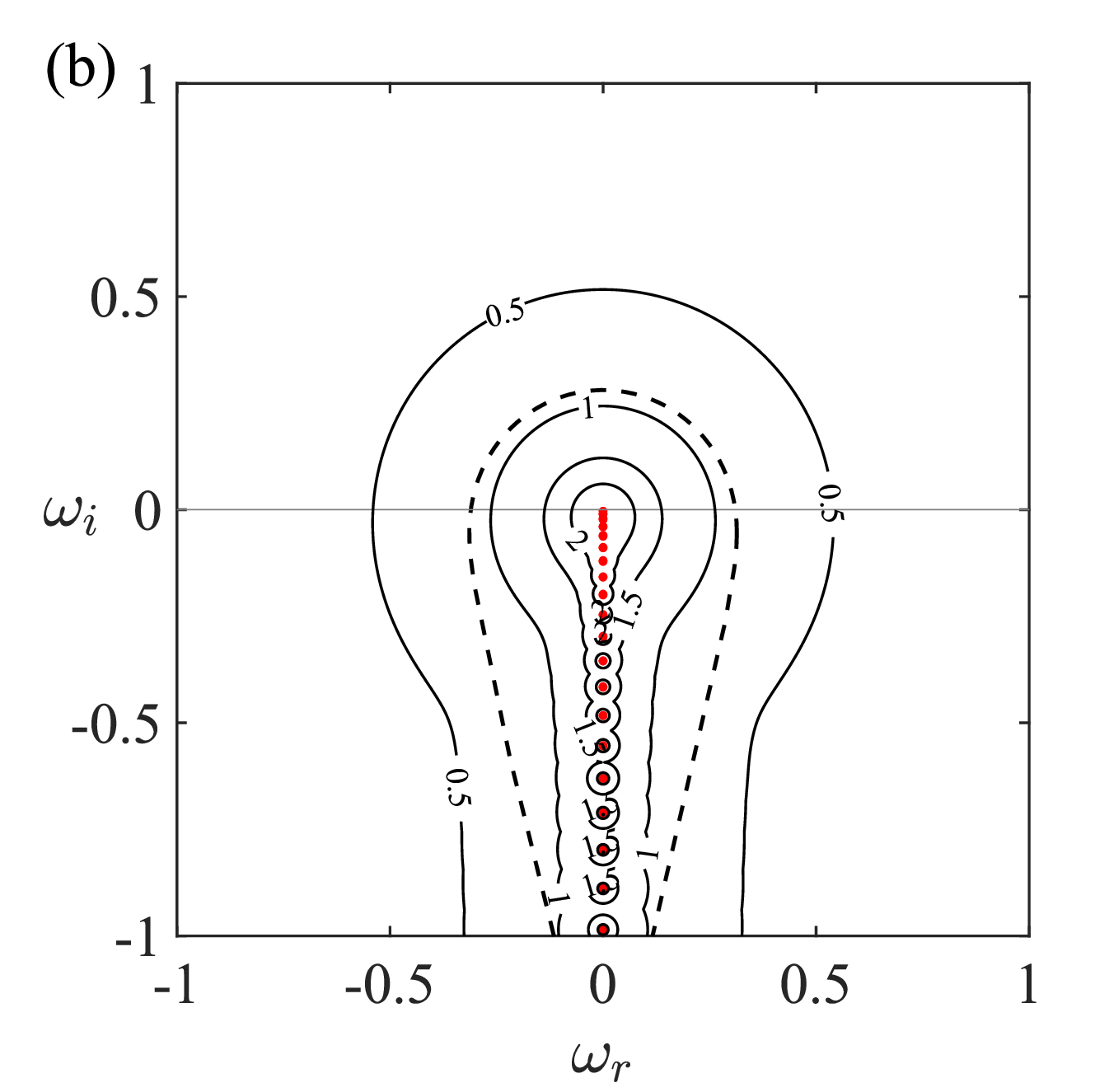}
    \caption{\small 
    Same as figure~\ref{fig:pseudo_anisotropy2} but for $\alpha=0$, $\beta=1$ (streamwise independent perturbation).
    Numerical abscissa $\varphi=$0.5747 and 0.2807 for panels a and b, respectively. 
    }
    \label{fig:pseudo_anisotropy2_1}
\end{figure}

Figures~\ref{fig:pseudo_anisotropy2} and~\ref{fig:pseudo_anisotropy2_1} depict the spectrum (red dots), $\varepsilon$-pseudospectra (solid lines), and boundary of numerical range (dashed line) for spanwise (figure~\ref{fig:pseudo_anisotropy2}) and streamwise (figure~\ref{fig:pseudo_anisotropy2_1}) independent disturbances, respectively, at $Re=1000$ and $d=1$ for two sets of anisotropic permeabilities $(\sigma,\,\xi)=$ $(0.006, 0.001)$ and $(0.002,10)$. The $p$ contours delineate the boundaries within the complex plane of an $\varepsilon$-pseudospectrum, where $\varepsilon = 10^{-p}$. 
It is seen that 
the entire spectrum remains in the stable lower half-plane ($\omega_i<0$), ensuring that the flow is linearly stable.
Nevertheless, the numerical range boundary (dashed line) extends into the upper half of the complex plane ($\omega_i>0$). Consequently, the numerical abscissa is positive, indicating the short-term energy growth of the perturbation quantities.

Let us examine which forms of perturbations,
whether they manifest as independent in the streamwise or spanwise directions, exhibit the most substantial transient growth and the underlying reasons for this phenomenon.

We first compare panels of figure~\ref{fig:pseudo_anisotropy2}.
The spectrum shown in figure~\ref{fig:pseudo_anisotropy2}
is found to be ``Y"-shaped and composed of the A ($\omega_r \rightarrow 0.5$), P ($\omega_r \rightarrow 1.5$), and S ($\omega_r \approx 1$) branches as similar to classical PPF~\citep[see][]{schmid2001stability}. 
Notably, a concurrent decrease in $\sigma$ and an increase in $\xi$ reduces the growth rate of the most unstable mode in the A branch (see figure~\ref{fig:pseudo_anisotropy2}).
It is seen 
that the variation in $\sigma$ and $\xi$ has minimal impact on the boundary of the numerical range and the numerical abscissa ($\omega_i$ co-ordinate of the peak point in a dashed curve).
 Recall from \S~\ref{subsec:energy_stability} that the rate of disturbance kinetic energy is proportional to the numerical abscissa.
For spanwise independent perturbations, the rate numerical abscissa is only marginally influenced by the anisotropic permeability for a given set of parameters.

{By comparing panels of figure~\ref{fig:pseudo_anisotropy2}, it is seen that pseudospectrum for $\varepsilon=10^{-3}$ 
(thick contour with label 3) shrinks substantially in the complex $\omega$-plane. }
This result indicates that as $\sigma$ increases and $\xi$ decreases (panel b to panel a), the sensitivity of the S branch to perturbations increases considerably, while the other branches are not significantly sensitive. Hence, analogous to the classical plane Poiseuille flow, the eigenfunctions associated with the modes in the S-branch exhibit non-orthogonality. As a result, the combined influence of these modes leads to an amplification of transient or short-term energy amplifications.

In the presence of streamwise independent disturbances (figure~\ref{fig:pseudo_anisotropy2_1}), all modes are located along the imaginary axis $(\omega_r = 0)$.  
%despite changes in anisotropic permeability $\sigma$, $\xi$ 
The $\varepsilon$-pseudospectra clearly demonstrate that even slight perturbations can lead to 
larger transient growth.
This is evident from the observation that the 
$\varepsilon$-pseudospectra
extend into the unstable half-plane, even at moderate contour levels.
Nevertheless, the spectrum becomes more continuous as $(\sigma,\,\xi)$ alters from (0.002,\,10) to (0.006,\,0.001) in panel (b) and (a). %Note that 
The boundary of numerical range (dashed line) and the value of numerical abscissa $\varphi$ (see the caption) are considerably affected by a change in $(\sigma,\,\xi)$. Hence, the 
rate of disturbance kinetic energy rises when $\sigma$ increases and $\xi$ decreases 
for streamwise independent disturbances. 
Moreover, these changes significantly increase the area of $\varepsilon$-pseudospectra.
As a result, for the streamwise independent ($\alpha=0$) perturbation, eigenvalues are likewise more sensitive than that of spanwise independent ($\beta=0$) perturbations. 
Similarly to the mean permeability, the depth ratio increases the numerical abscissa and the eigenvalue sensitivity, thereby positively contributing to the disturbance kinetic energy growth (figure not shown).

The non-normality 
of operator $\mathscr{L}$ 
becomes more pronounced as $\xi$ decreases and $\sigma$ or $d$ increases simultaneously. This 
effect is amplified for streamwise independent ($\alpha=0$) perturbations.
Since the numerical abscissa for streamwise independent disturbances is greater than that for spanwise independent disturbances ($\beta=0$), 
streamwise independent ($\alpha=0$) disturbances will likely cause more transient energy growth.

% ------------------------------------
\subsubsection{Maximum response and optimal response}
\label{subsubsec:response_function}
% ------------------------------------

{
In figures~\ref{fig:pseudo_anisotropy2} and~\ref{fig:pseudo_anisotropy2_1}, $\varepsilon$-pseudospectra is shown for various system parameters in the complex frequency plane ($\omega$-plane) to understand the 
pseudo-resonance phenomenon. 
In this section, we illustrate the same phenomenon that occurs for the real frequencies of the  external excitations.   
}
%
% {\color{blue}
% We have displayed the $\varepsilon$-pseudospectra in the complex $\omega$-plane for several typical parameters in figures~\ref{fig:pseudo_anisotropy2} and~\ref{fig:pseudo_anisotropy2_1}. Our goal is to understand the pseudo-resonance phenomenon that occurs along the real axis in the complex $\omega$-plane, which corresponds to external excitations at real frequencies.}
% In order to do so, 
% we focus on the response to external excitations by examining harmonic excitations with varied wavelengths at different real frequencies $f$. 
To do so, 
%we focus on the  by 
we 
examine the response to external harmonic excitations with varied wavelengths at different real frequencies $f$. 
Such response is quantify using the maximum $R_{\max}$ and optimal $R_{\rm{opt}}$ responses  and, respectively, as follows 
\begin{align}
    R_{\max} (f)=
    \max_{\alpha,\beta}  R(f,\alpha,\beta),
    %\label{eqn:R_max},
    \quad \mbox{and}\quad
       R_{\rm{opt}}=\max_{f} R_{\max}(f)  \,\, \, \text{at} \,\, \, f=f_{\rm{opt}},
    \label{eqn:R_opt}
\end{align}
where $R$ is the response function (resolvent norm)~\eqref{eqn:responce_function}.

\begin{figure}
    \centering
    \includegraphics[scale=0.22]{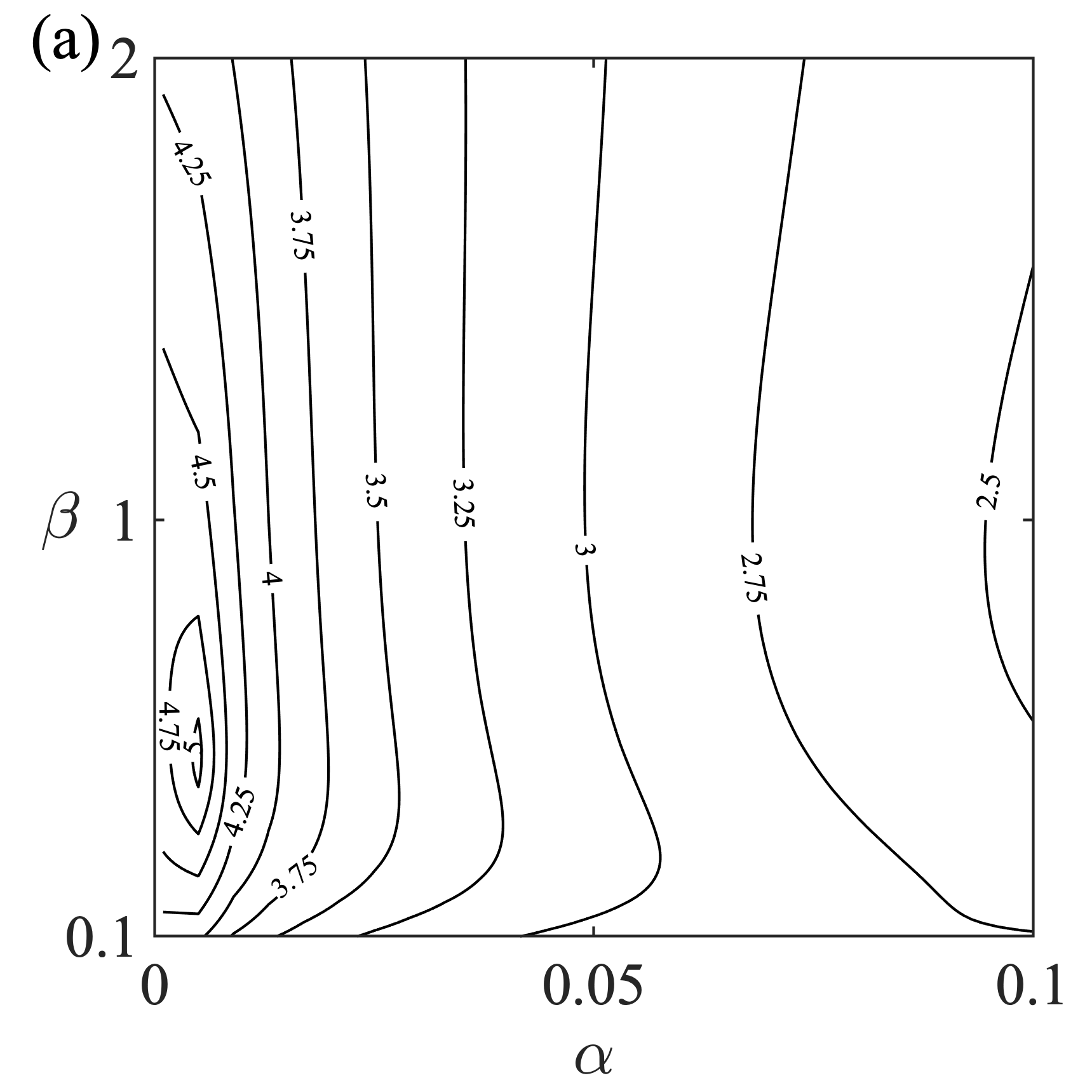}
    \includegraphics[scale=0.22]{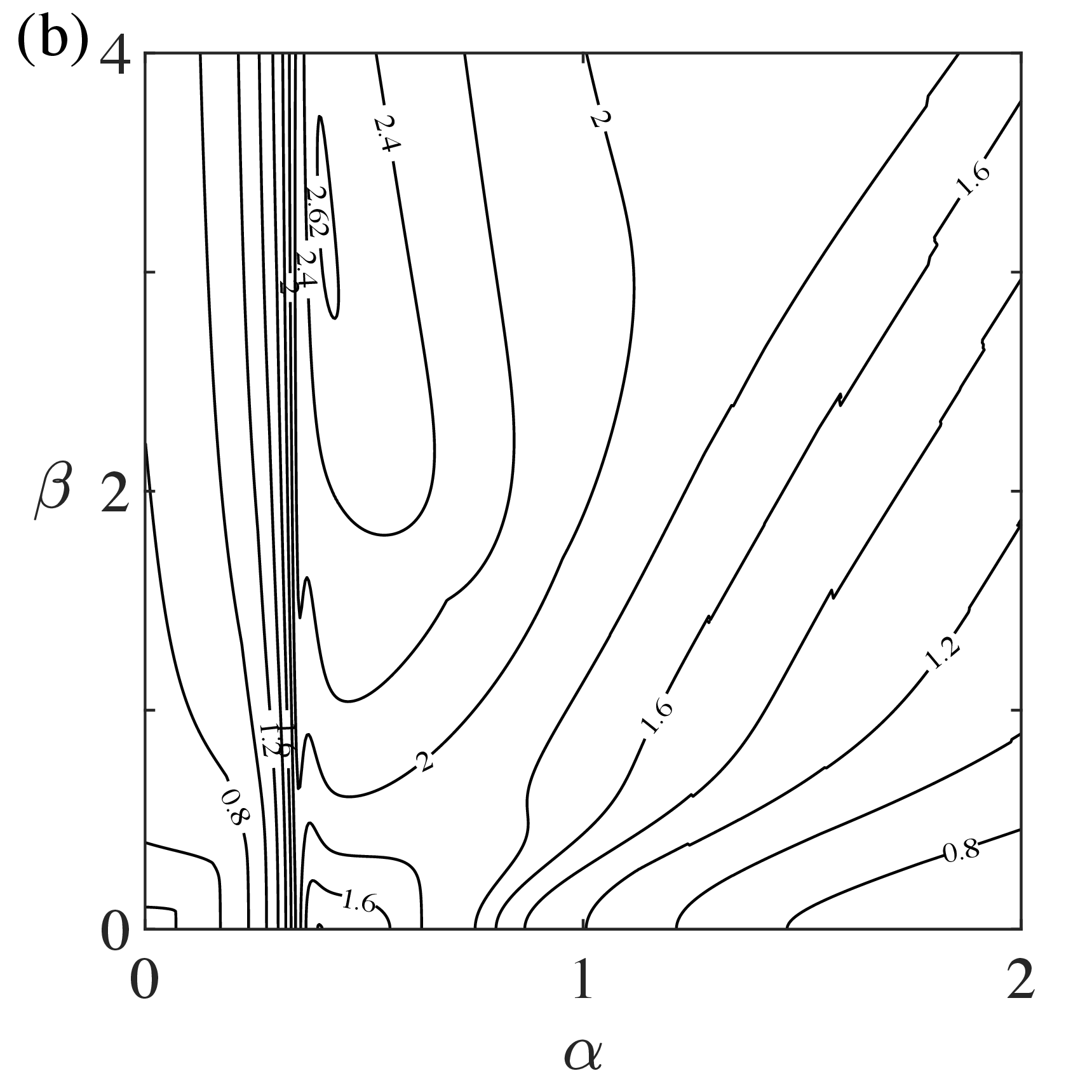}
    \includegraphics[scale=0.22]{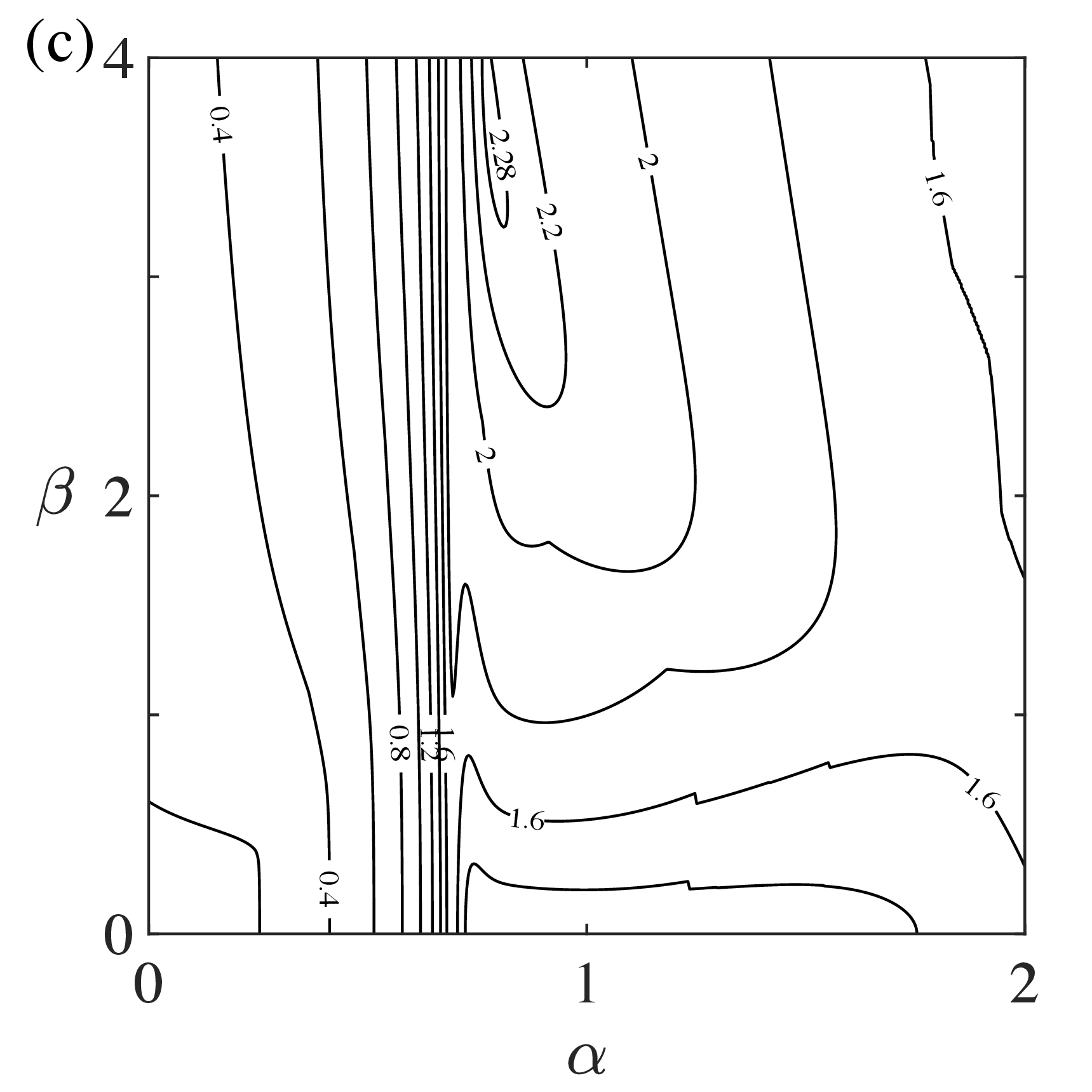}
     \includegraphics[scale=0.22]{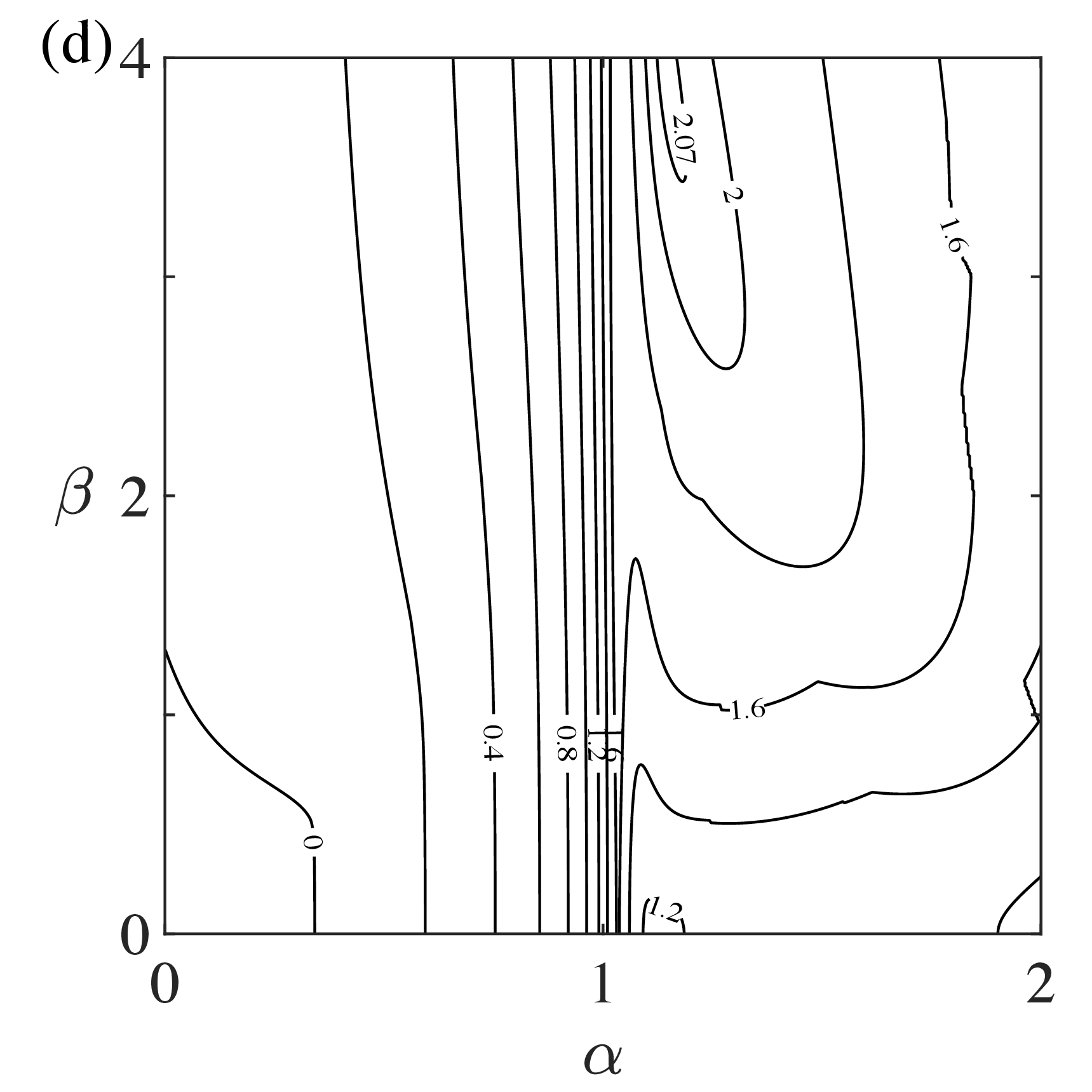}
    \caption{\small [$Re=1000$, $\sigma=0.02$, $\xi=0.001$, and $d=1$] Isolines of the logarithm of the response function $\log_{10}(R)$ in $(\alpha, \beta)$-plane when $f=$ (a) 0, (b) 0.5, (c) 1, and  (d) 1.5.}
    \label{fig:maximum_response}
\end{figure}
To determine the optimal frequency $f_{\rm opt}$, we plot the level curve of the logarithm of the response function $\log_{10}(R)$ in the $(\alpha,\beta)$-plane in figure~\ref{fig:maximum_response} for four different frequencies, 
with other parameters being fixed as $Re=1000,\,d=1$ and $(\sigma,\xi)=(0.02,0.001)$.
It is seen that at $f=0$, $(\alpha,\, \beta)\approx (0.005,\,0.51)$ produces the maximum response $R_{\max} \approx 10^{5}$ (panel~a). 
Therefore, the maximum response observed near the $\beta$-axis for long-wave disturbance $\alpha \approx 0.005$. For $f=0.5$ the maximum response shifts to 
$(\alpha,\beta)\approx (0.42,\,3.38)$, with  $R_{\max}$ being equal to $10^{2.6}$ (panel b). Further increasing the forcing frequency $f$ to 1 and 1.5, the maximum response $R_{\max}$ becomes $10^{2.28}$ and $10^{2.07}$ for $(\alpha,\,\beta)\approx (0.78,\,3.74)$ and $(1.17,\,3.88)$, respectively. Consequently, as forcing frequency $f$ increases, the maximum response $R_{\max}$ reduces and tends to occur in the short-wave regime. 
Therefore, 
the optimum response $R_{\operatorname{opt}} = 10^{5}$ is obtained at oblique wavenumber $(\alpha,\beta) = (0.005,0.51)$, with the optimal forcing frequency being $f_{\operatorname{opt}}=0$ (corresponding to panel a). It has also been verified that the response curves exhibit similar structures and behaviors at various Reynolds numbers (either critical or sub-critical) with different depth ratios and anisotropic permeabilities.
Furthermore, 
the observed trend persists for various values of $(\sigma,\,\xi)$, and $d$, as the simultaneous decrease of $\sigma$ and $d$ with an increase of the anisotropy parameter $\xi$ result in a reduction of the optimal growth in the $(\alpha,\beta)$-plane. In each instance, the optimal response occurs at the harmonic force frequency $f=0$, and as $|f|$ increases, the maximal response decreases and occurs at a comparatively large wavenumbers regime.

\begin{figure}
    \centering
    \includegraphics[scale=0.31]{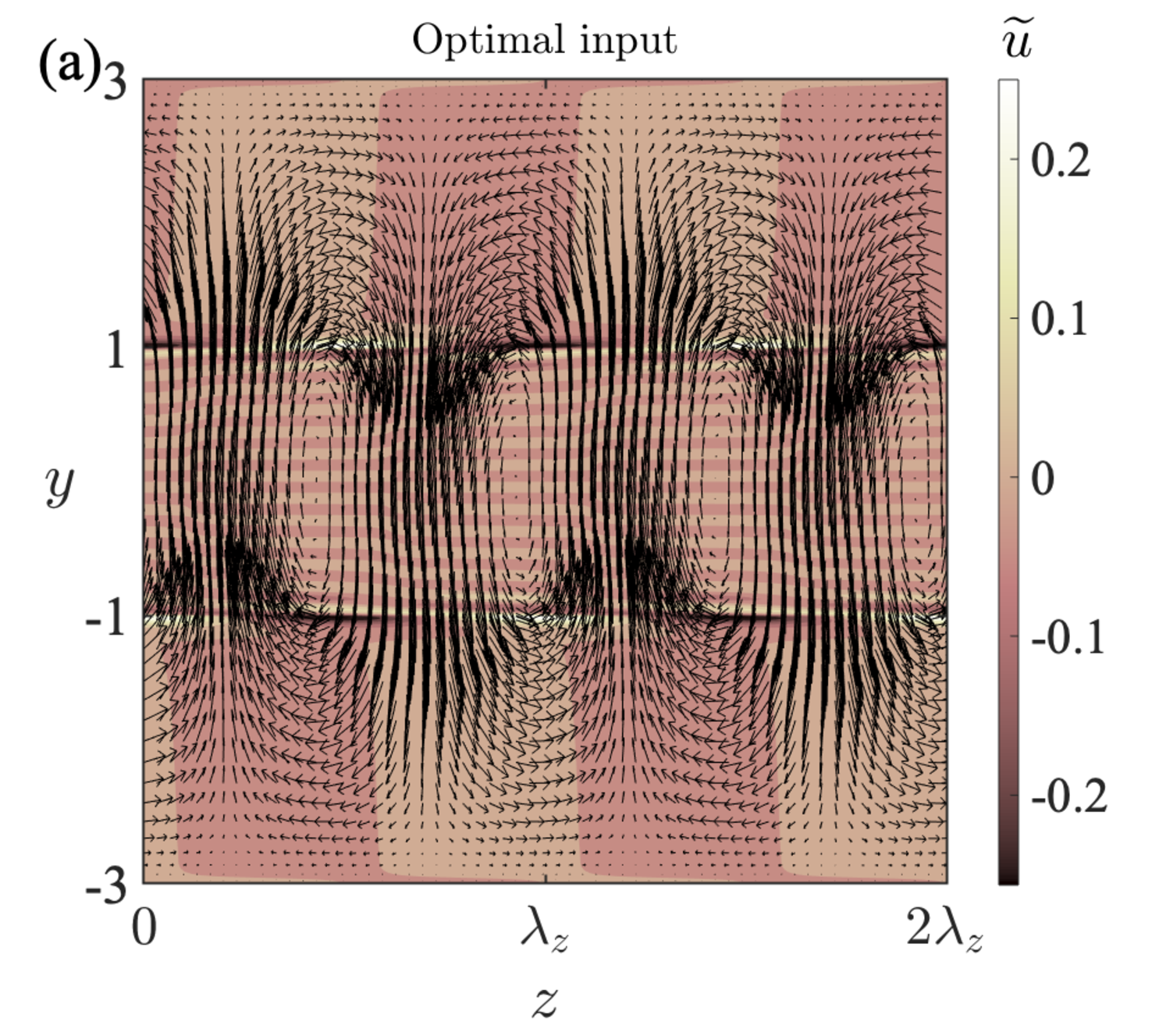}
    \includegraphics[scale=0.31]{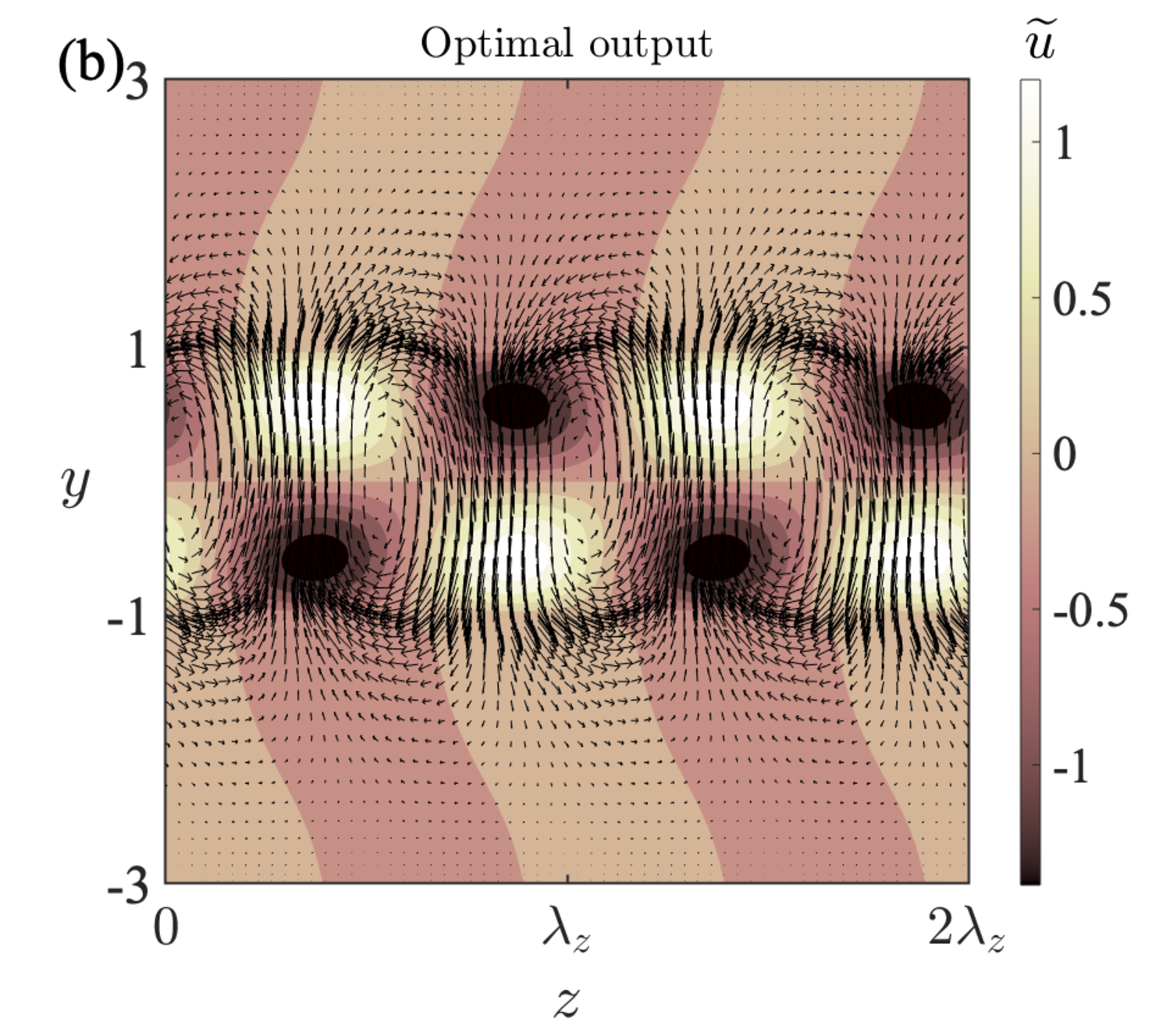}
    \caption{
    Variation of optimal response field---optimal input (panel~a) and optimal output (panel~b): Contours of streamwise velocity $\widetilde{u}$ and cross-stream velocity field $(\widetilde{v},\widetilde{w})$ in $(y,z)$-plane for $Re=1000,\,\sigma=0.02,\,\xi=0.001$.
    For panels~a and b, 
    $(|\widetilde{u}|, |\widetilde{v}|, |\widetilde{w}|) =(0.256, 0.556, 0.648)$ and 
    $(1.385, 0.011, 0.018)$, respectively.
    }
    \label{fig:opti_response_structure}
\end{figure}

We next focus on the shape of the forcing which produces the largest response in the flow,
together with the output flow response. 
Each point on the curve $R(f)$ are listed by optimizing the ratio $\nicefrac{\|\widehat{\boldsymbol{q}}_p\|}{\|\widehat{\boldsymbol{F}}_q\|}$ over sets of external forces (with various amplitudes). The optimal forcing that yields optimal response at a given frequency $f_{\operatorname{opt}}$ satisfies
\begin{align}
    \left( f_{\operatorname{opt}}\mathcal{I}-\mathscr{L} \right)^{-1}\, \boldsymbol{q}_p^0=\| \left(f_{\operatorname{opt}}\mathcal{I}-\mathscr{L}\right)^{-1}\|\,\boldsymbol{q}_p^{\operatorname{opt}},
    \label{eqn:optimal_frequency}
\end{align}
where the optimal forcing (or input) $\boldsymbol{q}_p^0$ is advanced by $\left( f_{\operatorname{opt}}\mathcal{I}-\mathscr{L} \right)^{-1}$ producing the optimal response (or output) $\boldsymbol{q}_p^{\operatorname{opt}}$ that is amplified by $R_{\operatorname{opt}}=\|\left(f_{\operatorname{opt}}\mathcal{I}-\mathscr{L}\right)^{-1} \|$~\citep[see][]{Schmid2014analysis}.

The optimal frequency $f_{\operatorname{opt}}$ for parameters of figure~\ref{fig:maximum_response} is zero. Thus ~\eqref{eqn:optimal_frequency} reduces to 
$\mathscr{L}^{-1}\, \boldsymbol{q}_p^0=\| \mathscr{L}^{-1}\|\,\boldsymbol{q}_p^{\operatorname{opt}}$. 
At $f_{\operatorname{opt}}=0$, 
the velocity fields of external optimal excitation (or input) and optimal response (or output) 
are depicted in figure~\ref{fig:opti_response_structure}, with other parameters being the same as figure~\ref{fig:maximum_response}. 
The flow fields comprising $(\widetilde{v},\widetilde{w})$ velocity vectors and contours of streamwise velocity $\widetilde{u}$ are shown for both input (panel a) and output (panel b) optimal response fields.
For the input (panel~a), the magnitude of streamwise velocity is much smaller than those of cross-stream velocities, whereas the opposite is true for the output field (see panel~b). 
Thus, in the optimal input, the transverse and spanwise velocities contribute most to the perturbation kinetic energy, while in the optimal output, the streamwise velocity has a larger impact.
A spanwise periodic array of streamwise counter-rotating vortices or rolls characterizes the optimal input. 
Furthermore, the dominant $(\widetilde{v},\,\widetilde{w})$-velocity field for optimal input extends into the porous layer. This explains the importance of a porous layer in determining the flow transition characteristics due to the external forcing. 
On the other hand, the $\widetilde{u}$-contours for optimal response, confined significantly only in the fluid region, entails the presence of high-speed and low-speed streamwise velocity,  
referred to as streamwise streaks. These streaks contain higher perturbation kinetic energy, resulting in optimal response due to the external excitations~\citep{schmid2001stability,LIU_LIU_2011}.

% ------------------------------------------
\subsection{Response to initial conditions: Growth function}
\label{subsec:response_initialCond}
% ------------------------------------------

The response to the initial disturbances governs by the growth function $G(t)$. More precisely, the maximum possible amplification of an initial disturbance at any instant $t$ is given by the growth function
\begin{align}
    G(t)\equiv G(t,\alpha,\beta)=\max_{\boldsymbol{q}_0 \ne 0} \frac{\| \boldsymbol{q}(y,t)\|^2}{\| \boldsymbol{q}_0\|^2}=\| \exp{\left( -\mathrm{i}\mathscr{L}t \right)}\|^2,
    \label{eqn:growth_function}
\end{align}
see~\eqref{eqn:initial_value_problem}--\eqref{eqn:Solinitial_value_problem}. 
Similarly to $R_{\max}$ and $R_{\rm opt}$, the maximum growth $G_{\max}$ at $t=t_{\max}$ is defined as
\begin{align}
    G_{\max}\equiv G_{\max}(\alpha,\beta) =\max_{t\ge 0} G(t,\alpha,\beta), 
\end{align}
and the optimum energy growth $G_{\rm opt}$ at $t=t_{\rm opt}$ is defined as
\begin{align}
    G_{\text{opt}}=\max_{\alpha,\beta} G_{\max} \left( \alpha, \beta \right).
    %\quad \text{at} \quad t=t_{\rm{opt}}.
    \label{eqn:optimal_pert}
\end{align}

\begin{figure}
    \centering
   \includegraphics[scale=0.25]{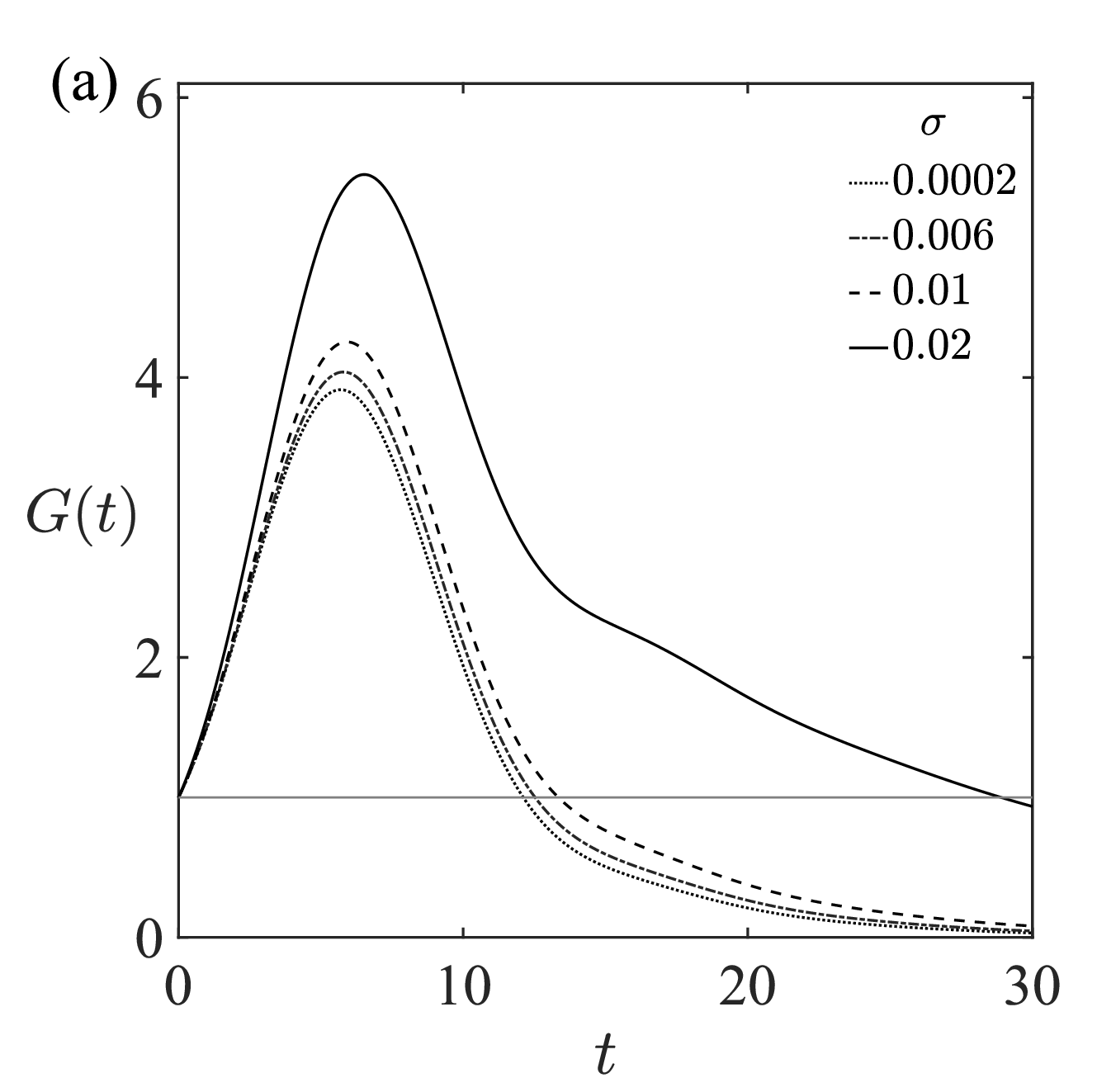} 
  \includegraphics[scale=0.25]{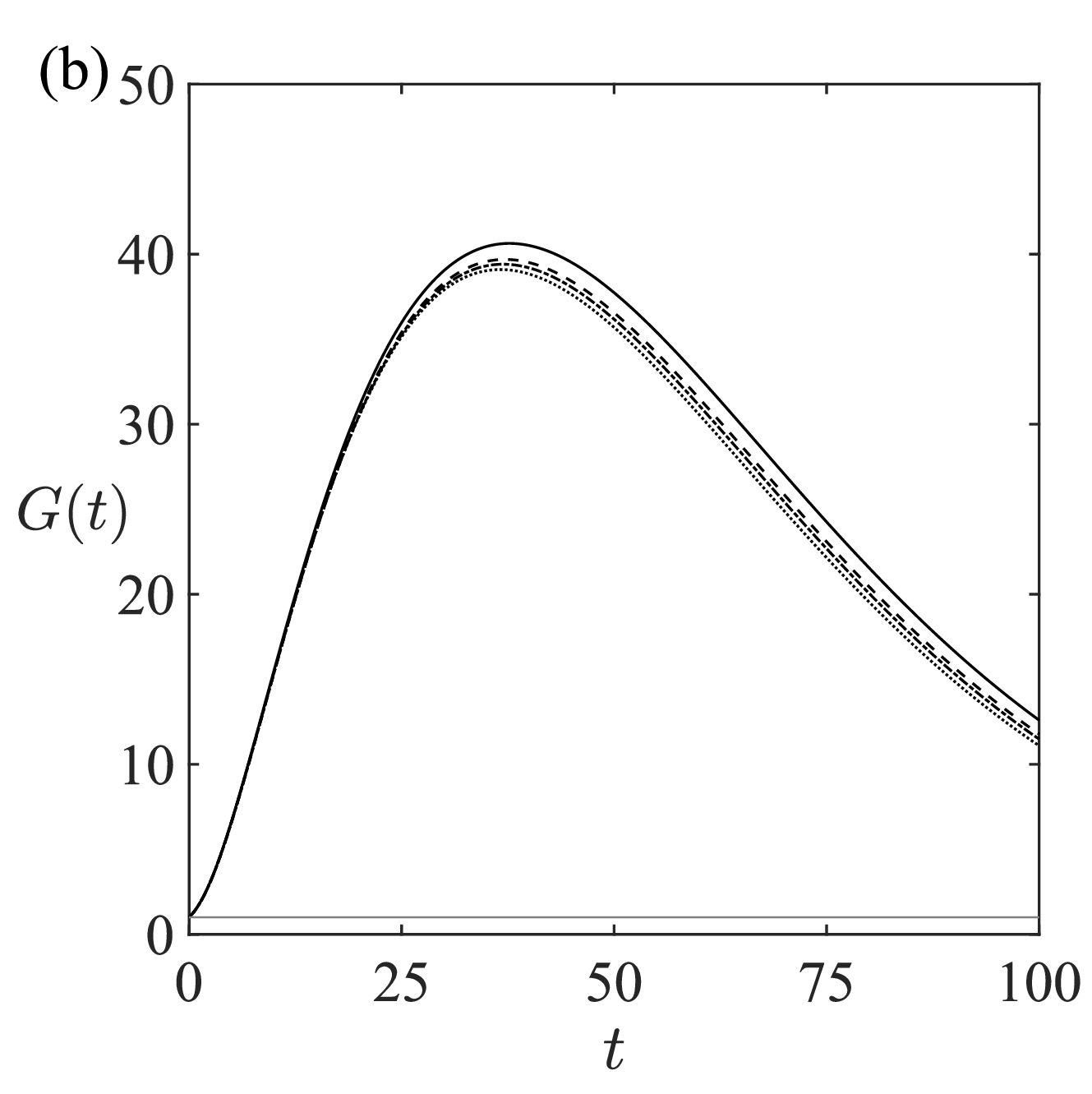} 
   \\
   \includegraphics[scale=0.25] {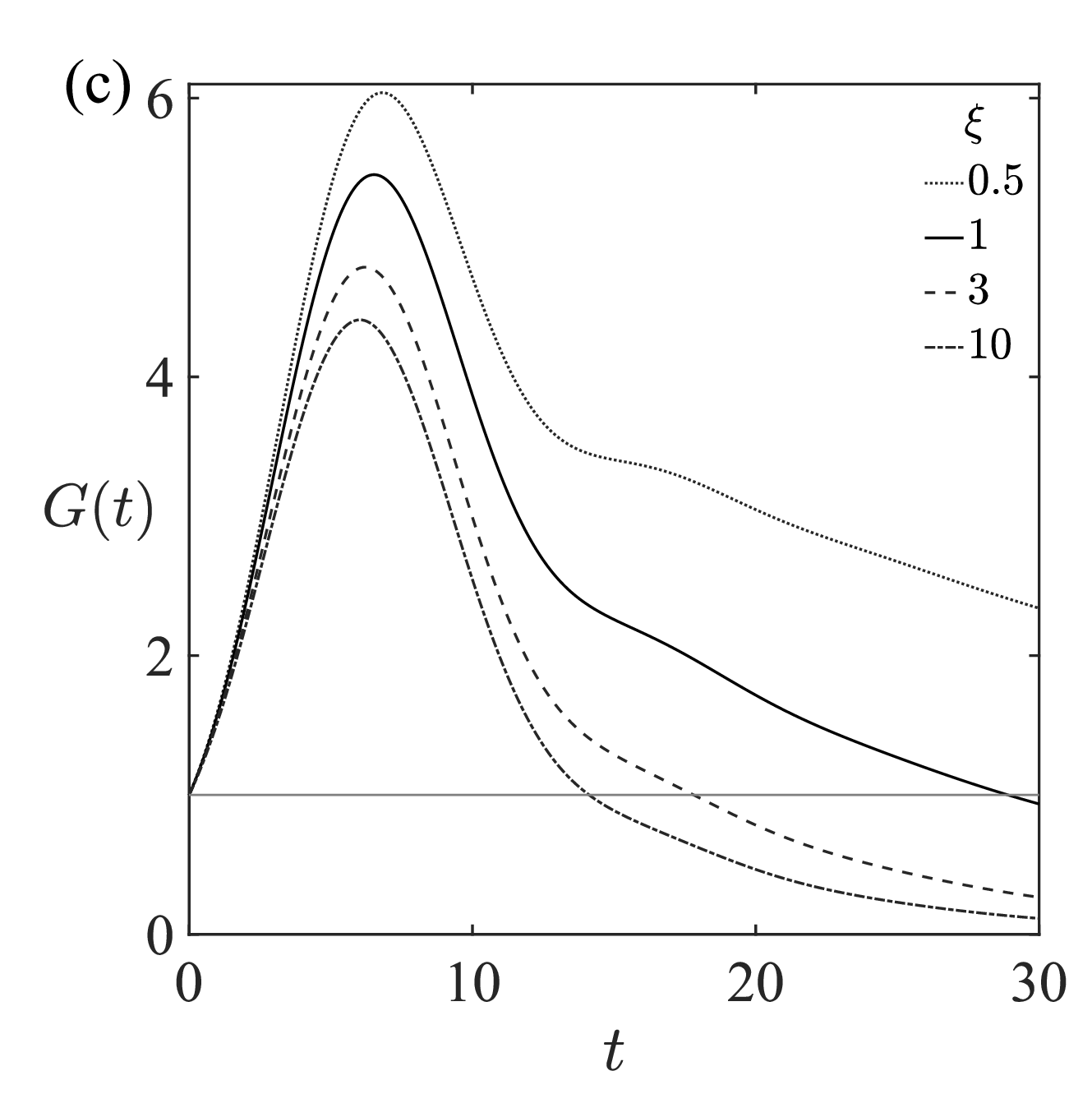}
   \includegraphics[scale=0.25]{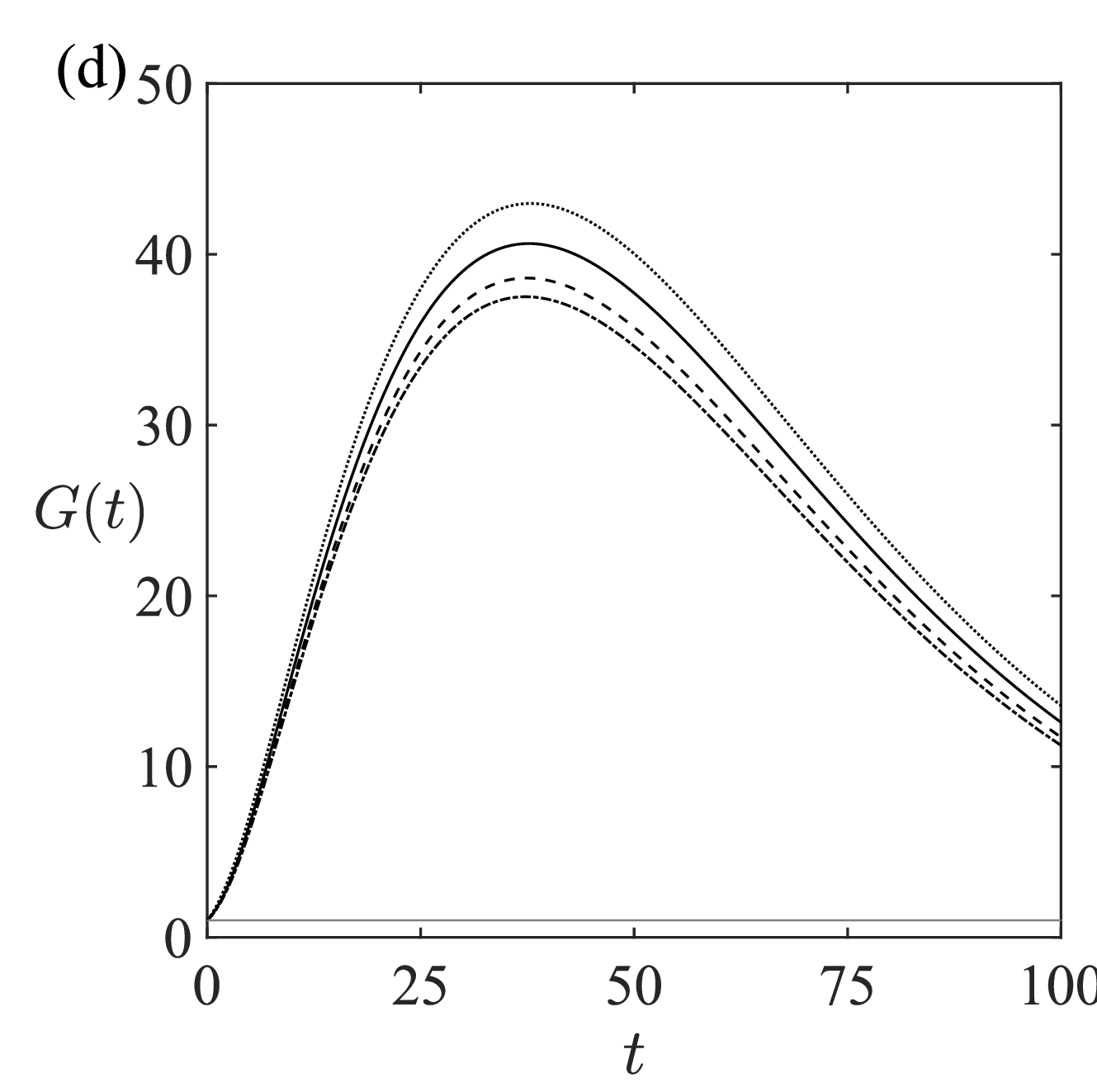}
\\
    \includegraphics[scale=0.25]{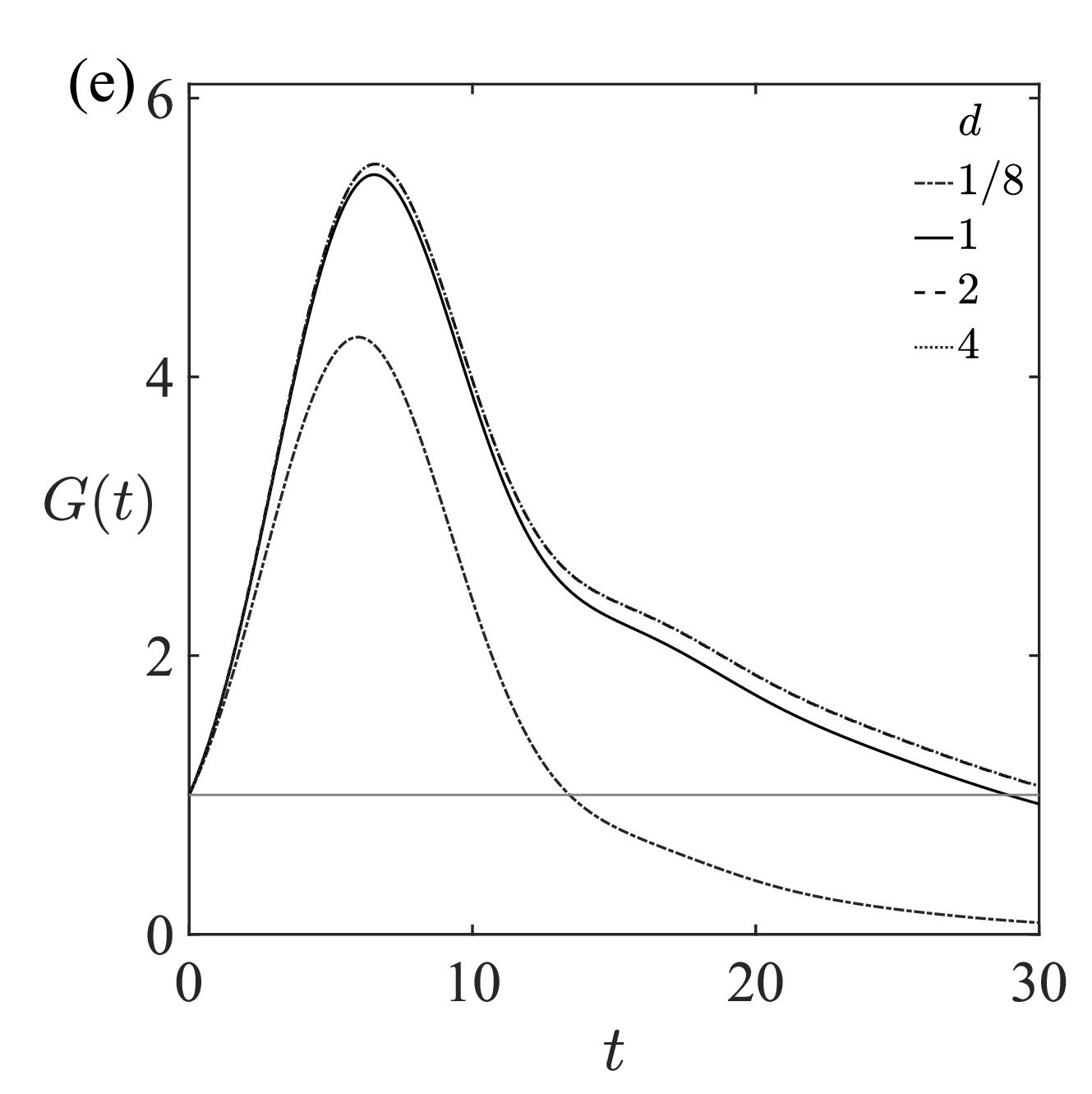}
   \includegraphics[scale=0.25]{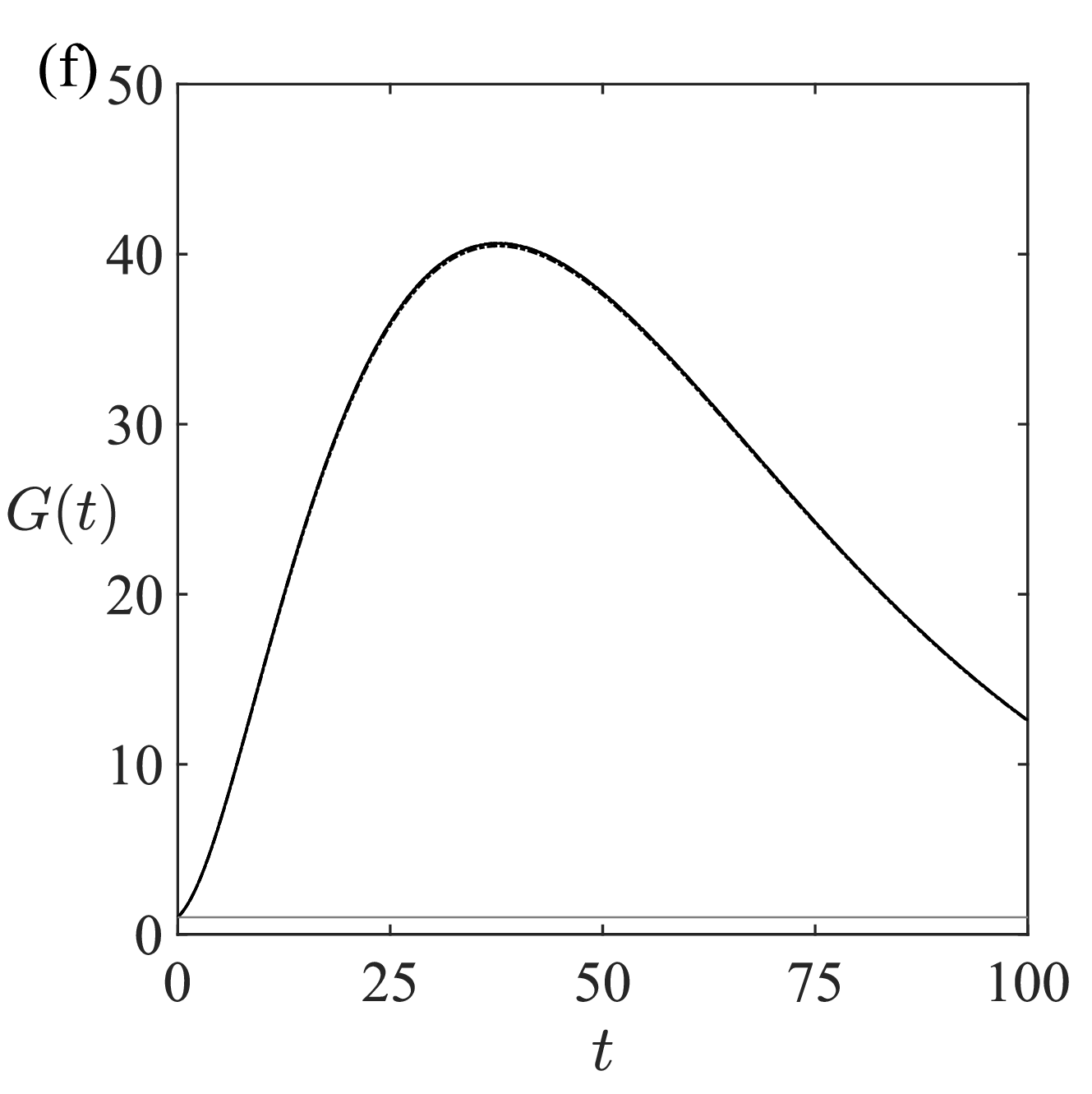}
    \caption{\small [$Re=400$] Growth function $G(t)$ for spanwise independent with $(\alpha, \beta)=(1,0)$ [first column] and streamwise independent with $(\alpha, \beta)=(0,1)$ [second column] perturbations. The effect of (a, c) the mean permeability $\sigma$ when $\xi=d=1$,  (b, d) the anisotropy parameter $\xi$ when $d=1$ and $\sigma=0.02$, and (e, f) the depth ratio $d$ when $\xi=1$, $\sigma=0.02$.
    }
    \label{fig:growth_function}
\end{figure}

In the context of classical PPF, the growth function is analyzed by~\citet{reddy_henningson_1993}. They showed that when $Re<Re_e$, $G(t)<1$ overall time, yielding $G_{\max}=1$ and $t_{\max}=0$.
However, when $Re_c<Re$, $G_{\max}$ is unbounded due to the unstable eigenvalues ($\omega_i>0$) of the linear operator $\mathscr{L}$. 
For Reynolds numbers $Re$ in the range $(Re_e,Re_c)$, the PPF is linearly stable but undergoes transient amplification, resulting in a finite 
$G_{\max}>1$. Note that we also observe similar behavior in the present set-up. However, the initial growth is controlled by the anisotropic permeability and depth ratios. For analyzing the growth function $G(t)$ and $G_{\max}$ in the present problem, we choose the $Re =400 \in (Re_e,\,Re_c)$.

Figure~\ref{fig:growth_function} displays the effect of physical parameters such as mean permeability $\sigma$ (panels a and b), anisotropy parameter $\xi$ (panels c and d), and depth ratio $d$ (panels e and f) on the initial growth $G(t)$ of the disturbance. In each panel, one physical parameter among three $\sigma$, $\xi$, and $d$ is varied while the other two are fixed. 
Panels in the first column (a,c,e) and second column (b,d,f) correspond to spanwise $\beta=0$ and streamwise $\alpha=0$ independent disturbances. 
It is seen that the growth function starts to increase and achieve a peak value $G_{\max}$ at $t=t_{\max}$,
which then decreases sharply. 
Both $G_{\max}$ and $t_{\max}$ are amplified with increasing mean permeability (see panels a and b). However, this amplification is considerably higher for streamwise independent (panel b) perturbations than spanwise independent (panel a)  perturbations. 
In contrast to mean permeability, the growth function $G(t)$ decreases with increasing anisotropy parameter $\xi$ at any time incident (see panels c and d). Consequently, $G_{\max}$ and $t_{\max}$ both decrease with an increase in $\xi$. 
For the depth ratio $d$ (see panels e and f), the maximum growth $G_{\max}$ 
%time corresponding to maximum growth 
and corresponding 
$t_{\max}$ are increasing. However, the increment beyond $d=1$ is small. 
%It is worth noticing that 
The growth function $G(t)$ is nearly identical for all $d \ge 1$. 
Moreover, $G(t)$ 
is seen to be overlapped for streamwise independent perturbations (see panel~f). 
The growth function for spanwise independent (first column) perturbation is 
%approximately 7 to 8 time larger than
always much lower than spanwise independent (second column) perturbations. 
To what follows, there exists notable transient growth $G(t)$ at a sub-critical Reynolds number of $Re=400<Re_c$, and this amplification of transient growth becomes more pronounced as the Reynolds number increases (figure not shown).

Increasing mean permeability $\sigma$ or decreasing anisotropic parameter $\xi$ enhances permeability by accumulating more fluid in the porous layers. Furthermore, the enhanced permeability thus intensifies transient growth. Also, an increasing depth ratio allows more fluid into the porous layer, similar to permeability, enhancing the maximum growth. Thus, the fluid and porous layers jointly contribute to the transient growth.

% --------------------------------
\subsubsection{Maximum growth $G_{\max}$}
% ---------------------------------------

\smallskip

\begin{figure}
   \centering
    {\includegraphics[scale=0.3]{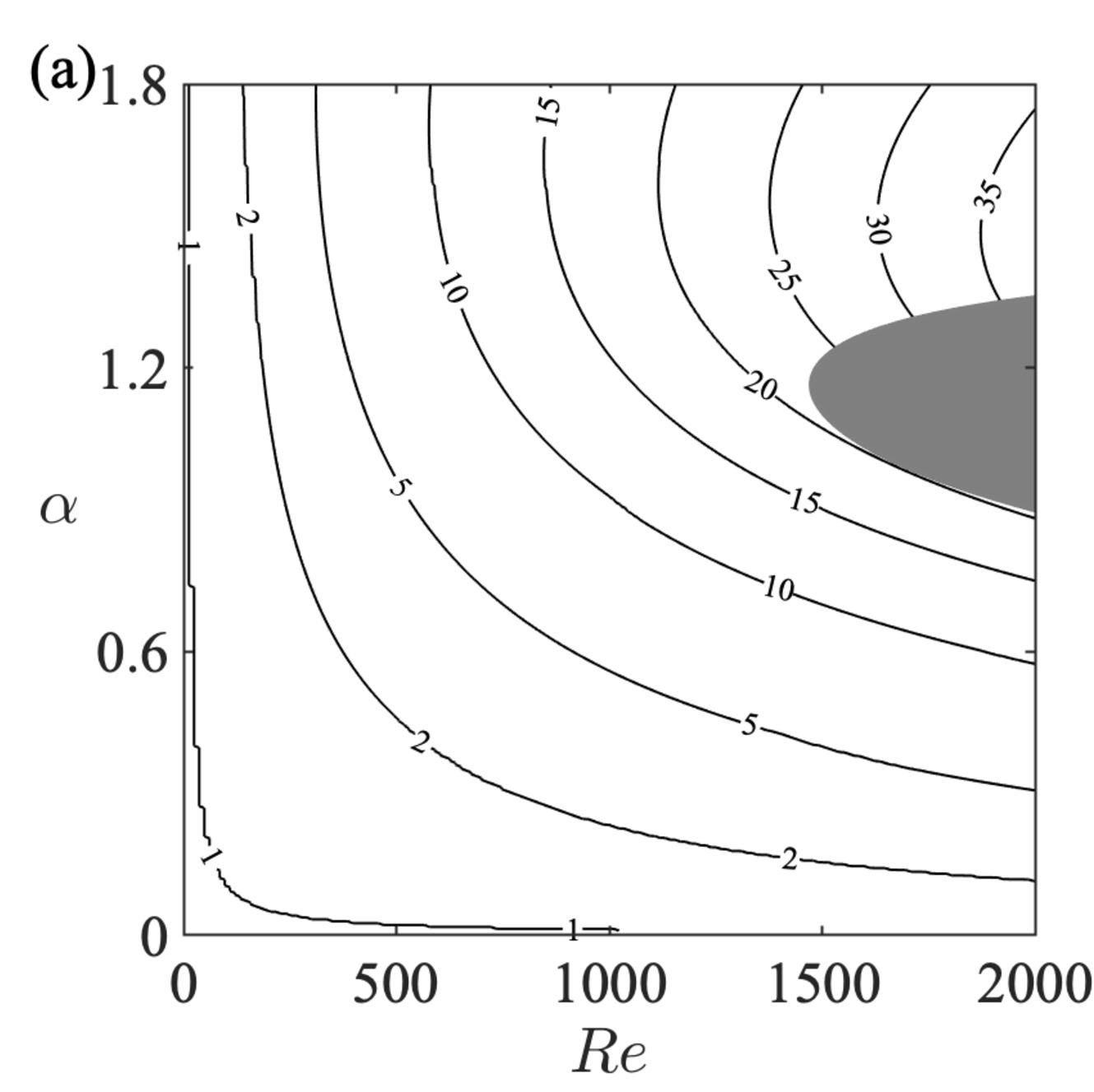}}\quad
{\includegraphics[scale=0.3]{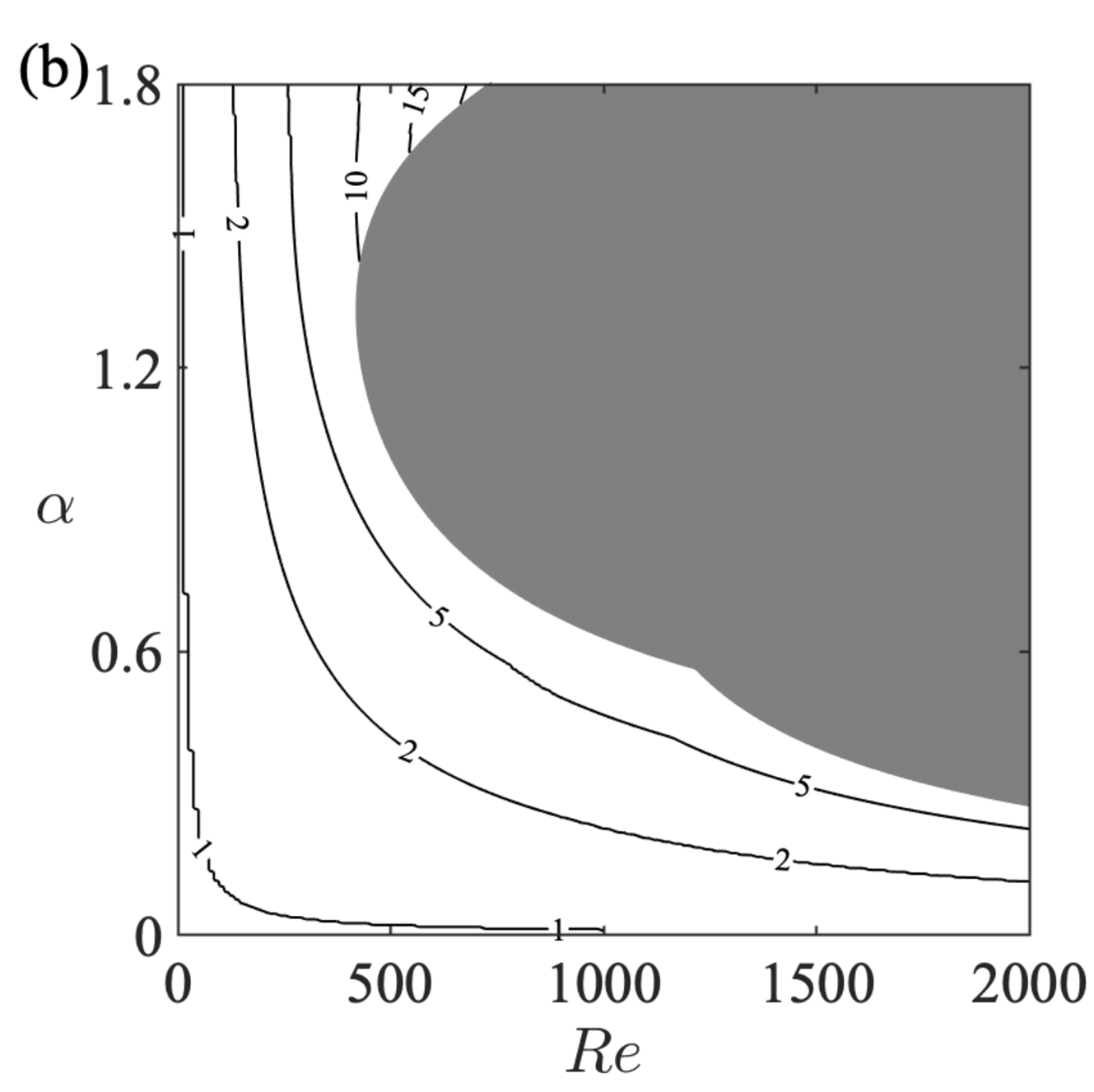}}
    \caption{\small [$\xi=d=1$] Contours of maximum energy growth $G_{\max}$ 
    in the $(Re, \alpha)$-plane for spanwise independent ($\beta=0$) perturbation with two values of $\sigma$: (a) $\sigma=0.006$, and (b) $\sigma=0.02$. 
    The shaded part represents unstable modes (where $G_{\max} \to \infty$). 
    }
    \label{fig:growth_curves_Sigma}
\end{figure}

Next, in figure~\ref{fig:growth_curves_Sigma}, we display the contours of the maximum growth $G_{\max}$ in the $(Re, \alpha)$-plane for two values of mean permeability with spanwise independent perturbations; other parameters are fixed with $\xi=d=1$. 
The shaded region represents parameters where the flow is asymptotically unstable $(\omega_i>0)$, 
allowing unbounded amplification due to exponential growth. Thus, $G_{\max}$ is infinite in this region.
Interestingly,
maximum growth $G_{\max}$ is significant, which
increases as the Reynolds number $Re$ increases for both low and high mean permeability. 
This increase in $G_{\max}$ can also be observed in figure~\ref{fig:growth_function}(a).
Furthermore, at a fixed $Re$, the short-wave modes are more favorable to achieve maximum transient growth than the long-wave modes ($\alpha \approx 0$). 
Similarly to the mean permeability, increasing the depth ratio $d$ or decreasing the anisotropy parameter $\xi$ enhances the maximum growth $G_{\max}$.
The modal analysis determines the mechanism for the transition to turbulence at larger $\sigma$ and lower $\xi$ as the exponential growth zone advances to a lower Reynolds number, decreasing the possibility of transient energy amplification.

\begin{figure}
    \centering
    {\includegraphics[scale=0.3]{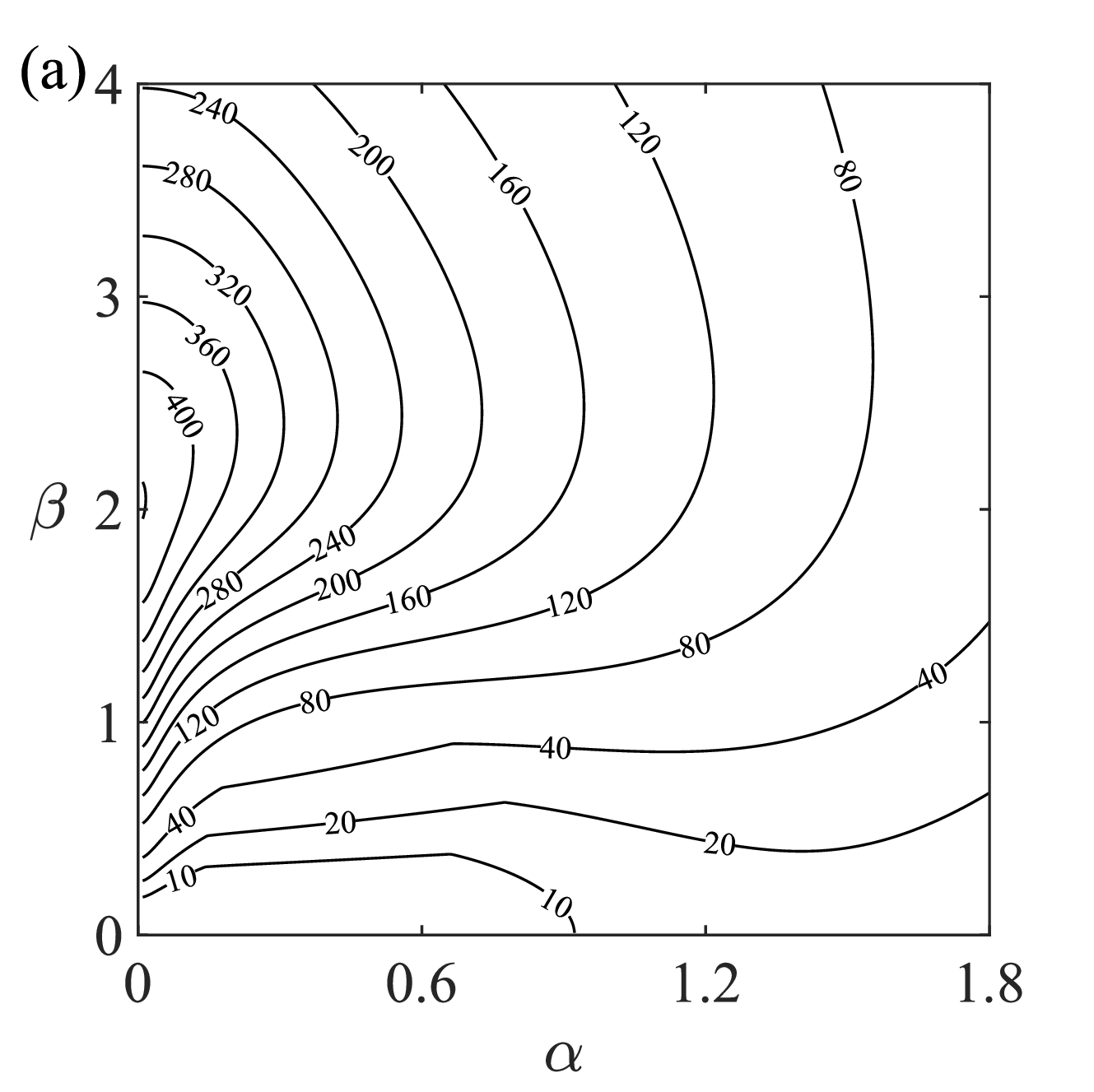}}
    {\includegraphics[scale=0.3]{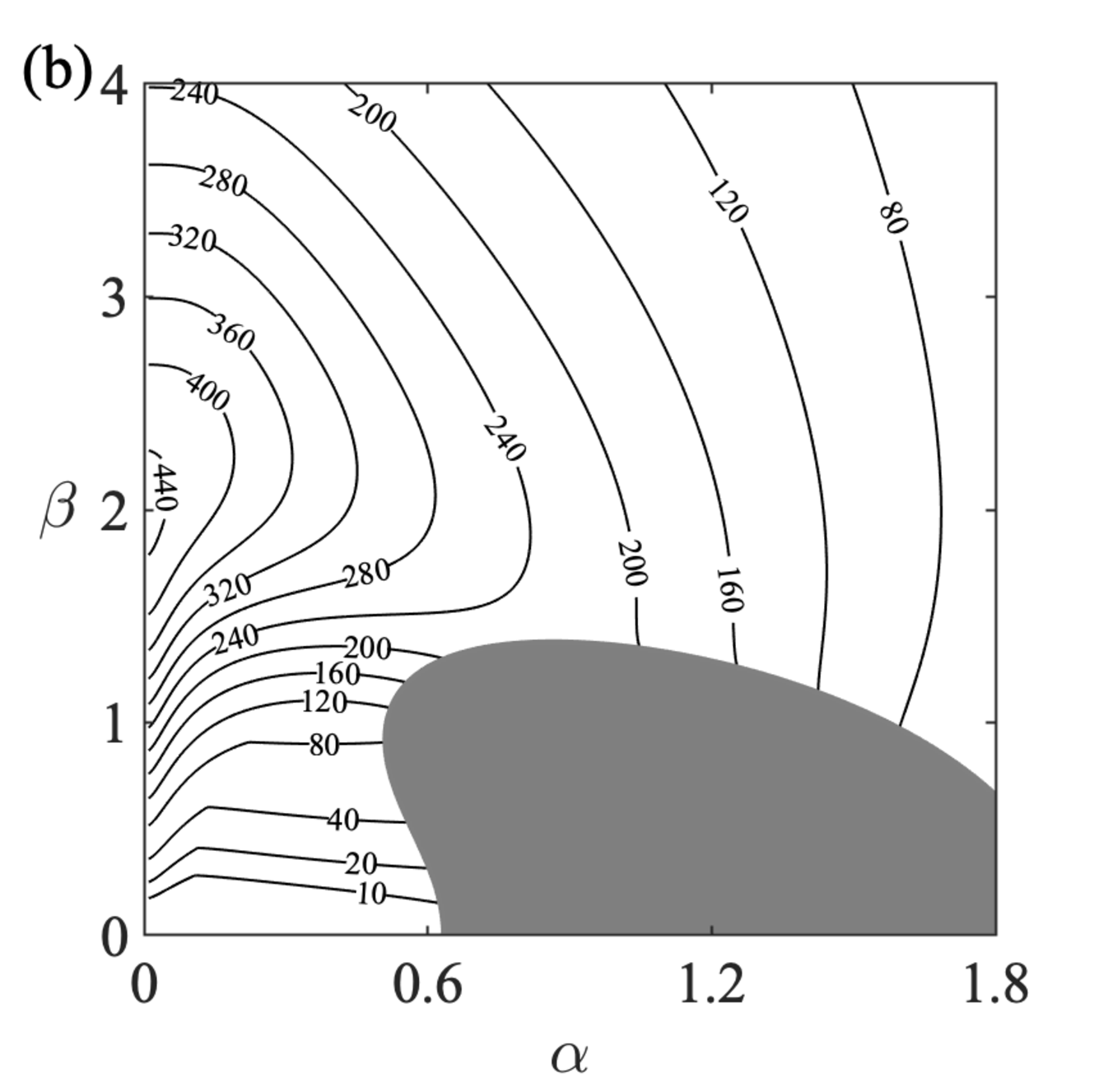}}\\
    {\includegraphics[scale=0.3]{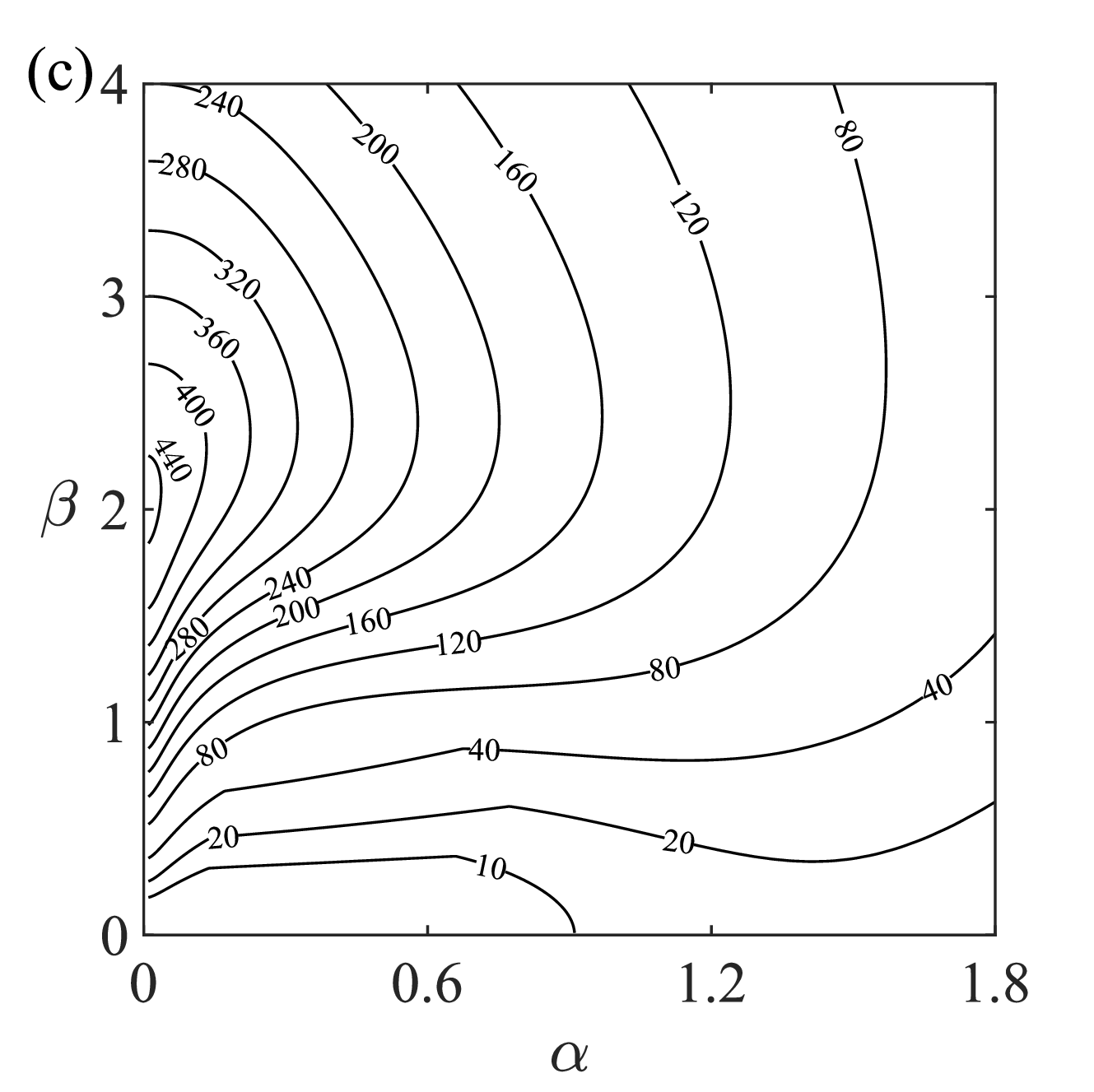}}
    {\includegraphics[scale=0.3]{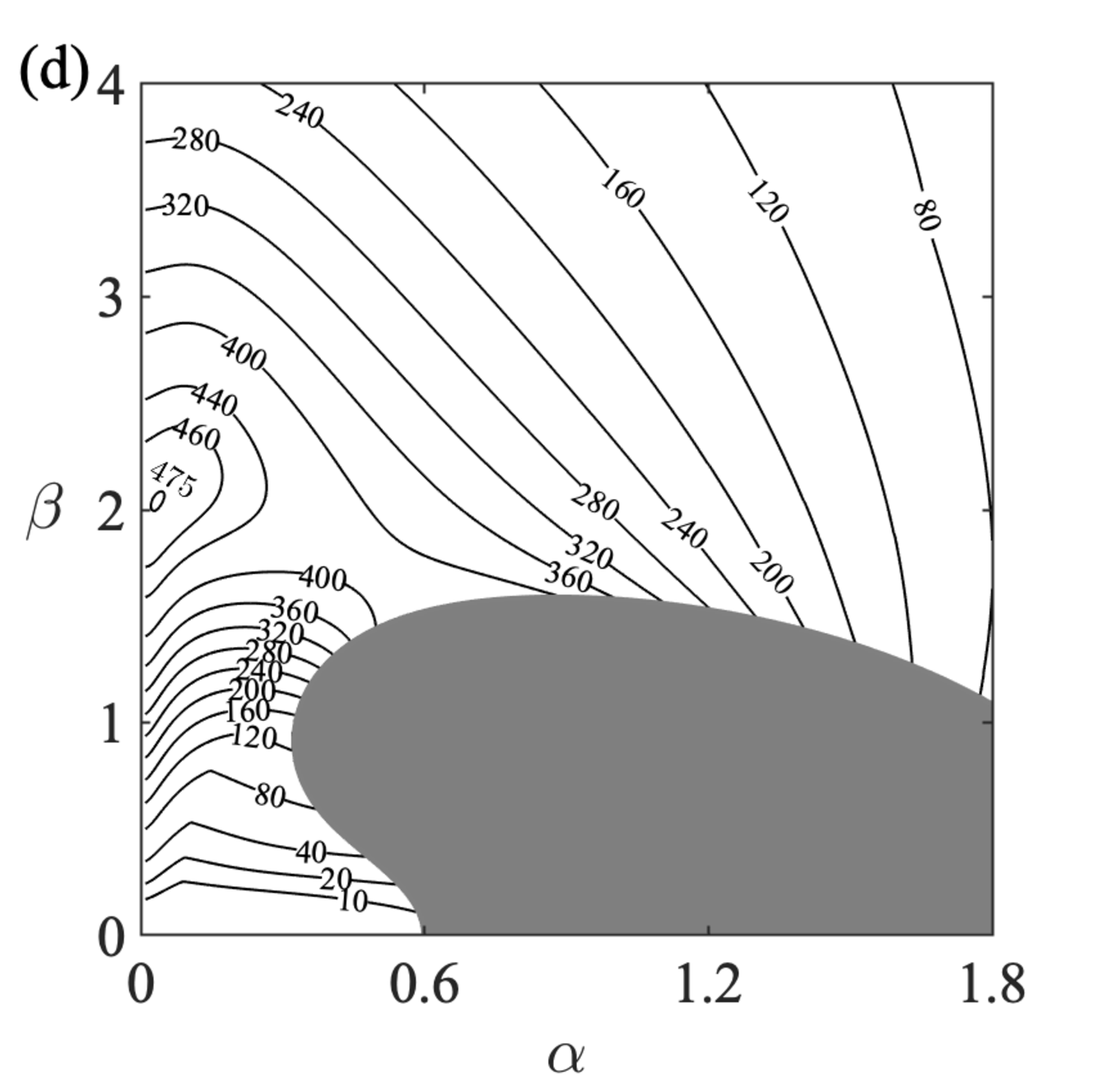}}
    \caption{\small [$Re=1000$,\,$d=1$] Contours of the maximum energy growth $G_{\max}$ in the $(\alpha, \beta)$-plane for $(\sigma, \xi)=$ (a) (0.006,1), (b) (0.02,1), (c) (0.006,0.5) and (d) (0.02,0.5). 
    The flow is linearly unstable in the shaded region ($\omega_i>0$). 
    }
    \label{fig:growth_curves_Sigma2}
\end{figure}

To understand the role of oblique perturbations ($\alpha, \beta \ne 0$) on transient growth, 
the contours of $G_{\max}$ 
in $(\alpha,\,\beta)$-plane 
for various $(\sigma,\,\xi)$ values are displayed in figure~\ref{fig:growth_curves_Sigma2} at $Re=1000$. 
%where the shaded area denotes the asymptotically unstable $(\omega_i>0)$ region. 
The mean permeability varies from $0.006$ to $0.02$ (row-wise), and anisotropy parameter from $1$ to $0.5$ (column-wise). Clearly, significant transient growth is observed throughout the $(\alpha, \beta)$-plane despite the flow being asymptotically stable. 
It is seen in panels a--d that for a fixed wavenumber $\beta$, $G_{\max}$ 
increases (decreases) with decreasing (increasing) $\alpha$, which implies that the transient growth is suppressed (enhanced) in the short-wave (log-wave) regime.  
%Similar features are seen in all other parameters (see panels b--d).
While contours of maximum growth do not vary much with increasing $\sigma$ (compare left and right panels), exponential growth region (shaded area) increases when increasing $\sigma$. 
% It is seen that as $\sigma$ ($\xi$) increases (decreases), the optimal growth $G_{\rm opt}$ increases (decreases) along with the associated optimal time $t_{\rm opt}$, see table~\ref{table:transient_growth_values}. 
As $\sigma$ increases or $\xi$ decreases, the optimal growth $G_{\rm opt}$ increases along with the optimal time $t_{\rm opt}$; see also table~\ref{table:transient_growth_values}. 
%It is worth noticing from figure~\ref{fig:growth_curves_Sigma2} that 
For $\xi=1$ (isotropic case), 
$\alpha_{\rm opt}=0$ 
%corresponding to $G_{\rm opt}$ 
is always zero
% , irrespective of $\xi$ value 
(see panels a and b),
even for small $\sigma$ and $\xi <1$, %optimal %streamwise wavenumber 
$\alpha_{\rm opt}$ remains zero with $\beta_{\rm opt}$ being non-zero (panel c). For sufficient large $\sigma$ with $\xi < 1$ (panel d), $G_{\rm opt}$ is achieved at  oblique perturbation with $(\alpha, \beta)$ being approximately $(0.04, 2)$, wherein 
%Furthermore, 
the exponential growth zone (shaded region) is more expanded (also see neutral stability curve shown in figure~\ref{fig:temporal_growth}). 
Note that 
%As discussed earlier in \S~\ref{subsec:modal_stability}, 
for very small anisotropic parameters and sufficiently large mean permeability, the exponential growth zone
encompasses the entire $(\alpha,\,\beta)$-plane (see figure~\ref{fig:temporal_growth}d), leading to infinite energy growth.
%$G_{\max}, G_{\rm opt} \to \infty$.  
% {\color{red} ASK SUPRIYO: Therefore, for very small $\xi$ and higher $\sigma$, the instability is fully understood by the modal stability. }
By comparing figures~\ref{fig:growth_curves_Sigma}(a) and~\ref{fig:growth_curves_Sigma}(b) with figures~\ref{fig:growth_curves_Sigma2}(a) and~\ref{fig:growth_curves_Sigma2}(b), we can conclude that
the transient energy growth of 
three-dimensional 
oblique disturbances is significantly higher than the two-dimensional $\alpha \neq 0, \beta=0$ ones.

 \begin{figure}
    \centering
    \includegraphics[scale=0.33]{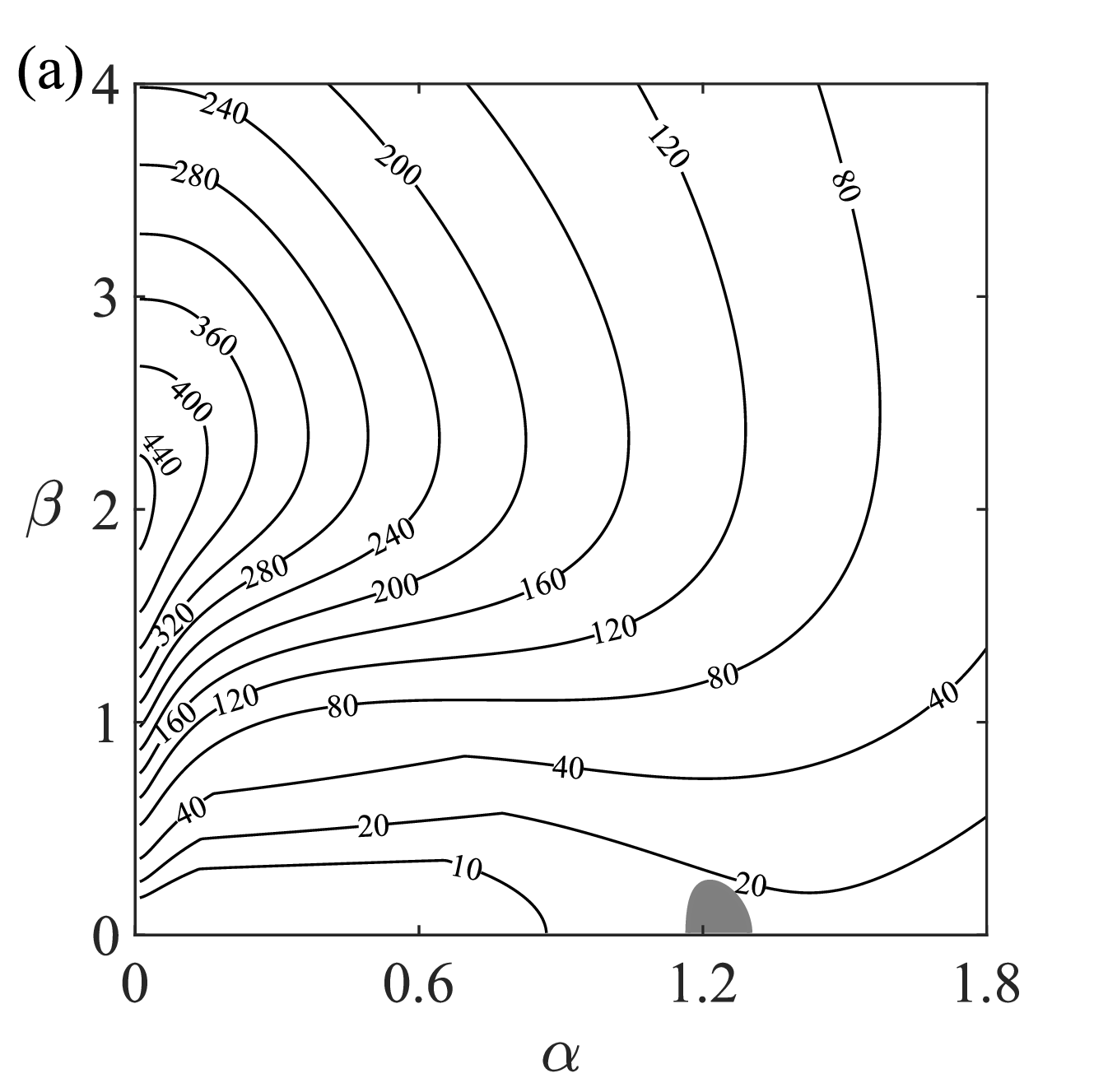}\quad
    \includegraphics[scale=0.33]{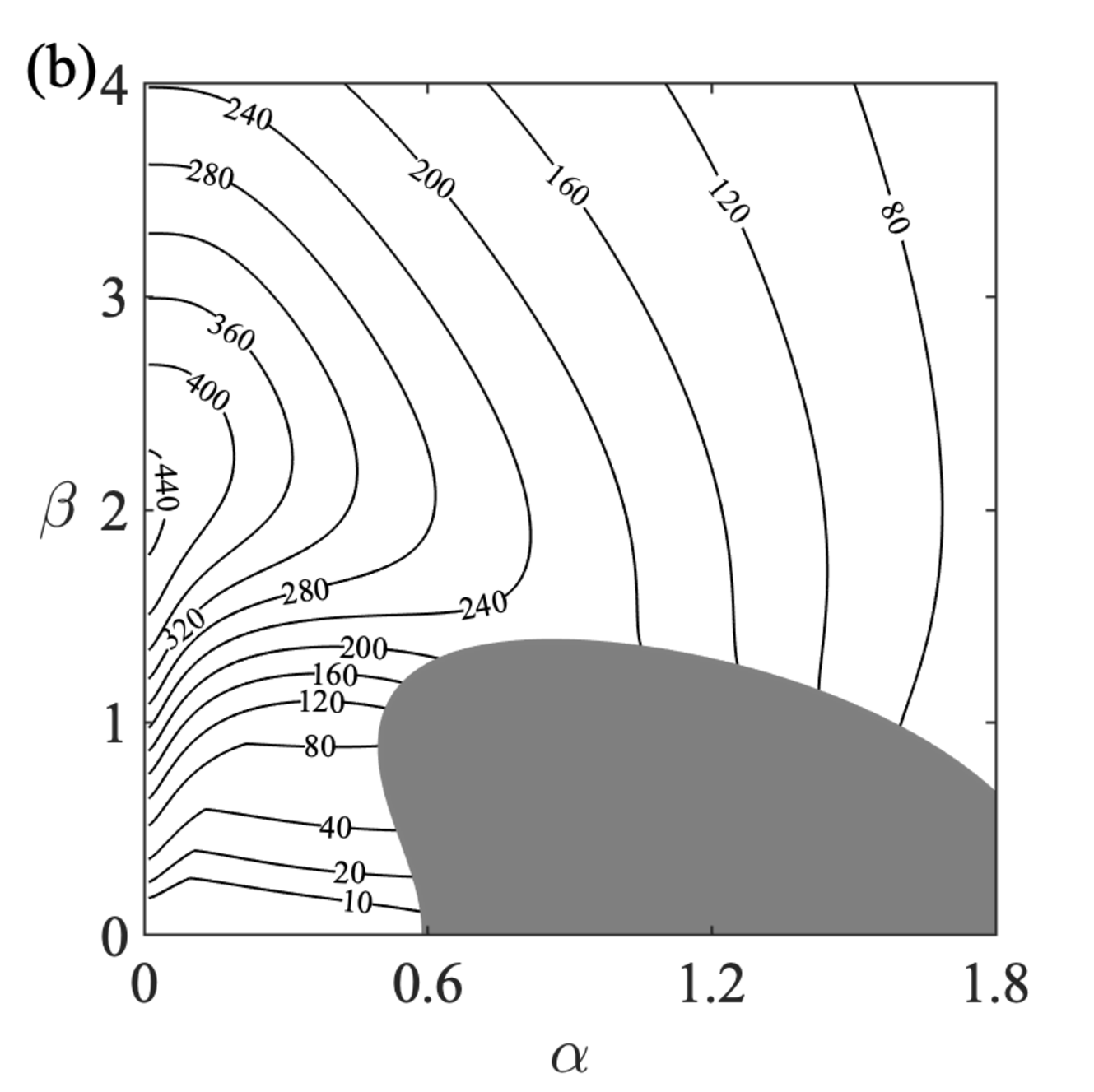}
    \caption{\small [$Re = 1000, \sigma= 0.02, \xi = 1$] Same as figure~\ref{fig:growth_curves_Sigma2}(b) but for $d\neq 1$: (a) $d=1/8$ and (b) $d=2$. 
     }
    \label{fig:growth_curves_Depth}
\end{figure}

\begin{table}
    \centering
  \begin{tabular}{@{}|lll|l|l|l|l|@{}}
         \hline
          $\sigma$ & $\xi$ & $d$ & $G_{\rm{opt}}$ & $t_{\rm{opt}}$ & $\alpha_{\rm opt}$ & $\beta_{\rm opt}$\\
       \hline
        % \toprule
          0.006 & 1 & 1 & 441.78 & 76.88 & 0 & 2.017 \\
          0.02  & 1 & 1 & 449.86 & 77.88 & 0 & 2.017 \\
          0.006 & 0.5 & 1 & 446.82 & 76.48 & 0 & 2.041 \\
          0.02 & 0.5 & 1 & 475.26 & 73.27 & 0.04 & 2.041 \\[0.2em]
          0.02 & 1 & 0.125 & 448.26 & 78.08 & 0 & 2.017 \\
          0.02 & 1 & 2     & 449.86 & 77.68 & 0 & 2.025 \\
          \hline
    % \bottomrule
    \end{tabular}
 \caption{The optimum growth and time $(G_{\operatorname{opt}},t_{\operatorname{opt}})$, and the corresponding wavenumbers $(\alpha,\beta)$ at $Re=1000$ for various sets $(\sigma,\xi,d)$.}
 \label{table:transient_growth_values}
\end{table}

Figure~\ref{fig:growth_curves_Depth} illustrates the effects of the depth ratio on 
transient growth in $(\alpha,\,\beta)$-plane
%associated with oblique perturbations, 
with other parameter being the same as
figure~\ref{fig:growth_curves_Sigma2}(b). 
%for $(\sigma,\xi)=(0.02,1)$ at $Re=1000$. 
% It should
% be noted that a smaller depth ratio $d<1 $ corresponds to thinner porous walls, while a higher
% depth ratio $d>1$ corresponds to thicker porous walls.
The $G_{\max}$ contours for smaller and larger depth ratios look nearly similar to that of equal-width layer $d=1$ case (figure~\ref{fig:growth_curves_Sigma2}b). 
In contrast the exponential growth regime 
increases with increasing depth ratio.

The optimal energy growth and other related to optimal parameters~(7.9) are listed in Table~\ref{table:transient_growth_values}.  
For $\sigma=0.02$, $G_{\rm{opt}}$ increases with $d\leq 1$ (see 2nd and 5th rows of Table~\ref{table:transient_growth_values}), but remains unchanged for $d>1$ (see 2nd and 6th rows of Table~\ref{table:transient_growth_values}).
For isotropic porous walls ($\xi=1$), 
%it is seen that for all values of $d$, 
the optimal growth $G_{\rm{opt}}$ is achieved 
at $\alpha=0$ (streamwise independent perturbations). 
%when considering streamwise independent perturbations ($\alpha=0$) 
%for all considered values of $d$.  
It has also been verified that 
when channel has
thicker porous walls ($d\geq 1$) with significant mean permeability along with increased cross-stream permeability $\xi<1$,  $G_{\rm{opt}}$ is attained at oblique wavenumber.  %($\alpha, \beta \neq 0$). 
To summarize, the transient growth map at a sub-critical Reynolds number ($Re<Re_c$) demonstrates that the perturbation kinetic energy can increase substantially due to changes in mean permeability, anisotropy parameter, and depth ratio before it finally decays.

% -----------------------------------------
\subsubsection{Perturbation flow field}
\label{subsec:optimum_flow_field}
% --------------------------------------------------------------

% \begin{center}
%     Perturbation flow field
% \end{center}
% \smallskip

% \begin{figure}
%     \centering
%     \includegraphics[scale=0.24]{opti_input_l1_sg0p02_s1.eps}
%     \includegraphics[scale=0.24]{opti_output_l1_sg0p02_s1.eps}
%     \includegraphics[scale=0.24]{opti_input_l1_sg0p02_s0p5.eps}
%     \includegraphics[scale=0.24]{opti_output_l1_sg0p02_s0p5.eps}
%     \caption{\small
%      Cross-stream ($(y,z)$-plane) representation of the optimal initial condition [first column; $(\widetilde{v},\widetilde{w})$-velocity field] and optimal response [second column; contours of $\widetilde{{u}}$] 
%     % of the initial value problem 
%     % for the streamwise independent disturbance 
%     at $\sigma=0.02$ and  for $\xi=$ 
%     (a) $1$ and (b) $0.5$. 
%     %[Figure~\ref{fig:growth_curves_Sigma2}]. 
%     Other parameters are the same as figure~\ref{fig:growth_curves_Sigma2} (b).}
%     \label{fig:optimal_anisotropy}
% \end{figure}

\begin{figure}[!htbp]
    \centering
    \includegraphics[scale=0.25]{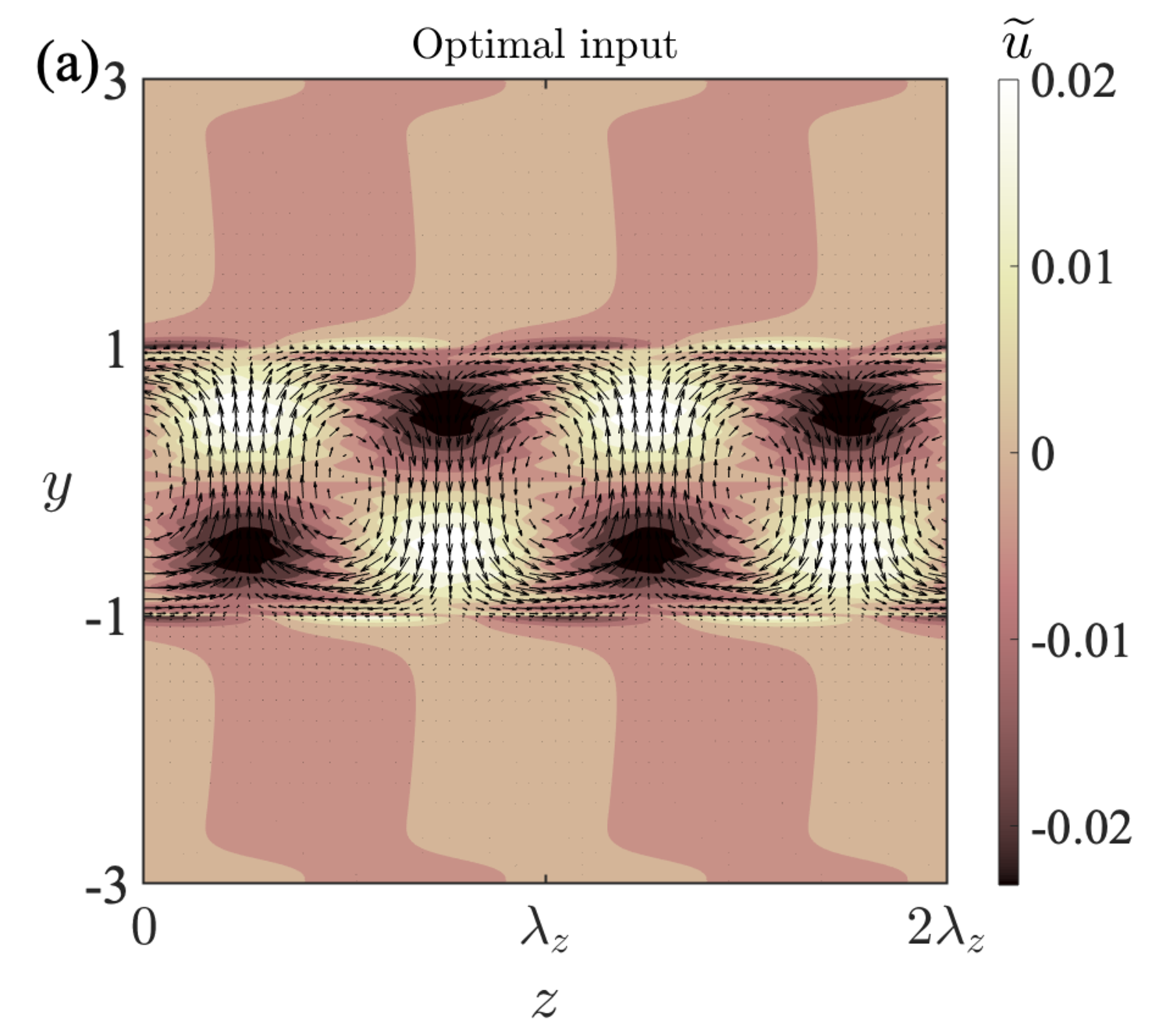}
    \includegraphics[scale=0.25]{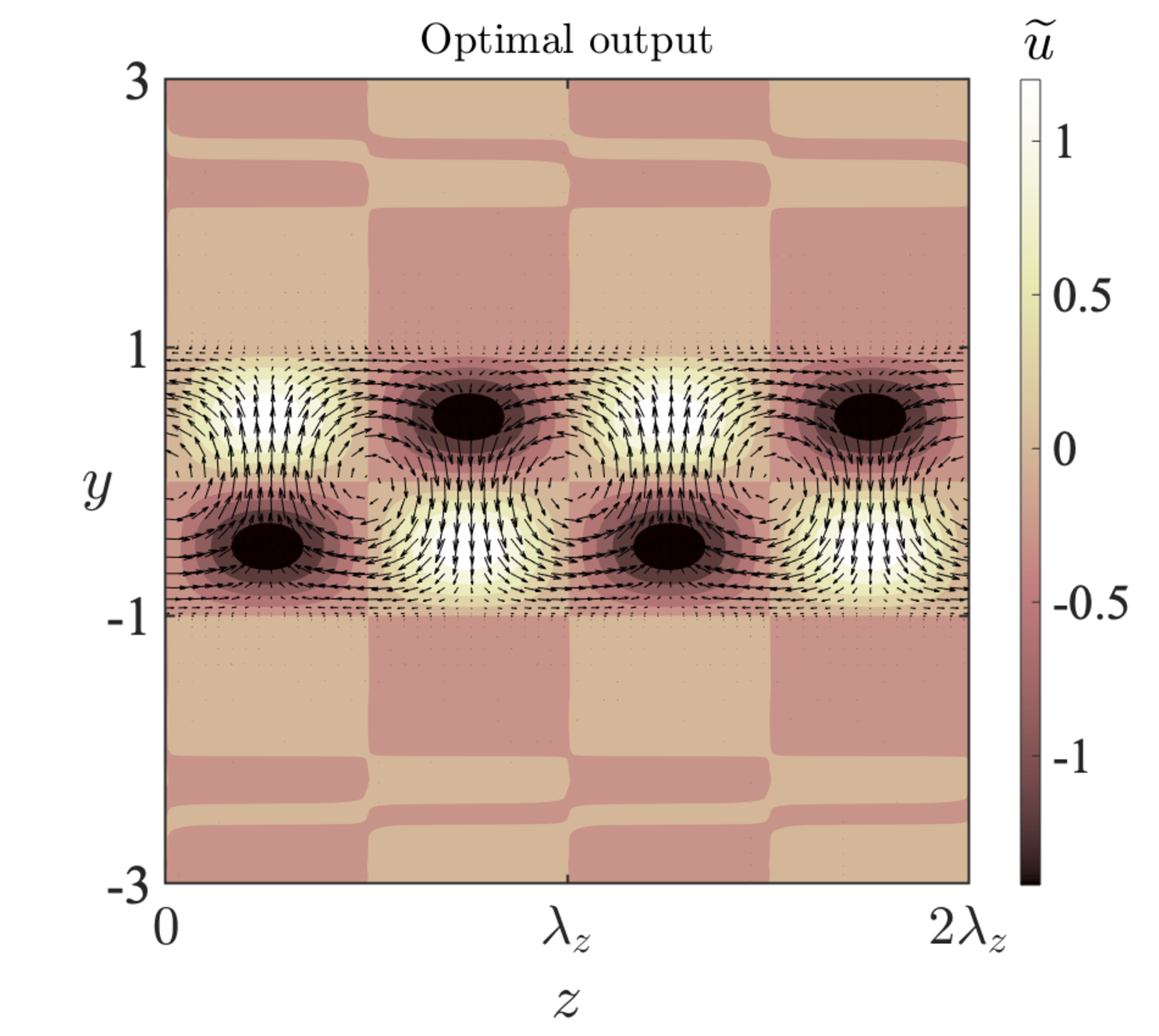}
    \includegraphics[scale=0.25]{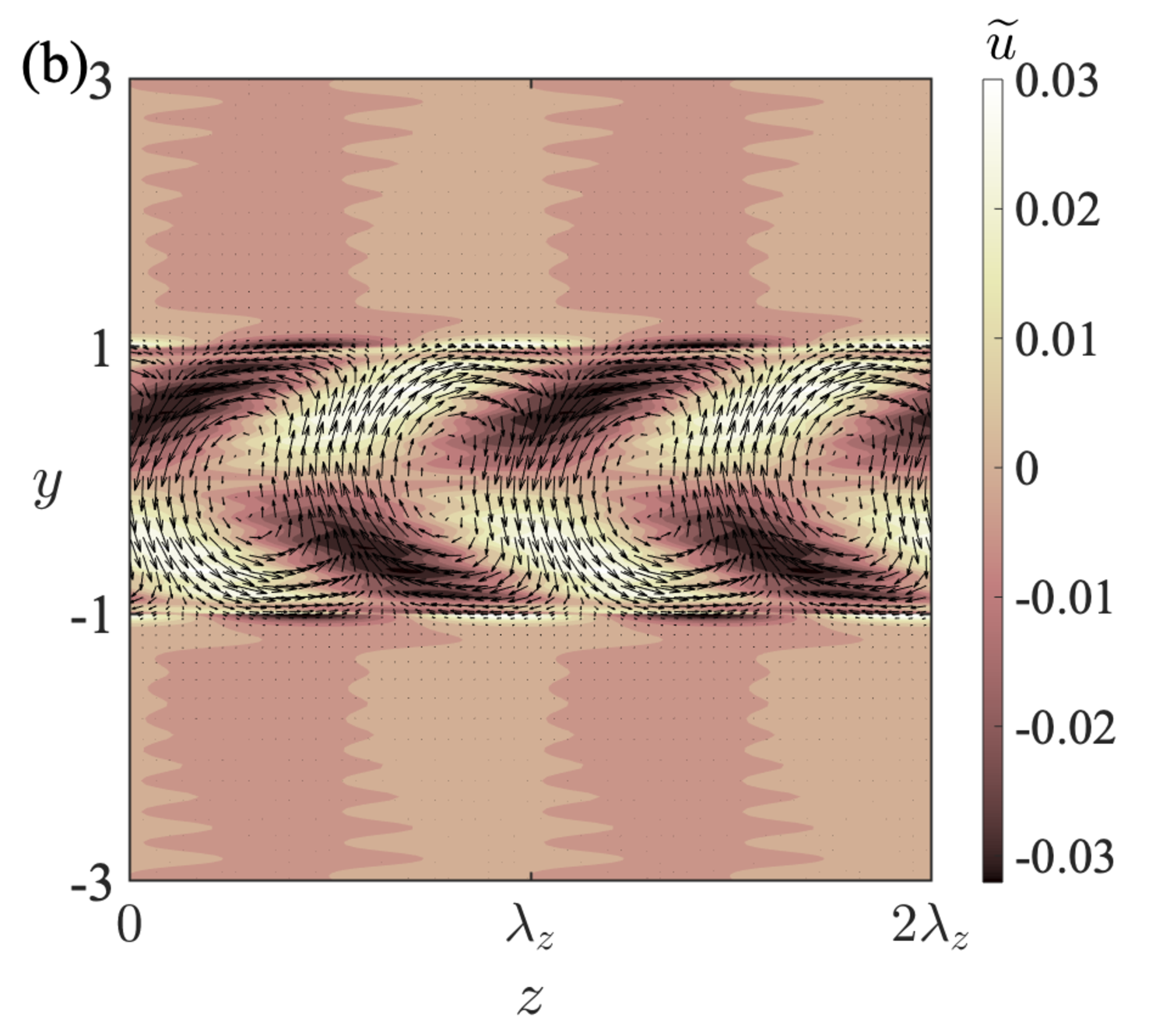}
    \includegraphics[scale=0.25]{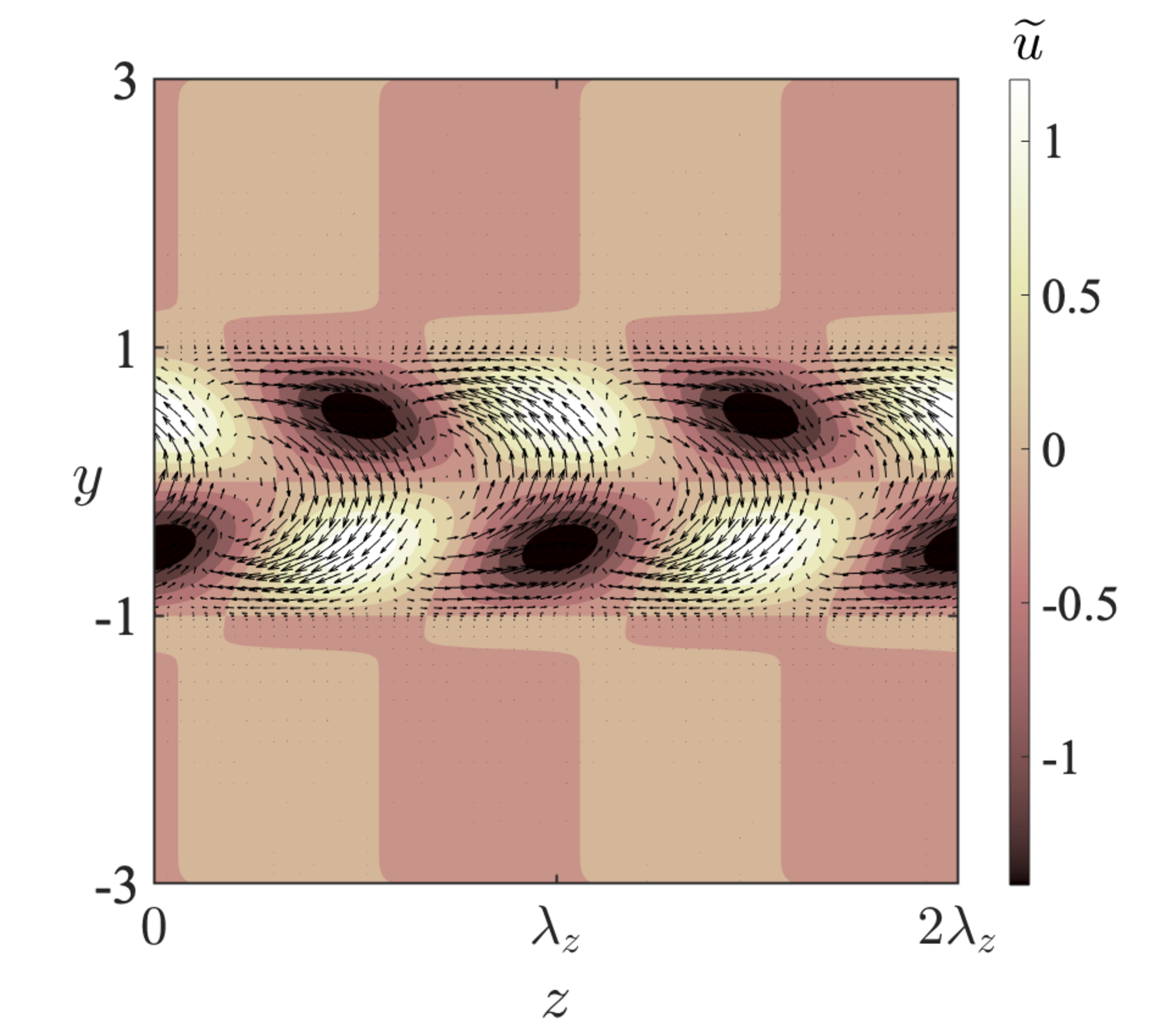}
    \caption{\small Variation of optimal perturbation field---optimal input (left) and optimal response (right): 
    Contours of streamwise velocity $\widetilde{u}$ and cross stream velocity field $(\widetilde{v},\widetilde{w})$ 
    in $(y,z)$-plane for $Re=1000$, $d=1$,  $\sigma=0.02$, when $\xi=$ (a) 1 and (b) 0.5.
   % Cross stream representation ($(y,z)$-plane) of the perturbation field [$\widetilde{u}$-contours and $(\widetilde{v},\widetilde{w})$-vector field] corresponding to the optimal growth at $Re=1000,\,d=1,\; \sigma=0.02$, when $\xi=$ (a) 1 and (b) 0.5. 
   % The velocity amplitudes  
    %of the input
    % transverse and spanwise velocity 
    %shown in 
    For panel~a: optimal input
    $(|\widetilde{u}|, |\widetilde{v}|, |\widetilde{w}|)$ =$(0.023, 1.051, 1.015)$, %$|\widetilde{v}|=1.051$ and $|w|=1.015$, 
   % while for the output are 
   and optimal output 
    $(|\widetilde{u}|, |\widetilde{v}|, |\widetilde{w}|)$=$(1.421, 0.025, 0.019)$.    
    % $|\widetilde{u}|=1.421$, $|\widetilde{v}|=0.025$ and $|\widetilde{w}|=0.019$. 
    %At $\xi=0.5$ (panel~b), amplitude of the input 
    % transverse and spanwise 
    %velocity are 
     For panel~b: optimal input
 $(|\widetilde{u}|, |\widetilde{v}|, |\widetilde{w}|)$ =$(0.032, 0.981, 0.991)$, and optimal output  $(|\widetilde{u}|, |\widetilde{v}|, |\widetilde{w}|)$ =$(1.421,0.027,0.030)$. See figures~\ref{fig:growth_curves_Sigma2}(b),~\ref{fig:growth_curves_Sigma2}(d), and 2nd and 4th rows of Table~\ref{table:transient_growth_values}. 
% $|\widetilde{u}|=0.032$, $|\widetilde{v}|=0.981$ and $|w|=0.991$, while for the output are $|\widetilde{u}|=1.421$, $|\widetilde{v}|=0.027$ and $|\widetilde{w}|=0.030$. 
    }
    % \caption{\small
    %  Cross-stream ($(y,z)$-plane) representation of the optimal initial condition [first column; $(\widetilde{v},\widetilde{w})$-velocity field] and optimal response [second column; contours of $\widetilde{{u}}$] 
    % % of the initial value problem 
    % % for the streamwise independent disturbance 
    % at $\sigma=0.02$ and  for $\xi=$ 
    % (a) $1$ and (b) $0.5$. 
    % %[Figure~\ref{fig:growth_curves_Sigma2}]. 
    % Other parameters are the same as figure~\ref{fig:growth_curves_Sigma2} (b).}
    \label{fig:optimal_anisotropy}
\end{figure}

% We determine the evolution of the perturbation structure as time goes.

This section portrays the structure of optimal perturbations and their responses that maximize transient growth, thus focusing on the disturbances that achieve the highest transient growth $G_{\operatorname{opt}}$ in time $t=t_{\rm{opt}}$ for a given parameter. 
% In this section, we determine the structure of 
% optimal perturbations and responses that amplify the transient growth function.
% To do this, we focus on the disturbance that achieves the largest amplification $G_{\operatorname{opt}}$ in time $t_{\operatorname{opt}}$ for a given set of parameters. 
Similarly to the optimal forcing~\eqref{eqn:optimal_frequency}, each point on the growth curve $G(t)$ is listed by optimizing the ratio $\nicefrac{\| \boldsymbol{q}(y,t)\|}{\|\boldsymbol{q}_0\|}$ over initial conditions. Thus, each point in $G(t)$ curve corresponds to different initial conditions, and  
%Therefore, 
if an initial condition $\boldsymbol{q}_0$ yields optimal energy amplification at a given time $t_{\operatorname{opt}}$, one can write
\begin{align}
    \exp{\left( \mathrm{-i\mathscr{L}t_{\operatorname{opt}}} \right)}\boldsymbol{q}_0=\|  \exp{\left( \mathrm{-i\mathscr{L}t_{\operatorname{opt}}} \right)} \| \boldsymbol{q}_{\operatorname{opt}},
\end{align}
where the initial condition (or input) $\boldsymbol{q}_0$ is advanced in time by $\exp{\left(-\mathrm{i}\mathscr{L}t_{\operatorname{opt}}\right)}$ producing the optimal output (response) vector $\boldsymbol{q}_{\operatorname{opt}}$ that is amplified by $\|  \exp{\left( \mathrm{-i\mathscr{L}t_{\operatorname{opt}}} \right)} \|$~\citep[see][]{Schmid2014analysis}. 

In numerous experimental studies for wall-bounded shear flows, streamwise streaks and rolls are observed~\citep[e.g., see][]{elofsson1999experiments,KITOH_NAKABYASHI_NISHIMURA_2005,Liu_Semin_Klotz_Godoy-Diana_Wesfreid_Mullin_2021,Tao2024liftup}. The nonmodal analysis of these flows also revealed that the initial perturbation field (optimal input) had streamwise rolls. These rolls then created streamwise streaks in the optimal perturbation field (optimal response), yielding optimal growth~\citep{reddy_henningson_1993,schmid_henningson_1994,NOUAR_KABOUYA_DUSEK_MAMOU_2007,LIU_LIU_2011,Schmid2014analysis}. For the classical PPF, optimal energy growth is realized in the vicinity of the $\beta$-axis, particularly for streamwise independent perturbations with $(\alpha,\,\beta) \approx (0,2)$~\citep{reddy_henningson_1993}.
Here, we identify an analogous phenomenon to the classical PPF when the mean permeability is minimal or when the anisotropy parameter is large. 
% The flow fields exhibiting the highest degree of transient growth are determined for various flow parameters (anisotropic permeability, depth ratio), and their impact on it are identified in the following paragraphs.
% The flow fields associated with the optimal transient growth are computed.
%The subsequent paragraphs 
In the following, we analyze the influence of  different flow parameters, such as anisotropic permeability and depth ratio, on the flow fields associated with optimal transient growth.

{ For parameters corresponding to figures~\ref{fig:growth_curves_Sigma2}(b) and~\ref{fig:growth_curves_Sigma2}(d), the optimal energy growth parameters  $\left(G_{\operatorname{opt}},\,t_{\operatorname{opt}}\right)$ are $(449.86,77.88)$ and $(475.26,73.27)$, respectively (see also 2nd and 4th rows of Table~\ref{table:transient_growth_values}). 
For these parameters, 
%We displayed 
the optimal perturbation (optimal initial condition and response) flow fields are illustrated in figure~\ref{fig:optimal_anisotropy} to understand the effects of 
% anisotropic permeability $(\sigma,\xi)$ 
the anisotropy parameter $\xi$. The figure presents the contours of optimal streamwise velocity component $\widetilde{u}$ superimposed with the optimal cross-stream components of velocity $(\widetilde{v},\widetilde{w})$
at $t=0$ (optimal input $\boldsymbol{q}_0$, left panels) together with $t=t_{\rm{opt}}$
(output response $\boldsymbol{q}_{\rm opt}$, right panels). 
% The amplitudes of the velocity components 
% % for different anisotropy parameters $\xi=1$ and 0.5, 
% are listed in the caption of figure~\ref{fig:optimal_anisotropy}. 
Similar to~\S\ref{subsec:response_external_force}, it is observed that (i) the streamwise disturbance velocity of optimal input is minimal compared to the normal and spanwise components, (ii) 
%Also, 
the normal and spanwise disturbance velocities of optimal output are very weak compared to the streamwise component, see the caption of figure~\ref{fig:optimal_anisotropy}. 
Hence, the perturbation kinetic energy is concentrated 
mainly along the cross-stream velocities for optimal input.
%in the transverse and spanwise velocities. 
In contrast, the streamwise velocity has a greater influence on the perturbation kinetic energy for optimal output. 
%From the figure~\ref{fig:optimal_anisotropy}(a), 
It is seen when $\xi=1$ optimal input comprises a spanwise periodic pattern of streamwise counter-rotating vortices (or rolls). The optimal response consists of spanwise alternative high and low-velocity streamwise streaks (panel a). 
Similar portraits are seen 
% as figure~\ref{fig:optimal_anisotropy}(a)
in cases of $(\sigma,\xi)=(0.006,1)$ and (0.006,0.5) (parameters corresponding to figures~\ref{fig:growth_curves_Sigma2}(a) and~\ref{fig:growth_curves_Sigma2}(c), figures not shown). However, these rolls and streaks become oblique in the spanwise direction for $(\sigma,\xi)=(0.02,0.5)$ (panel~b).
When comparing optimal output (right-side figures) in (a) and (b) panels, 
%the optimal output for $\xi=1$ and $\xi=0.5$ at $\sigma=0.02$, 
we may analyze the anisotropy parameter $\xi$,
% factors {\color{red}what factors?} 
leading to greater optimal growth in the latter case. It should be noted that in both $\xi$ cases, the amplitude of $\widetilde{u}$ ($=1.421$) remains unchanged. However, when $\xi = 0.5$, due to the oblique perturbations, the amplitudes of cross-stream velocity components $\widetilde{v}$ and $\widetilde{w}$ increase, resulting in a larger disturbance kinetic energy.
}

\begin{figure}[!htbp]
    \centering
    \includegraphics[scale=0.25]{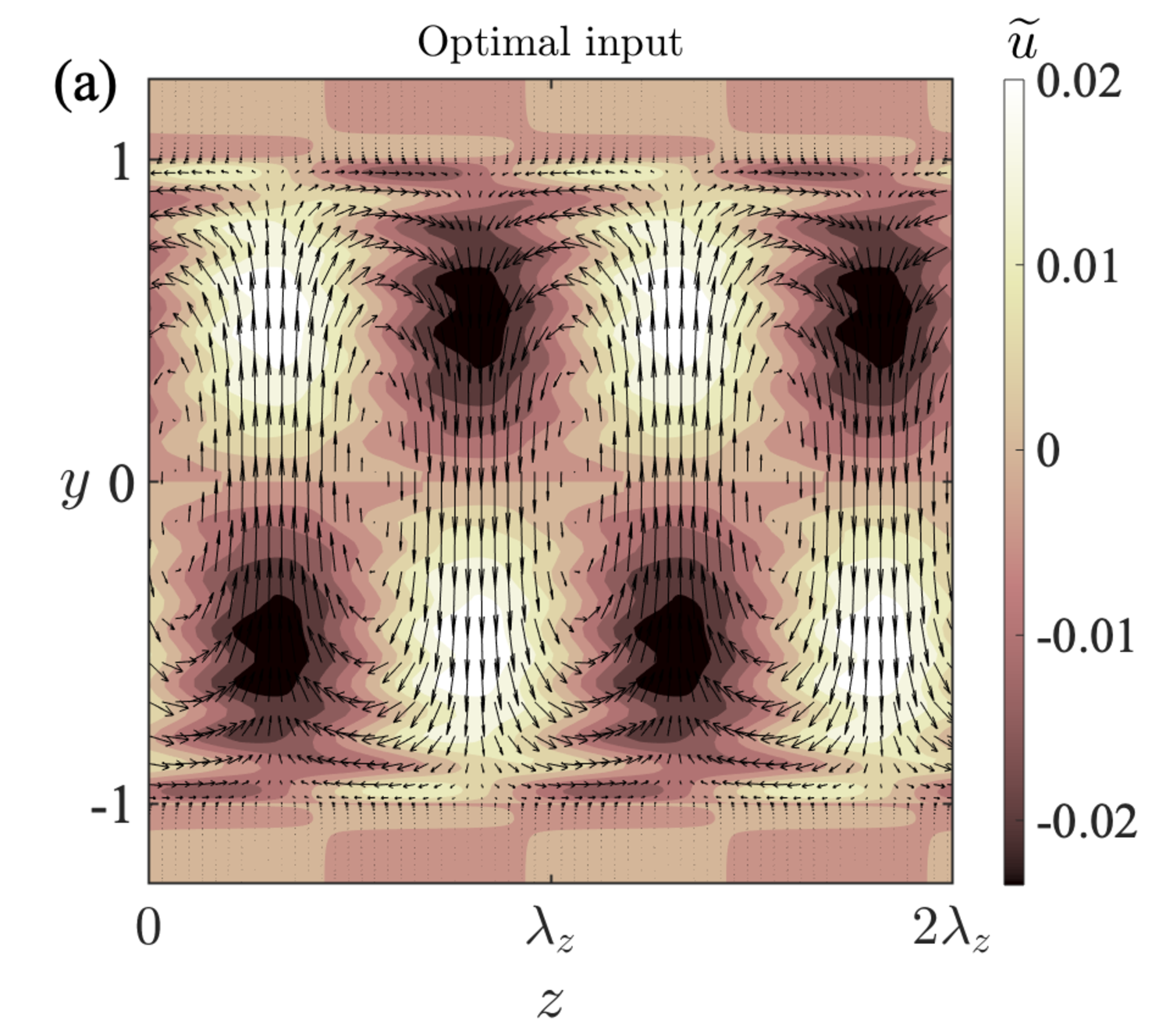}
    \includegraphics[scale=0.25]{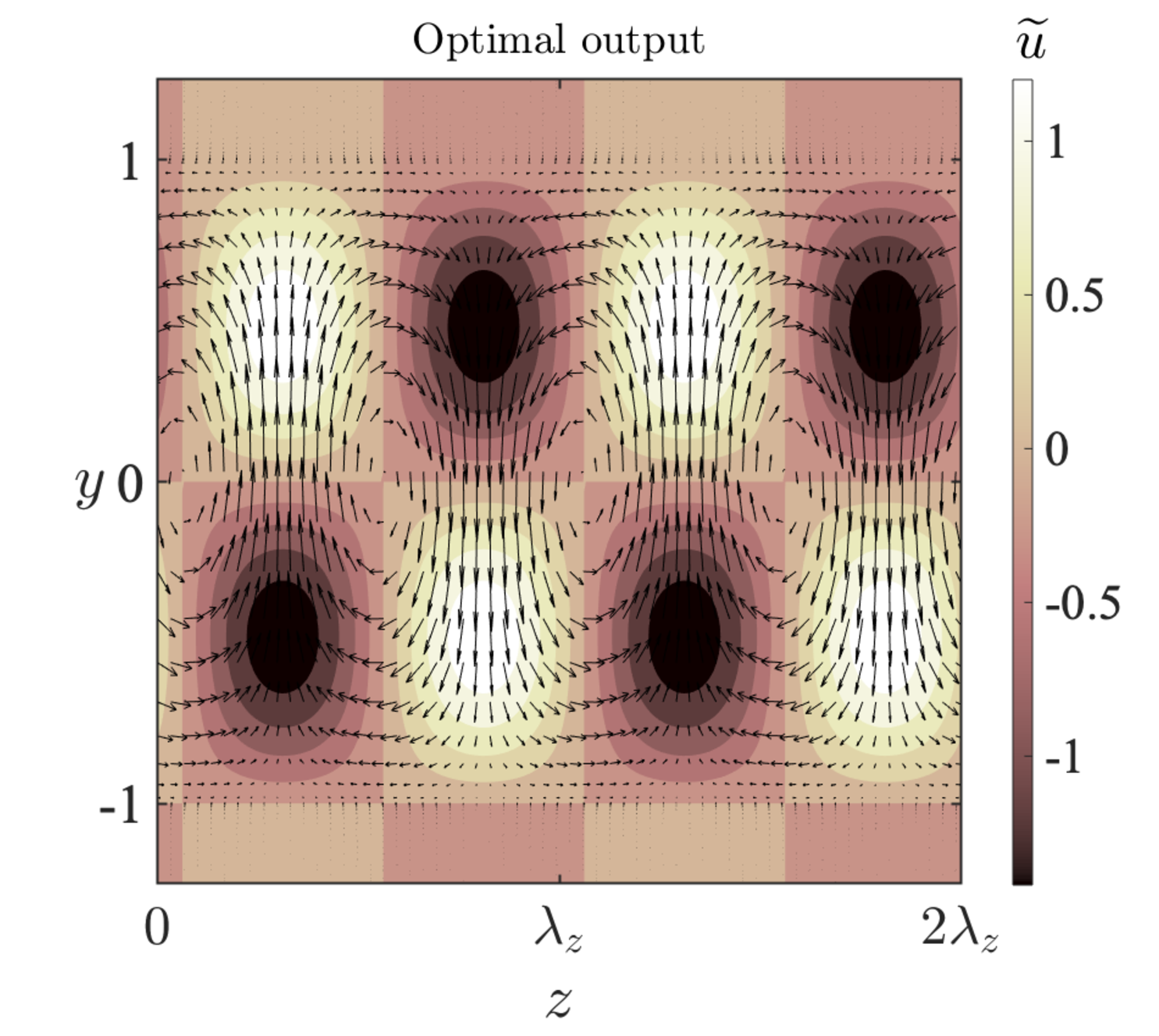}
    \includegraphics[scale=0.25]{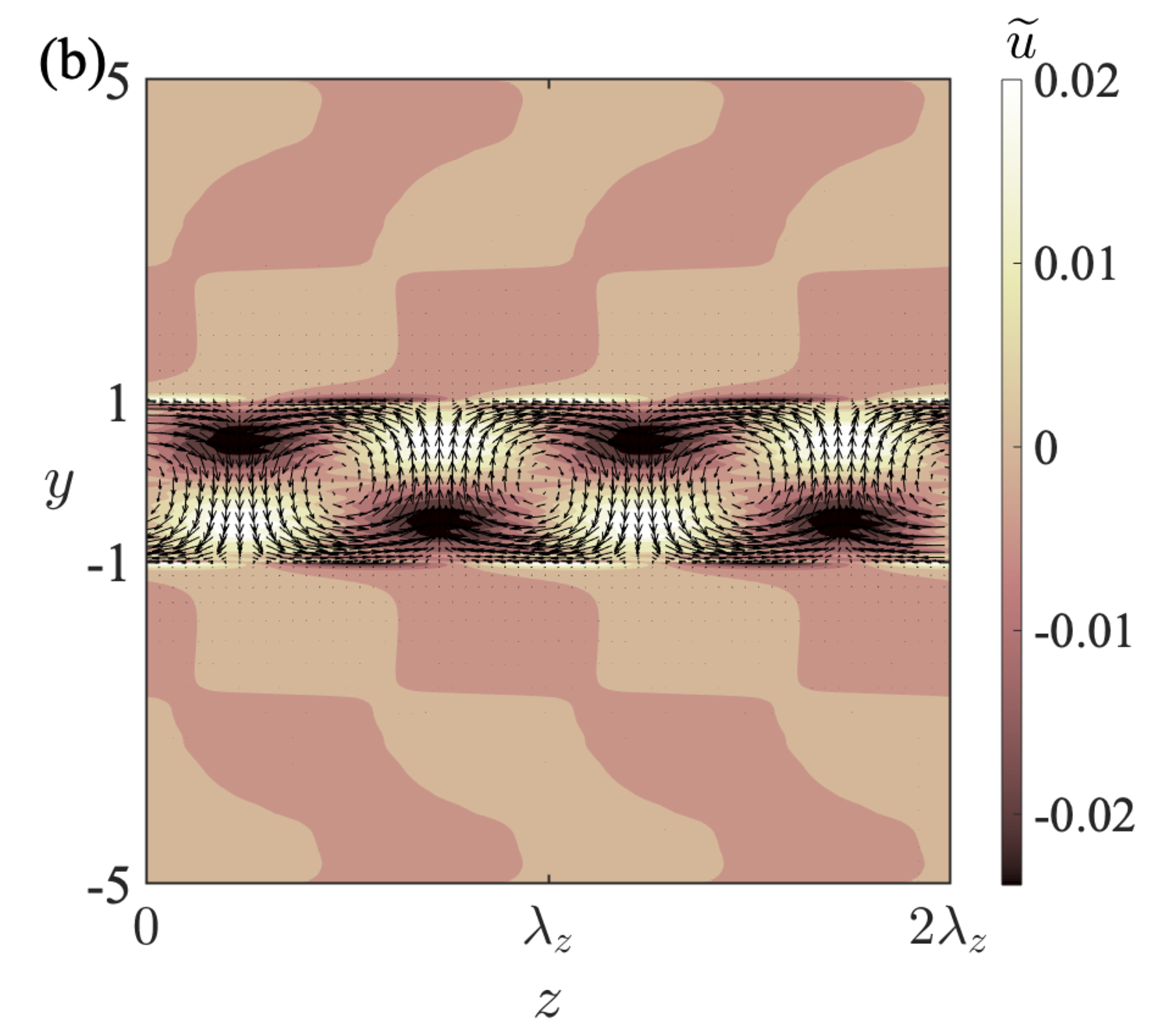}
    \includegraphics[scale=0.25]{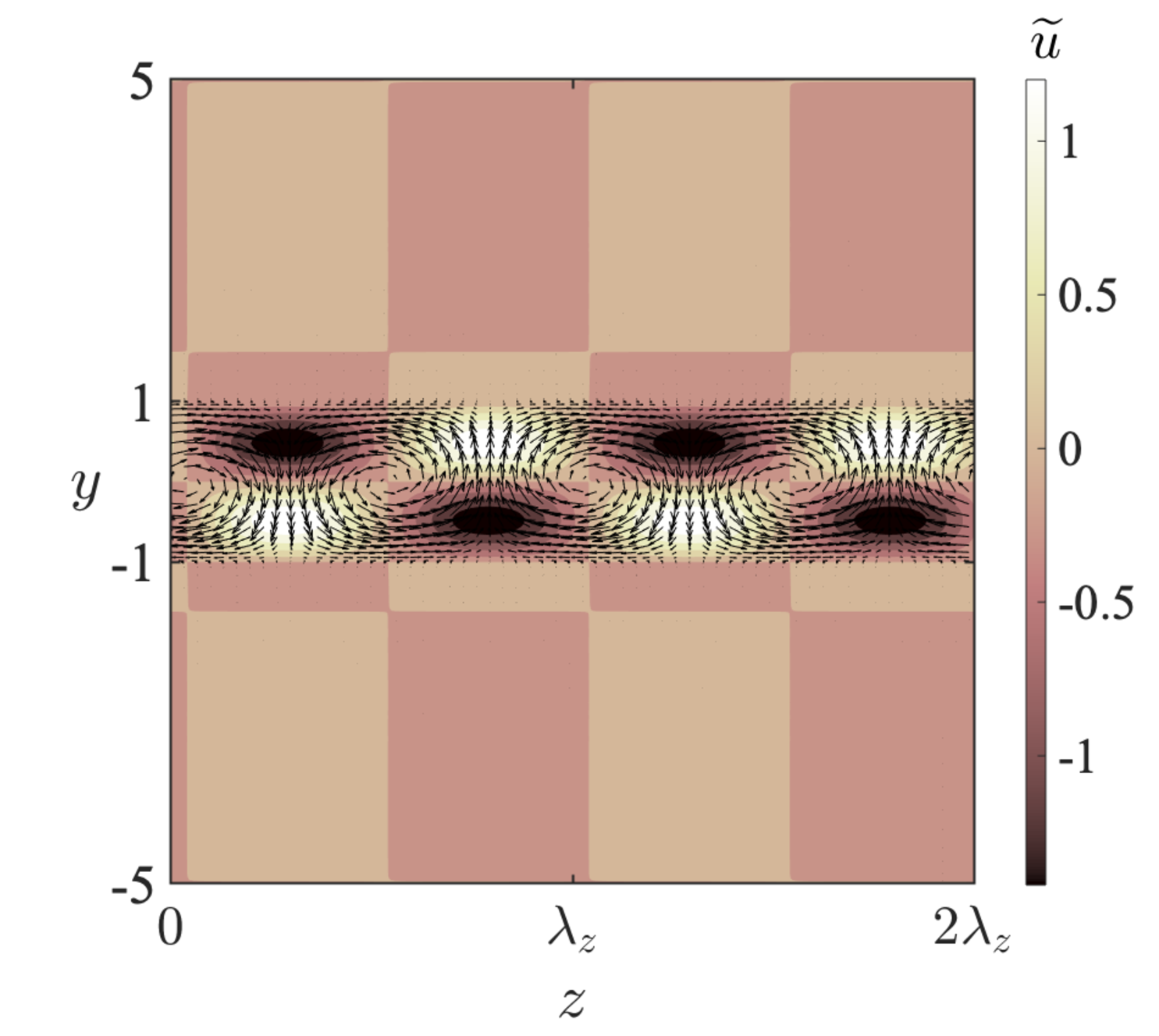}
    \caption{\small 
    {\small Same as figure~\ref{fig:optimal_anisotropy} but for (a) $d=1/8$, (b) $d=2$. } For panel~a: optimal input
    $(|\widetilde{u}|, |\widetilde{v}|, |\widetilde{w}|)$ =$(0.024, 1.049, 1.024)$, %$|\widetilde{v}|=1.051$ and $|w|=1.015$, 
   % while for the output are 
   and optimal output 
    $(|\widetilde{u}|, |\widetilde{v}|, |\widetilde{w}|)$=$(1.419, 0.025, 0.019)$.    
     For panel~b: optimal input
 $(|\widetilde{u}|, |\widetilde{v}|, |\widetilde{w}|)$ =$(0.024, 1.033, 1.011)$, and optimal output  $(|\widetilde{u}|, |\widetilde{v}|, |\widetilde{w}|)$ =$(1.421,0.025,0.019)$. See figures~\ref{fig:growth_curves_Depth}(a),~\ref{fig:growth_curves_Depth}(b), and 5th and 6th rows of Table~\ref{table:transient_growth_values}
    }
    \label{fig:optimal_depth}
\end{figure}

Increasing the anisotropic permeability (with higher $\sigma$ and lower $\xi$) leads to a higher velocity through the porous walls, 
resulting in a larger volume of fluid entering the porous layer and a significant exchange of flux at the interface between the porous and the fluid layers. Consequently, the amplitudes of $\widetilde{u},\,\widetilde{v}$ and $\widetilde{w}$ slightly enhanced, the optimum transient growth become more pronounced.

Let us see the effect of depth ratio on optimal profiles. Similar to figure~\ref{fig:optimal_anisotropy}, the optimal input and output perturbations are displayed in figures~\ref{fig:optimal_depth}(a) and~\ref{fig:optimal_depth}(b) for the parameters
same as figures~\ref{fig:growth_curves_Depth}(a) and~\ref{fig:growth_curves_Depth}(b), respectively. The corresponding values of optimal growth $G_{\operatorname{opt}}$, optimal time $t_{\operatorname{opt}}$ and optimal wavenumber $(\alpha_{\operatorname{opt}},\beta_{\operatorname{opt}})$ are listed in 5th and 6th rows of Table~\ref{table:transient_growth_values}. 
%(see 5th and 6th rows of Table~\ref{table:transient_growth_values}). 
Similar to $d=1$ (figure~\ref{fig:optimal_anisotropy}a), the optimal initial conditions, and optimal response are represented by
streamwise vortices and streaks, respectively.Comparing both panels, it is seen that the structure of the flow field remains unchanged even if 
the depth ratio is changed---rolls and streaks are not crooked due to the changes in the depth ratio from $d=\nicefrac{1}{8}$ to 2.  
For porous walls that are relatively thicker, with a thickness of $d \ge 1$, the rolls (and streaks) become oblique in the spanwise direction when the cross-stream permeability is increased ($\xi < 1$) (figure not shown). While the amplitude of cross-stream velocity ($\widetilde{v}$, $\widetilde{w}$) of optimal output remains the same with the change in depth ratio, streamwise velocity component $\widetilde{u}$ changes slightly (see figures~\ref{fig:optimal_anisotropy}a and~\ref{fig:optimal_depth}). 
Note that, $|\widetilde{u}|=1.419$ for $d=\nicefrac{1}{8}$ and $|\widetilde{u}|=1.421$ for $d=1$ and 2. Since optimal outputs (more specifically, $\widetilde{u}$ contours for straight streaks) are related to the optimal energy growth, this explains why $G_{\operatorname{opt}}$ is less for $d=\nicefrac{1}{8}$ and equal for $d=2$ compared to $d=1$.
%}

The results of~\S\ref{subsec:response_external_force} and~\S\ref{subsec:response_initialCond} demonstrate that the amplification of external forcing is more evident than the initial perturbations. Since the transient growth function $G(t)=\sup_{\boldsymbol{q}_0 \ne 0} \displaystyle{\nicefrac{\|\boldsymbol{q}(y,t)\|^2}{\|\boldsymbol{q}_0\|^2}}$ and the response function to external excitations $R(f)=\max_{\widehat{\boldsymbol{F}}_q \ne 0} \displaystyle{\nicefrac{\Vert \widehat{\boldsymbol{q}}_p\Vert }{\Vert \widehat{\boldsymbol{F}}_q \Vert }}$, it is %necessary to carry out a comparison between
more instructive to compare 
$|R(f)|^2$ and $G(t)$. As an example, when $(\sigma, \xi,d)=(0.006,0.5,1)$ for the flow at $Re=1000$, $\alpha=0$, $\beta=1$: $G_{\max} \approx 247.99$ and $|R_{\max}|^2 \approx 1.79 \times 10^7$. Furthermore, at any fixed Reynolds number in the subcritical range, $R_{\max}$ is much higher than $G_{\max}$ for wide ranges of wavenumbers.
These results indicate that sub-critical flows can experience significant amplifications and are more sensitive to external noise than initial disturbances.

% -----------------------------------------
\section{Instability mechanisms}
\label{sec:growth_mechanism}
% -----------------------------------
\subsection{Exponential growth}
% ----------------------------------
The perturbation kinetic energy budget analysis provides information on the physical mechanisms of modal instability~\citep{hooper_boyd_1983,BOOMKAMP1996,Lin_Chen1998themechanism}. Since Squire's theorem is valid in the present case, we consider only two-dimensional disturbances for the following energy budget analysis. The energy budget equations are obtained by taking the inner product of linearized momentum equations~\eqref{eqn:pert_channel}-\eqref{eqn:pert_bd_interface} with the disturbance velocity vector ($\widetilde{\boldsymbol{q}}$) and integrating over the transverse flow domain in each layer. The resulting equations in each layer are averaged over one axial wavelength of $\lambda_x=2\pi/\alpha$ in the streamwise direction, yielding the following disturbance kinetic energy balance equation,  
\begin{align}
  \frac{dE_V}{dt} =   \frac{d}{dt}\sum\limits_{j=0}^2 {\textsl{KE}}_j= \sum\limits_{j=0}^2 \textsl{REY}_j +  \sum\limits_{j=0}^2 D_j +  \sum\limits_{j=1}^2 {E}_j + E_s,
    \label{eqn:ke_balance}
\end{align}
where index $j=0$ denotes the fluid layer and $j=1, 2$ represents lower and upper porous layers; $\textsl{KE}_{j}$ 
represents the 
kinetic energy of the infinitesimal perturbations, given as
% Here $\textsl{KE}_{1,2}$ and $\textsl{KE}_0$ represent the 
% % rate of 
% kinetic energy of the infinitesimal perturbations in porous layers and the fluid layer
\begin{align}
    \begin{aligned}
         \textsl{KE}_j= \frac{1}{2\epsilon_j\lambda_x}\int_{0}^{\lambda_x} dx \int_{a_j}^{b_j}  
    \widetilde{u}_j^2+\widetilde{v}_j^2 \,\, dy, 
    %     \right],
    % \\
    \quad \,\,
        \textsl{KE}_{0}=\frac{1}{2 \lambda_x}\int_{0}^{\lambda_x} dx \int_{-1}^{1}  
        \widetilde{u}^2+\widetilde{v}^2  \,\,
        dy,
      %  \right].
    \end{aligned}
    \label{eqn:ke}
\end{align}
The right-hand side production terms 
$ \textsl{REY}_j$
%$P_{1,2}$ and $P_0$ 
signify the exchange of energy with the basic flow via Reynolds stresses
\begin{align}
    \textsl{REY}_j=\frac{1}{\epsilon_j^2\lambda_x}\int_{0}^{\lambda_x} dx \int_{a_j}^{b_j} 
   -\widetilde{u}_j\widetilde{v}_j\frac{d U_j}{d y} \, dy,
    %\right] ,
      \quad
    \textsl{REY}_{0}=\frac{1}{\lambda_x}
    \int_{0}^{\lambda_x} dx
   % \bigintsss_{0}^{\lambda_x} dx %\bigintsss_{-1}^{1} 
    \int_{-1}^{1} 
    -\widetilde{u}\,\widetilde{v}\frac{d U}{d y}\, dy,
   % \right], 
\end{align}
and $D_{j}$ represents viscous dissipation
\begin{align}
\begin{aligned}
D_{j} &=
    -\frac{1}{ \epsilon_jRe \lambda_x}\int_{0}^{\lambda_x} dx \int_{a_j}^{b_j}  
    2\left( \frac{\partial \widetilde{u}_j}{\partial x}\right)^2+ 2\left( \frac{\partial \widetilde{v}_j}{\partial y}\right)^2+\left(  \frac{\partial \widetilde{u}_j}{\partial y} + \frac{\partial \widetilde{v}_j}{\partial x}\right)^2 \, dy, 
    %\right] ,
    \\
D_0 &=-
\frac{1}{Re \lambda_x} %\bigintsss
\int_{0}^{\lambda_x} dx \int_{-1}^{1}   
2\left( \frac{\partial \widetilde{u}}{\partial x}\right)^2+ 2\left( \frac{\partial \widetilde{v}}{\partial y}\right)^2+\left(  \frac{\partial \widetilde{u}}{\partial y} + \frac{\partial \widetilde{v}}{\partial x}\right)^2  \,
dy.
%\right].
\label{eqn:dissipation}
    \end{aligned}
\end{align} %}
The terms $E_{j}$ and $E_s$ in~\eqref{eqn:ke_balance} represent the energy loss due to Darcy drag, and the work-done by shear stresses at porous-fluid interfaces,
\begin{align}
    E_j&=-\frac{1}{\sigma_j^2 Re}\frac{1}{\lambda_x}\int_{0}^{\lambda_x} \,dx \,\,\int_{a_j}^{b_j}   \widetilde{u}_j^2+\xi_{j1} \widetilde{v}_j^2\,\, dy ,\\
	E_s&=\frac{\tau_2}{\sigma_2 Re}\frac{1}{\lambda_x }\int_{0}^{\lambda_x}  \widetilde{u}_2^2 \big|_{y=1} \, dx +\frac{\tau_1}{\sigma_1 Re}\frac{1}{\lambda_x}\int_{0}^{\lambda_x}  \widetilde{u}_1^2 \big|_{y=-1} \, dx\nonumber\\
	& +\frac{1}{Re \lambda_x}%\bigintsss
 \int_{0}^{\lambda_x}  
 \left( \widetilde{u} \frac{\partial \widetilde{v}}{\partial x}-\frac{1}{\epsilon_2} \widetilde{u}_2 \frac{\partial \widetilde{v}_2}{\partial x}\right)\bigg|_{y=1}- \left( \widetilde{u} \frac{\partial \widetilde{v}}{\partial x}-\frac{1}{\epsilon_1} \widetilde{u}_1 \frac{\partial \widetilde{v}_1}{\partial x}\right)\bigg|_{y=-1}
 \,
 dx,
\end{align}
where integration limits $(a_1, b_1)=(-1-2d_1,-1)$ and $(a_2, b_2)=(1,1+2d_2)$. 
% Here $j=1$, $a_j=-1-2d_1$, $b_j=-1$ for the lower porous wall and $j=2$, $a_j=1$, $b_j=1+2d_2$ for the upper porous wall. 
Energy balance equation~\eqref{eqn:ke_balance} is used to determine the mechanism underlying instability, as explained below.

Note that~\eqref{eqn:ke_balance} can be reformulated 
with the help of eigenfunctions~\citep{karmakar2024linear}, by assuming disturbance quantities as linear combinations of wave-like modes $\widetilde{\boldsymbol{u}}_j=\operatorname{Re}\left\lbrace  \widehat{\boldsymbol{u}}_j(y) e^{\mathrm{i}\left( \alpha x -\omega t\right)} \right\rbrace$ and $\widetilde{\boldsymbol{u}}=\operatorname{Re}\left\lbrace  \widehat{\boldsymbol{u}}(y) e^{\mathrm{i}\left( \alpha x -\omega t\right)} \right\rbrace$ and then normalized by dividing $E_V$. 
After some algebra, it can be shown that the growth rate $\omega_i$ is indeed equal to the $\frac{1}{E_V}\frac{d E_V}{dt}$ when expressed in terms of eigenfunctions. 
Therefore, large magnitude terms of the normalized energy balance equation have a dominant effect in enhancing the growth rate of the corresponding mode.
The sign of these dominant terms indicates whether their effects on the flow are stabilizing (if negative) or destabilizing (if positive). 
The largest positive term in the energy equation identifies the source of instability.
The integrations are performed using the Gauss--Chebyshev quadrature formula to obtain numerical results.

Note that 
%It is seen from the table that
(i) viscous dissipation terms $D_{j}$'s and Darcy drag terms $E_{j}$'s turn out negative  and thus exhibit stabilizing effects, and (ii) due to the symmetric base velocity with $\tau=0$, work done by shear stress $E_s \approx 0$. 
Moreover, $E_s$ 
%and %the value of 
%$E_s$ 
is barely affected by 
the physical parameters. 
%variations in the free parameters (such as the mean permeability $\sigma$, the anisotropy parameter $\xi$, and the depth ratio $d$) considered in this study. 
To what follows, the main factor determining the stability or instability of the flow is the energy production terms $\textsl{REY}_{j}$. Energy is transmitted from the base flow to the disturbance through the Reynolds stress, increasing the kinetic energies throughout all layers and accelerating the growth rates $\omega_i$ of the instability modes.

\begin{table}
    \centering
    \begin{tabular}{@{} lllllllllll@{}}
    \hline
       $Re_c$ & $\alpha_c$ & $\omega_i$ & $\partial_t (\textsl{KE}_j)$ & $\partial_t(\textsl{KE}_0)$ & $\textsl{REY}_j$ & $\textsl{REY}_0$  & $D_j$ & $D_0$  & $E_j$ & $E_s$\\
       \hline
      % \toprule
        2977.1 & 0.4& 0.0000 & 0.0000 &  0.0000 & 0.0026  &  0.0949    &-0.0019 &  -0.0133   & -0.0414  &  0.0000 \\
         833.45 & 1.22 & 0.0000 & 0.0000 & 0.0000 & 0.0005  &  0.0502 & -0.0015  & -0.0197
         & -0.0143  &  0.0000 \\
         \hline
         %\\
        % \bottomrule
    \end{tabular}
    \caption{The representations of different energy distributions for two distinct modes, illustrated in figure~\ref{fig:temporal_growth}(i).}
    \label{table:energy_modes}
\end{table}

The subsequent discussion aims to comprehend the bi-modal instability of fluid modes captured in figure~\ref{fig:temporal_growth}(i) and examine the impact of control parameters on the growth rate of the least stable fluid mode using the energy balance equation. It is seen in figure~\ref{fig:temporal_growth}(i) that there are two distinct modes of instability in the intermediate- and short-wavenumber regimes 
marked by 
`$(i)$' and `$(ii)$' pointing at critical points with $(Re,\,\alpha)= (2977.1,\,0.4)$ and $(833.45,\,1.22)$, respectively. 
The energy budgets associated with the points labeled `$(i)$' and `$(ii)$' are listed in Table~\ref{table:energy_modes}. 
As expected, production terms $\textsl{REY}_{j}$ alone 
positively accelerate the growth rate $\omega_i$ for both modes. We classify these fluid modes as mode~1 and mode~2 based on the characteristics of the eigenfunctions and secondary flow patterns.
For both modes, $\textsl{REY}_j<\textsl{REY}_0$, i.e., the energy production term in the fluid layer is much higher than the porous layers, which confirms the terminology of using fluid mode is appropriate.

Furthermore, it is observed in \S\,\ref{subsec:modal_stability} that as the mean permeability $\sigma$ and depth ratio $d$ increase and the anisotropy parameter $\xi$ decreases, the growth rate of the least stable mode likewise increases, leading to the destabilization of the flow. The rationale for these findings may be elucidated through the energy budget analysis, as altering these control parameters  amplify the energy production terms $\textsl{REY}_{j}$ and intensify the fluid layer mode.

\subsection{Transient growth}
% ---------------------------------------------------------

This section discusses various mechanisms %underlying 
to explain transient growth. \cite{Ellingsen1975Stability} showed that in the case of inviscid shear flows, the streamwise component of perturbation velocity grows linearly with time. This mechanism is commonly called {\it lift-up}, where streamwise vortices can induce the growth of streamwise velocity streaks~\citep{butler1992_three,reddy_henningson_1993,threfeten1993Hydrodynamic}. For viscous shear flows, the streamwise perturbation velocity increases linearly with time due to the lift-up effect before viscous dissipation dampens it~\citep{farrell1993optimal,butler1992_three,threfeten1993Hydrodynamic}. In addition to lift-up mechanisms, the shear flows exhibit a two-dimensional Reynolds stress mechanism known as the Orr mechanism~\citep{pedlosky2013geophysical}. While investigating the optimal excitation of 3D disturbances in the viscous constant shear flow, ~\citep{farrell1993optimal} suggested that the Orr mechanism involves the amplification of upstream tilting waves caused by the down-gradient Reynolds stress mechanism of two-dimensional shear instability. The authors also found that the lift-up and the Orr mechanisms can explain the optimal perturbation flow structures. Note that for streamwise independent disturbances ($\alpha = 0, \beta \ne 0$), the lift-up mechanism only contributes to the transient growth, giving rise to straight rolls and streaks. The combined effects of the lift-up and the Orr- mechanisms trigger transient growth in the case of oblique disturbances ($\alpha \ne 0,\,\beta \ne 0$), where the rolls and streaks are tilted. The interplay of both lift-up and Orr- mechanisms contribute to the transient growth in the present porous-fluid-porous PPF; see figures~\ref{fig:opti_response_structure},~\ref{fig:optimal_anisotropy} and~\ref{fig:optimal_depth}, which demonstrates alternating bands of streamwise velocity streaks caused by streamwise vortices.

It is  worth noticing that, despite considering various depth ratios $d$, the optimal perturbation has the same shape as that reported by~\citet{reddy_henningson_1993}, when $\sigma$ is small and $\xi$ is large [see figure~\ref{fig:optimal_anisotropy}a]. This is due to the fact that this combination of anisotropic permeability permits a lesser volume of fluid to enter the porous walls, making them nearly rigid. 
In the presence of such a streamwise independent perturbation, it is well known that the lift-up mechanism predominates transient growth. However, at larger $\sigma$ and smaller $\xi$, the shape of the perturbation becomes oblique [see figures~\ref{fig:opti_response_structure} and~\ref{fig:optimal_anisotropy}b]. As previously discussed, when the perturbation is oblique, both lift-up and Orr mechanisms influence the transient growth simultaneously.

Let us conclude by noting that~\citet{Waleffe1997on} postulated a nonlinear self-sustaining process (SSP) for the dynamics of flow patterns in shear flows. As a result of the lift-up effect, streamwise vortices that move fast fluid from the centre to the walls and simultaneously transfer slow fluid from the walls to the centre form streaks of high and low velocity. The spanwise inflections cause a wake-like instability in the streaks, which modulates the streamwise flow. The non-linear self-interaction of this streamwise modulated flow regenerates the streamwise vortices. This process leads to self-sustained 3D travelling waves, and the most significant structural characteristics of these waves are the streamwise vortices and streaks. Some experimental studies have observed this traveling wave structure (e.g.,~\citet{hof2004experimental}, for turbulent pipe flow;~\citet{Liu_Semin_Klotz_Godoy-Diana_Wesfreid_Mullin_2021,Tao2024liftup}, for plane Couette-Poiseuille flow).

% --------------------------------------------
\section{Conclusion and discussion}
\label{sec:conclusion}
% --------------------------------------------

This study has explored how instability causes changes in the flow dynamics of a viscous plane Poiseuille flow within a rigid, anisotropic porous channel. The present investigation has provided a comprehensive study of linear stability using modal analysis, the energy method, and nonmodal analysis, with a focus on how the anisotropic permeability and porous wall thickness affect the stability characteristics of the flow.

The modal stability analysis has been used to investigate the long-time behavior of the disturbances in the flow. Bi-modal instability has been found for specific parameter regimes when the flow is subjected to two-dimensional infinitesimal perturbation. The two instability modes responsible for bi-modal instability turn out to be fluid modes. One fluid mode is observed in the short wavenumber region and the other in the intermediate wavenumber region. The primary instability mode, occurring in the short-wave mode, significantly influences the long-term behavior of the flow as the perturbation amplitude grows or decays according to the growth rate of this mode. The present results have revealed that the mean permeability and thickness of the porous walls exert a destabilizing effect. At the same time, the anisotropy parameter acts as a stabilizing factor, enhancing the growth rate of the least stable fluid mode. Furthermore, an energy budget analysis has been studied to comprehend the physical mechanism of modal instability. It has been found that the energy production term transfers energy from the base flow to the disturbance via the Reynolds stress, enhancing the disturbance kinetic energies for the fluid and porous layers and accelerating the growth rate of the least stable modes. 

The energy stability has been analyzed, which gives the condition for no energy growth. The critical energy Reynolds number $Re_e$, below which there is no energy growth, has been identified by determining the value of $Re$ at which the numerical abscissa becomes positive. Regardless of any anisotropic permeability $\sigma$, $\xi$, and thickness ratio $d$, it has been observed that streamwise independent disturbances are more unstable than spanwise independent disturbances. As $\sigma$ or $d$ increases and $\xi$ decreases, the flow becomes more unstable. The disparity between the critical energy Reynolds number and the critical Reynolds number from modal analysis indicates transient energy amplification, prompting a nonmodal stability analysis. 

The responses to external excitation and initial conditions, characterized by response function $R(f)$ and growth function $G(t)$, respectively, have been studied in detail. For any combinations of anisotropic permeability $(\sigma,\,\xi)$ and thickness of the porous walls $d$, the maximum response to external excitation has been realized at zero frequency. For any fixed forcing frequency $f$, it has been shown that enhancing $\sigma$ or $d$ and decreasing $\xi$ amplify the optimal response $R_{\operatorname{opt}}$. The optimal wavenumber corresponding to the optimal response has been shifted from the long-wave to the short-wave regime as the forcing frequency is enhanced. Furthermore, we have determined the optimal flow structure subject to the optimal applied force, which verifies that the amplifications result from a combination of the lift-up and the Orr mechanisms.

A parametric study has been performed to understand the effect of physical parameters on the transient growth $G(t)$ of initial conditions. Present results have revealed that when the Reynolds number $Re$ falls in $(Re_e,Re_c)$, the disturbances experience a transient growth and then an exponential decay $\left(1<G_{\max}<\infty\right)$.~\citet{reddy_henningson_1993} reported that the optimal disturbance for classical PPF is streamwise-independent with $\beta\approx 2$. Our analysis has confirmed that this result holds even in the current flow setup but for small $\sigma$, large anisotropy parameter, or small depth ratio $d$. It has also been observed that simultaneously increasing $\sigma$ and decreasing $\xi$ reduces the wall-blocking effect, enhancing optimal transient growth. This enhancement has been obtained at slightly oblique disturbances with small streamwise wavenumbers. The energy growth of the initial perturbations has been explained through the lift-up effect and the Orr mechanism. Additionally, it has been demonstrated that the flow is more sensitive to external stimuli than initial conditions.

There are several significant differences between the present study and previous investigations by~\citet{tilton2006destabilizing,tilton2008linear,karmakar2022stability,Karmakar2023instability} on the stability of PPF in a confined multi-layer porous channel. The primary difference lies in the scope of the current research, which includes nonmodal analysis and energy methods in addition to the conventional modal analysis. Nonmodal stability is crucial for examining the amplification of short-term perturbation kinetic energy in bounded flows due to the non-normal nature of the associated operator~\citep{schmid2001stability,reddy_henningson_1993}. As previously mentioned, modal analysis cannot accurately predict the occurrence of sub-critical transitions in these specific circumstances~\citep{schmid2001stability}. Furthermore, the physical mechanisms underlying the various modes of instability in modal stability
have not been well reported to the authors' knowledge. An energy
budget analysis has been conducted in the present work to address this gap.

It is important to note that the velocity profile is symmetric with respect to the origin when the depths of both porous walls are identical $d_1=d_2=d$. ~\cite{tilton2008linear,Karmakar2023instability} have shown that the symmetric channel geometry is more stable than the non-symmetric geometric channel ($d_1 \ne d_2$). We expect similar behaviors for the non-modal stability, verification of which has been left for the future. 
Furthermore, the present study has focused on a constant applied pressure gradient throughout the channel, resulting in a simplified linear base pressure profile. However, this assumption oversimplifies real-world situations, especially those involved in porous-fluid-porous systems with flexible or wavy porous walls, where a linear profile may not be sustainable. Recognizing that both modal and non-modal stability analyses under non-linear pressure profile variations will be conducted in the future. 
This theoretical work also motivates the experimental investigation of the PPF in a confined porous channel. Conducting such experiments would contribute to the verification of present theoretical and numerical predictions. While the current work focuses on iso-thermal PPF in a homogeneous porous medium, it is essential to note that our investigation is not restricted to this condition and can be readily extended to non-iso-thermal PPF in a heterogeneous porous medium. Future studies will explore these extended conditions to provide a more comprehensive understanding of the multi-layer porous PPF dynamics.

\section*{Acknowledgement}
P.S. acknowledges the financial support from SERB-DST through Grant Nos.~CRG/2023/002914 and MTR/2023/000892.

% ---------------------------------------------------------------
\appendix
% --------------------------------------------------

% ---------------------------------------------------------
\section{Transient growth and optimal perturbation}
\label{appn:TransientGrowth_and_optimal}
% ---------------------------------------------------------------------
\subsection{Kinetic energy density}
The kinetic energy $E_V$ of the perturbation $\widetilde{\boldsymbol{X}}$ is given by
\begin{align}
    E_V(t)= \frac{1}{2}\int_{V} \widetilde{\boldsymbol{X}}\cdot \widetilde{\boldsymbol{X}}\, dV,
\end{align}
where $V$ is the volume of the PFP flow geometry. If the Fourier mode of $\widetilde{\boldsymbol{X}}$, as defined in~\eqref{eqn:Fourier_mode}, is substituted, using continuity equations the expression for the disturbance kinetic energy in terms of our variables of interest ($\boldsymbol{q}$) is
\begin{align}
    E_V(t)=\frac{1}{2}\int_{\alpha}\int_{y}\int_{\beta}  \mathcal{E}_V(\boldsymbol{q};\alpha,\beta)\, d\alpha\, dy\, d\beta.
\end{align}
% In the context of non-modal stability theory, the energy norm is the physically relevant quantity for quantifying the growth of the disturbance. 
% For a given Fourier mode, the instantaneous disturbance kinetic energy is given by~\citep{gustavsson_1991}
where
\begin{align}
    \mathcal{E}_V(\boldsymbol{q};\alpha,\beta)=\frac{1}{k^2} & \left[ \int_{-1-2d_1}^{-1}  | D\overline{v}_1|^2+k^2|\overline{v}_1|^2+|\overline{\eta}_1|^2 \, dy+
    \int_{-1}^{1} | D\overline{v}|^2+k^2|\overline{v}|^2+|\overline{\eta}|^2 \, dy \right. \nonumber \\
    & \left. \hspace{4.5cm}+\int_{1}^{1+2d_2}  | D\overline{v}_2|^2+k^2|\overline{v}_2|^2+|\overline{\eta}_2|^2 \, dy\right].
\end{align}
% where $\boldsymbol{q}=\left( \overline{v}_1 , \overline{\eta}_1, \overline{v}, \overline{\eta}, \overline{v}_2, \overline{\eta}_2\right)^{\top}$. 
is commonly referred to as kinetic energy density. See~\citet{gustavsson1986excitation,gustavsson_1991} for more detail.

% -----------------------------------------------------------------------
\subsection{Energy inner product and energy norm}
% -----------------------------------------------------------------------
The methodology employed in this study to analyze the transient evolution of a perturbation is based on the works of~\citet{reddy_henningson_1993,schmid_henningson_1994} in case of classical plane Poiseuille and Hagen-Poiseuille flow, respectively.
% The transient evolution of a perturbation in the linear regime is determined by following the methodology of~\citet{reddy_henningson_1993,schmid_henningson_1994}. 
We introduce a scalar product based on the energy density
\begin{align}
    \left\langle \boldsymbol{q}_l,\boldsymbol{q}_m\right\rangle=&\int_{-1-2d_1}^{-1} \big(D{\overline{v}_1^l}^{\ast}D{\overline{v}_1^m}+k^2{\overline{v}_1^l}^{\ast} {\overline{v}_1^m} + {\overline{\eta}_1^l}^{\ast} {\overline{\eta}_1^m}\big)\, dy
    +\int_{-1}^{1} \big(D{\overline{v}^l}^{\ast}D{\overline{v}^m}+ k^2{\overline{v}^l}^{\ast} {\overline{v}^m}\nonumber \\ 
    & \hspace{2cm} + {\overline{\eta}^l}^{\ast} {\overline{\eta}^m}\big)\, dy + \int_{1}^{1+2d_2} \big(D{\overline{v}_2^l}^{\ast}D{\overline{v}_2^m}+k^2{\overline{v}_2^l}^{\ast} {\overline{v}_2^m} + {\overline{\eta}_2^l}^{\ast} {\overline{\eta}_2^m}\big)\, dy,
\end{align}
where $\ast$ denotes the complex-conjugate.  Hence, the corresponding norm, commonly known as the energy norm, can be expressed as
% \begin{align}
%     \| \boldsymbol{q} \|=\sqrt{\left\langle \boldsymbol{q},\boldsymbol{q} \right\rangle}.
%     \label{eqn:energy_norm}
% \end{align}
\begin{align}
    \| \boldsymbol{q}\|^2=\left\langle \boldsymbol{q}, \boldsymbol{q} \right\rangle
    &= \int_{-1-2d_1}^{-1}  \big(| D\overline{v}_1|^2+k^2|\overline{v}_1|^2+|\overline{\eta}_1|^2 \big) \,dy+
    \int_{-1}^{1}  \big(| D\overline{v}|^2+k^2|\overline{v}|^2+|\overline{\eta}|^2 \big)\,dy \nonumber \\
    & \hspace{5 cm}+\int_{1}^{1+2d_2}  \big(| D\overline{v}_2|^2+k^2|\overline{v}_2|^2+|\overline{\eta}_2|^2 \big) \,dy \nonumber \\
    & = \int_{-1-2d_1}^{1+2d_2}\, \boldsymbol{q}^{\mathcal{H}} \mathcal{M} \boldsymbol{q} \, dy,
    \label{eqn:energy_norm}
\end{align}
where $\mathcal{M}=\operatorname{diag}{\left( k^2-D^2,\, 1,\, k^2-D^2,\,1,\, k^2-D^2,\,1\right)}$ and $\mathcal{H}$ is complex-conjugate transpose.

% -----------------------------------------------------------------------
\subsection{Procedure of finding the transient growth function \texorpdfstring{$G(t)$}{Lg}}
% -----------------------------------------------------------------------
Let $\lbrace \omega_j \rbrace$ are collections of eigenvalues that have been arranged in descending order based on their imaginary parts
and $\left\lbrace \widehat{q}_j\right\rbrace$ are corresponding eigenfunctions of $\mathscr{L}$. The solution of~\eqref{eqn:initial_value_problem} can be written as,
\begin{align}
    \boldsymbol{q}(y,t)=\sum_{j} \kappa_j^0 \exp{\left(-\mathrm{i} \omega_j t\right)} \widehat{q}_j(y)= \sum_{j} \kappa_j(t) \widehat{q}_j(y).
\end{align}
As $\operatorname{Im}{\left( \omega_j \right)} \ll 0$ for sufficiently large $j$, the term  $\kappa_j(t)  \widehat{q}_j(y)$ will be negligible for $t>0$. Hence for sufficient large $K$, we have
\begin{align}
    \boldsymbol{q} \approx \boldsymbol{q}_K(y,t)= \sum_{j=1}^K \kappa_j(t) \widehat{q}_j(y).
    \label{eqn:ivp_soln}
\end{align}
Now by assuming $\widehat{\boldsymbol{q}}_K(y)=\left[  \widehat{q}_1(y), \widehat{q}_2(y), \cdots, \widehat{q}_K(y) \right]$ and, 
\[\boldsymbol{\kappa}(t)=[\kappa_1(t),\kappa_2(t),\cdots,\kappa_K(t)]^\top=\mbox{diag}\left( e^{-\mathrm{i}\omega_1t}, e^{-\mathrm{i}\omega_2t},\cdots, e^{-\mathrm{i}\omega_K t} \right) \left[\kappa_1^0,\kappa_2^0,\cdots,\kappa_K^0 \right]^\top,\]
equation~\eqref{eqn:ivp_soln} can be recast as
\begin{align}
    \boldsymbol{q}_K(y,t)=\widehat{\boldsymbol{q}}_K(y) \boldsymbol{\kappa}(t).
    \label{eqn:ivp_soln_compact}
\end{align}
The energy norm of $\boldsymbol{q}_K$ by using~\eqref{eqn:ivp_soln_compact} and~\eqref{eqn:energy_norm}, can be expressed as
\begin{align}
    \| \boldsymbol{q}_K(y,t) \|^2=&\int_{-1-2d_1}^{1+2d_2} \boldsymbol{q}_K^{\mathcal{H}} \mathcal{M} \boldsymbol{q}_K\, dy =  \int_{-1-2d_1}^{1+2d_2} \left( \widehat{\boldsymbol{q}}_K \boldsymbol{\kappa} \right)^{\mathcal{H}} \mathcal{M}  \left( \widehat{\boldsymbol{q}}_K \boldsymbol{\kappa} \right)\, dy \nonumber \\
    =& \boldsymbol{\kappa}^{\mathcal{H}} \left( \int_{-1-2d_1}^{1+2d_2} \widehat{\boldsymbol{q}}_K^{\mathcal{H}} \mathcal{M} \widehat{\boldsymbol{q}}_K\, dy  \right) \boldsymbol{\kappa}=\boldsymbol{\kappa}(t)^{\mathcal{H}} \mathcal{A} \,\boldsymbol{\kappa}(t),
    \label{eqn:ivp_soln_norm}
\end{align}
where
\begin{align}
    \mathcal{A}&=\int_{-1-2d_1}^{1+2d_2} \widehat{\boldsymbol{q}}_K^{\mathcal{H}} \mathcal{M} \widehat{\boldsymbol{q}}_K \; dy =
    \bigints_{-1-2d_1}^{1+2d_2} \begin{bmatrix}
        \widehat{q}_1^{\mathcal{H}} \\  \vdots \\ \widehat{q}_K^{\mathcal{H}}
    \end{bmatrix} 
    \begin{bmatrix}
        \mathcal{M}\widehat{q}_1 & \cdots & \mathcal{M}\widehat{q}_K
    \end{bmatrix}\, dy
  \nonumber  \\
    &=\begin{bmatrix}
        \int_{-1-2d_1}^{1+2d_2} \widehat{q}^{\mathcal{H}}_1\mathcal{M}\widehat{q}_1\; dy  & \cdots & \int_{-1-2d_1}^{1+2d_2} \widehat{q}^{\mathcal{H}}_1\mathcal{M}\widehat{q}_K \;dy \\
    \vdots & \ddots & \vdots\\
        \int_{-1-2d_1}^{1+2d_2} \widehat{q}^{\mathcal{H}}_K \mathcal{M}\widehat{q}_1\; dy & 
        \cdots & \int_{-1-2d_1}^{1+2d_2} \widehat{q}^{\mathcal{H}}_K \mathcal{M}\widehat{q}_K\; dy
    \end{bmatrix} =
  \begin{bmatrix}
    \langle \widehat{q}_1,\widehat{q}_1 \rangle  & \cdots &  \langle\widehat{q}_1,\widehat{q}_K \rangle \\
   \vdots & \ddots & \vdots \\
 \langle \widehat{q}_K,\widehat{q}_1 \rangle  & \cdots &  \langle\widehat{q}_K,\widehat{q}_K \rangle 
     \end{bmatrix}.
    \label{eqn:A_innerproduct_matrix}
\end{align}
In other words $\mathcal{A}=[a_{ij}]$ where $a_{ij}=\langle \widehat{q}_i, \widehat{q}_j \rangle= \int_{-1-2d_1}^{1+2d_2} 
\widehat{q}^{\mathcal{H}}_i \mathcal{M}\widehat{q}_j\; dy$.  
Since $\mathcal{A}$ is Hermitian ($\mathcal{A}^{\mathcal{H}}=\mathcal{A}$), can be decomposed as $\mathcal{A}=\mathcal{F}^{\mathcal{H}} \mathcal{F}$, where $\mathcal{F}$ is a matrix of order $K$. From~\eqref{eqn:ivp_soln_norm}
\begin{align}
    \| {\boldsymbol{q}}_K(y,t) \|^2&=\boldsymbol{\kappa}^{\mathcal{H}}(t) \mathcal{F}^{\mathcal{H}} \mathcal{F} \boldsymbol{\kappa}(t)=\left( \mathcal{F} \boldsymbol{\kappa}(t) \right)^{\mathcal{H}} \left( \mathcal{F} \boldsymbol{\kappa}(t) \right) %\nonumber\\
   =\|\mathcal{F} \boldsymbol{\kappa}(t) \|_2^2,
    \label{eqn:energy_growthtwo}
\end{align}
where the subscript 2 denotes the $2$-norm.
From~\eqref{eqn:growth_function} using~\eqref{eqn:energy_growthtwo} the growth function $G(t)\approx G_K(t)$ where
\begin{align}
   % G(t)&\approx 
   G_K(t)=\sup_{\boldsymbol{\kappa}(0)\ne 0} \frac{\| \mathcal{F} \boldsymbol{\kappa}(t) \|_2^2}{\| \mathcal{F} \boldsymbol{\kappa}(0)\|_2^2} &=\sup_{\boldsymbol{\kappa}(0)\ne 0} \frac{\| \mathcal{F} \exp{(-\mathrm{i} \boldsymbol{\Omega}_K t)\boldsymbol{\kappa}(0)} \|_2^2}{\| \mathcal{F} \boldsymbol{\kappa}(0)\|_2^2} \nonumber \\ 
  & =  \sup_{\mathcal{F}\boldsymbol{\kappa}(0)\ne 0} \frac{\| \mathcal{F} \exp{(-\mathrm{i} \boldsymbol{\Omega}_K t)  \mathcal{F}^{-1} \mathcal{F}\boldsymbol{\kappa}(0)} \|_2^2}{\| \mathcal{F} \boldsymbol{\kappa}(0)\|_2^2}
    \nonumber \\
   & =\| \mathcal{F} \exp{(-\mathrm{i} \boldsymbol{\Omega}_K t)} \mathcal{F}^{-1} \|_2^2,
\end{align}
where $ \boldsymbol{\Omega}_K=\textnormal{diag}{\left(\omega_1,\omega_2,\cdots,\omega_K\right)}$. 
Therefore, in order to calculate the growth function G(t), it is sufficient to have the eigenvalues $\left\lbrace  \omega_j\right\rbrace$ and the corresponding eigenfunctions $\left\lbrace \widehat{q}_j \right\rbrace$ of the operator $\mathscr{L}$.

% -------------------
%   Bibloography
% ---------------------
% \bibliographystyle{rspublicnatwithsort_implicitdoi}
%\bibliographystyle{apsrev-titles.bst}
%\bibliographystyle{apsrev4-1}
%\bibliographystyle{plainnat}
\bibliography{refer}

\end{document}